\documentclass[11pt,a4paper]{article}
\pdfoutput=1
\usepackage{jheppub}
\usepackage{amsmath}
\usepackage{hyperref}
\usepackage{amssymb}
\usepackage{mathtools}
\usepackage{mathrsfs}
\usepackage{amsfonts}
\usepackage{bbm}
\usepackage{enumerate}
\usepackage{array}
\usepackage{booktabs}
\usepackage{textcomp}
\usepackage{caption}
\usepackage{cancel}
\usepackage{units}
\usepackage{multirow}
\usepackage{longtable}
\usepackage{relsize}
\usepackage{pdflscape}

\usepackage[latin1]{inputenc}
\usepackage{boxedminipage}
\usepackage{enumerate}
\usepackage[usenames,dvipsnames,table]{xcolor}
\usepackage{amssymb,latexsym}
\usepackage{graphicx}
\usepackage{subfigure}
\usepackage{multicol}
\usepackage{float}
\usepackage{slashbox}

\newcommand{\zsix}[0]{\ensuremath{\mathbbm{Z}_{6-\text{II}}}}
\newcommand{\ztf}[0]{\ensuremath{\mathbbm{Z}_{2}\times\mathbbm{Z}_{4}}}
\newcommand{\ztt}[0]{\ensuremath{\mathbbm{Z}_{2}\times\mathbbm{Z}_{2}}}
\newcommand{\ee}[0]{\ensuremath{E_8}}
\newcommand{\eexee}[0]{\ensuremath{E_8 \times E_8}}
\newcommand{\es}[0]{\ensuremath{E_6}}
\newcommand{\SU}[1]{\ensuremath{SU(#1)}}
\newcommand{\SO}[1]{\ensuremath{SO(#1)}}
\newcommand{\mmod}[0]{\ensuremath{\text{ mod }}}

\newcommand{\ts}[0]{\ensuremath{\mathbf{27}}}
\newcommand{\st}[0]{\ensuremath{\mathbf{16}}}

\title{A Zip-code for Quarks, Leptons and Higgs Bosons}

\author{Dami\'an Kaloni Mayorga Pe\~na,}
\author{Hans Peter Nilles and}
\author{Paul-Konstantin Oehlmann}

\affiliation{Bethe Center for Theoretical Physics, Physikalisches Institut der Universit\"at Bonn, \\
Nussallee 12, 53115 Bonn, Germany}

\emailAdd{damian@th.physik.uni-bonn.de}
\emailAdd{nilles@th.physik.uni-bonn.de}
\emailAdd{oehlmann@th.physik.uni-bonn.de}

\begin{document}

\abstract{The location of matter fields and the pattern of gauge symmetry in extra dimensions are crucial ingredients for string model building. We analyze realistic MSSM models from the heterotic \zsix\ Mini--Landscape and extract those properties that are vital for their success. We find that Higgs bosons and the top-quark are not localized in extra dimensions and live in the full $D=10$ dimensional space-time. The first two families of quarks and leptons, however, live at specific fixed points in extra dimensional space and exhibit a (discrete) family symmetry. Within a newly constructed \ztf\ orbifold framework we further elaborate on these location properties and the appearance of discrete symmetries. A similar geometrical picture emerges. This particular Zip-code for quarks, leptons and Higgs bosons seems to be of more general validity and thus a useful guideline for realistic model building in string theory.
}

\maketitle
\flushbottom

\section{Introduction}
\label{sec:Introduction}

Both the standard model (SM) of particle physics and its supersymmetric extension
(MSSM) contain many parameters such as masses and coupling constants. To
understand the origin of these parameters one would need an ultraviolet (UV)
completion
such as grand unification (GUT) and/or string theory. In the latter the parameters of the
low energy effective theory in $D=4$ dimensions
depend crucially on the process of compactification
of $D=10$ to $D=4$. Both the location of fields and the pattern of
gauge symmetries in the extra dimensions are very important for the properties of
the effective $D=4$ theory. So the central question concerns the Zip-code of elementary
particles: where do quarks, leptons and Higgs-bosons live in extra dimensions?

In the present paper we are trying to analyze this question in the framework of
consistent global constructions based on heterotic string theory \cite{Gross:1984dd, Gross:1985fr}. Such global
constructions include gravitational and gauge interactions and thus provide a reliable
UV-completion. Our program tries to identify realistic theories of the MSSM from string
theory and then studies common features of the resulting models. A most promising
framework is the so-called heterotic brane-world \cite{Forste:2004ie,Nilles:2008gq}: orbifold compactifications \cite{Dixon:1985jw,Dixon:1986jc,Ibanez:1986tp} of the
\eexee\ heterotic string\footnote{Compactifications on smooth Calabi-Yau manifolds \cite{candelas,Anderson:2011ns} and the free fermionic formulation \cite{Faraggi:1992fa,Faraggi:1995yd} are other model building alternatives within the heterotic string. For other orbifold compactifcations different from those discussed in this paper see e.g. \cite{Kim:2007dx}.}. Up to now a small part of this landscape has
been analyzed
most notably in the context of the \zsix\ orbifold \cite{Kobayashi:2004ud,Kobayashi:2004ya}. Hundreds of realistic MSSM-like
models have been identified. Those models realize the SM-gauge group from grand unification, 3
families of quarks and leptons and a single pair of Higgs-doublets \cite{Buchmuller:2005jr,Buchmuller:2006ik,Lebedev:2006kn,Lebedev:2008un,Lebedev:2007hv}. Two questions
should be addressed at this point:

\begin{itemize}
\item which are the properties of the models that make them so successful?
\item are these results special for the \zsix\ orbifold or do they provide
a general pattern?
\end{itemize}

We attempt to address both questions here: the first through the interpretation of the \zsix\
and the second by means of a new analysis of the \ztf\
orbifold as well as a comparison with available \ztt\ constructions \cite{Blaszczyk:2009in}.

We find that the geographic location of the fields (in the extra dimensions) is of
utmost importance, most notably for a solution to the $\mu$-problem \cite{Lebedev:2007hv,Casas:1992mk, Antoniadis:1994hg}, the Yukawa-coupling of the top-quark \cite{Buchmuller:2006ik,Lebedev:2007hv}, the existence of discrete family symmetries \cite{Kobayashi:2006wq,Ko:2007dz} as well as the
pattern of soft supersymmetry breaking terms \cite{Lebedev:2006tr,Nilles:1982ik,Ferrara:1982qs,Nilles:1982xx}. Our analysis shows that the Higgs bosons
and the top-quark preferably live in the bulk ($D=10$) while quarks and leptons of
the first two families are localized at (different) fixed points/tori in the extra
dimensions. We shall argue that these results not only hold in the case of the \zsix\
orbifold, but seem to be valid more generally, even beyond the heterotic constructions.

The paper will be organized as follows: in section 2 we shall discuss the \zsix\ orbifold
and shall review the constructions of the so-called ``Mini--Landscape". Section 3 will be
devoted to the interpretation of these results: most notably the solution to the
$\mu$-problem through the presence of an $R$-symmetry and its consequences for the extra
dimensional properties of the Higgs multiplets. The large top-quark Yukawa
coupling and the presence of discrete family symmetries are shown to be the
consequences of the geographic location of the fields under consideration.

In section 4 we present the construction and
classification of the \ztf\ orbifold that has not been discussed in
the literature yet. This is then followed by a geometrical
exploration in this new framework to identify the rules for successful model
constructions in section 5. Section 6 is a comparison between the emerging picture
from \ztf\ and the \zsix\ and \ztt\ orbifold models.
The known properties of the MSSM
seem to indicate a very well defined picture (Zip-code) of the location of fields in the
extra--dimensional world, some aspects which might be of universal validity. In
section 7 we shall conclude and discuss strategies for further research along these
lines.

\section{The \zsix\ Mini--Landscape}
\label{sec:zsixtwo}
In this section we will review the Mini--Landscape searches using the \zsix\ orbifold geometry. Thus we will first analyze its fixed point structure and then we describe the realistic models found in that context.
\subsection{The \zsix\ Geometry}
Lets start by briefly discussing the construction of the \zsix\ orbifold model. Take the complexified coordinates $Z^i$ ($i=1,2,3$) to span the six real extra dimensions. Assume that such coordinates are equivalent upon lattice identifications, and take the lattice to be spanned by the simple roots of $G_2\times SU(3)\times SU(2)^2$, in such a way that the roots of each factor are aligned on each of the complex planes. Since the lattice possesses the symmetry $Z^i\rightarrow e^{2\pi {\rm i}v^i}Z^i$, with $v^i$ being the components of the twist vector
\begin{equation}
 v=\frac{1}{6}(0,1,2,-3)\,.
\end{equation}
One can build the orbifold by modding out such a symmetry of the torus previously constructed. This twist is compatible with the requirement of $\mathcal{N}=1$ SUSY in $4D$. Since the joint action of twist and lattice identifications is generally non free, one ends up having fixed points/tori which correspond to curvature singularities on the target space \cite{Dixon:1985jw}. From the string theory perspective, the presence of such fixed points/tori results in more alternatives for the strings to close: In addition to the bulk states one ordinarily has in the context of toroidal compactifications, new twisted states are required for consistency of the theory. The space group $S$ is defined as the semi-direct product of the multiplicative closure of the twist (point group) and the lattice. A certain string state is defined by means of the monodromy
\begin{equation}
Z(\tau,\sigma+\pi)=g Z(\tau,\sigma)=\theta^k Z(\tau,\sigma)+\lambda \, ,
\end{equation}
with $g=(\theta^k,\lambda)\in S$ being the generating element of the string, $\theta={\rm diag}(e^{2\pi {\rm i}v^1},e^{2\pi {\rm i}v^2},e^{2\pi {\rm i}v^3})$ and $\lambda$ some six dimensional lattice vector. The states can be grouped by sectors depending on the power $k$ in the previous equations. The states with $k=0$ belong to the untwisted sector, all other values of $k$ (i.e. $1,...,5$) define the $k-$th twisted sector $T_k$. It is convenient to describe now the type of singularities associated to each twisted sector, since they are in close relation to the multiplicities of the states. The twist itself acts non-trivially on all planes, so that for $T_1$ one obtains the fixed points depicted in table \ref{tab:T1}. For the sectors $T_2$ and $T_3$ one finds fixed tori due to the trivial action of the corresponding twist on one of the planes. On the $G_2$ plane there are some fixed points which are identified upon $\theta$ or $\theta^2$ so that they correspond to the same point on the orbifold space (see e.g. tables \ref{tab:T2} and \ref{tab:T3}).\\
\begin{table}[h!]
\centering
\renewcommand{\arraystretch}{1.4}
\begin{center}
\includegraphics[height=25mm,width=100mm]{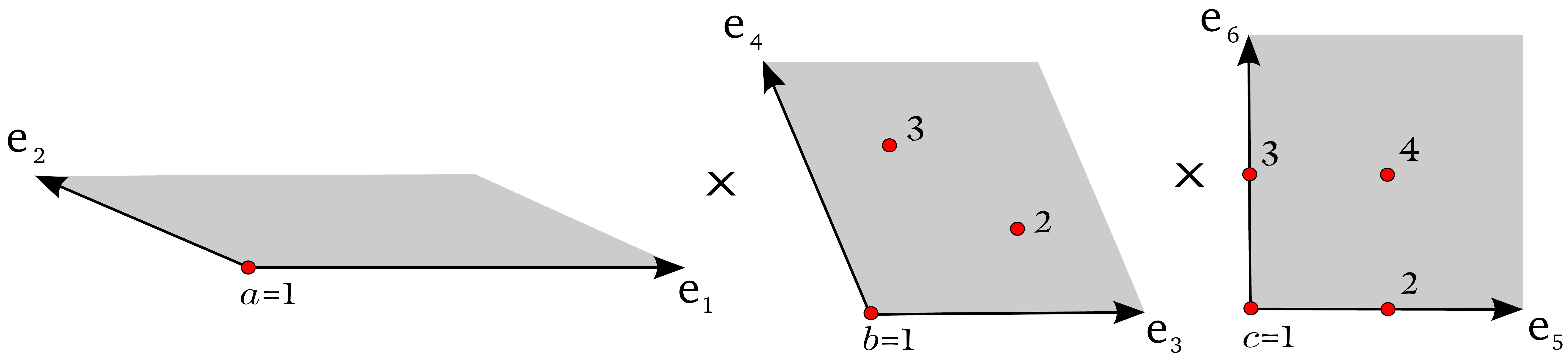}\\
\vspace{0.3cm}
\footnotesize{
\begin{tabular}{|c|c|c|c|c|}
\multicolumn{5}{c}{$T_1$}\\
\hline
\backslashbox{$b$}{$c$} & 1 & 2 & 3 & 4 \\\hline
1 & $0$ & $e_5$ & $e_6$ & $e_5+e_6$  \\
2 &  $e_3$ &  $e_3+e_5$ & $e_3+e_6$ & $e_3+e_5+e_6$ \\
3 &  $e_3+e_4$ & $e_3+e_4+e_5$ & $e_3+e_4+e_5$ & $e_3+e_4+e_5+e_6$   \\
\hline
\multicolumn{5}{c}{}\\
\multicolumn{5}{c}{$T_5$}\\
\hline
\backslashbox{$b$}{$c$} & 1 & 2 & 3 & 4 \\\hline
1 & $0$ & $e_5$ & $e_6$ & $e_5+e_6$  \\
2 &  $e_3+e_4$ &  $e_3+e_4+e_5$ & $e_3+e_4+e_6$ & $e_3+e_4+e_5+e_6$ \\
3 &  $e_4$ & $e_4+e_5$ & $e_4+e_6$ & $e_4+e_5+e_6$  \\
\hline
\end{tabular}}
\end{center}
\caption{Fixed points of the $T_1$ and $T_5$ sectors \cite{Nibbelink:2009sp}. Each combination $abc$ leads to an inequivalent singularity on the orbifold. The lattice part of the constructing element for each point is shown in the tables, depending on which sector one is considering. For example the fixed point $a=1$, $b=2$, $c=3$ is generated by the elements $(\theta,e_3+e_6)$ and $(\theta^5,e_3+e_4+e_6)$.}
\label{tab:T1}
\end{table}
\begin{table}[h!]
\centering
\renewcommand{\arraystretch}{1.4}
\begin{center}
\includegraphics[height=25mm,width=100mm]{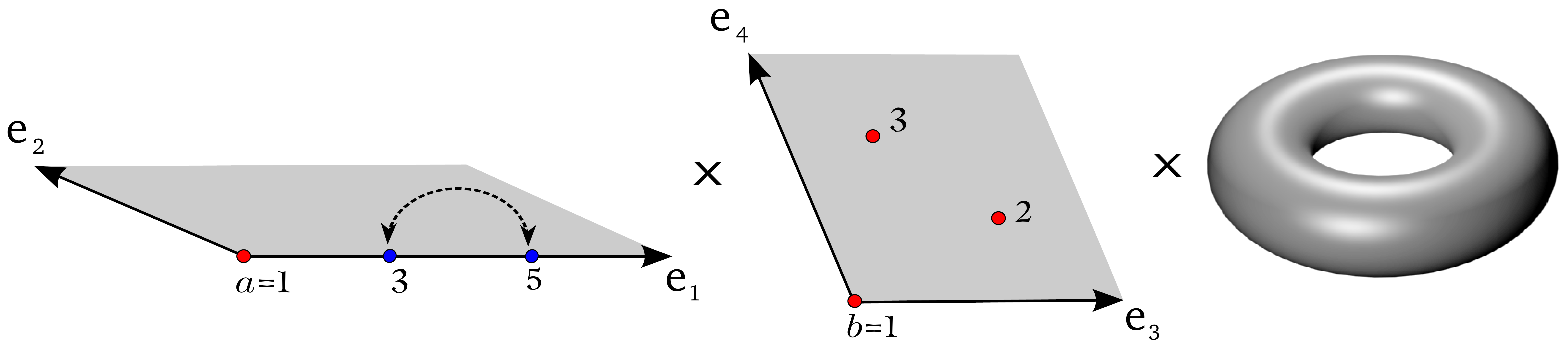}\\
\vspace{0.3cm}
\footnotesize{
\begin{tabular}{|c|c|c|c|}
\multicolumn{4}{c}{$T_2$}\\
\hline
\backslashbox{$a$}{$b$} & 1 & 2 & 3  \\\hline
1 & $0$ & $e_3+e_4$ & $e_4$   \\
3 & \cellcolor[gray]{0.7} $-e_2$ & \cellcolor[gray]{0.7} $-e_2+e_3+e_4$ & \cellcolor[gray]{0.7} $-e_2+e_4$  \\
\hline
\multicolumn{4}{c}{}\\
\multicolumn{4}{c}{$T_4$}\\
\hline
\backslashbox{$a$}{$b$} & 1 & 2 & 3  \\\hline
1 & $0$ & $e_3$ & $e_3+e_4$  \\
3 & \cellcolor[gray]{0.7} $e_1+e_2$ & \cellcolor[gray]{0.7} $e_1+e_2+e_3$ & \cellcolor[gray]{0.7} $e_1+e_2+e_3+e_4$ \\
\hline
\end{tabular}}
\end{center}
\caption{Fixed tori of the $T_2$ and $T_4$ sectors. The fixed points $a=3,5$ are identified under $\theta$, in the tables below we present only one constructing element per inequivalent fixed torus. The lattice vectors corresponding to such points are shaded to remark that they lead to special fixed tori.}
\label{tab:T2}
\end{table}
\begin{table}[h!]
\centering
\renewcommand{\arraystretch}{1.4}
\begin{center}
\includegraphics[height=25mm,width=100mm]{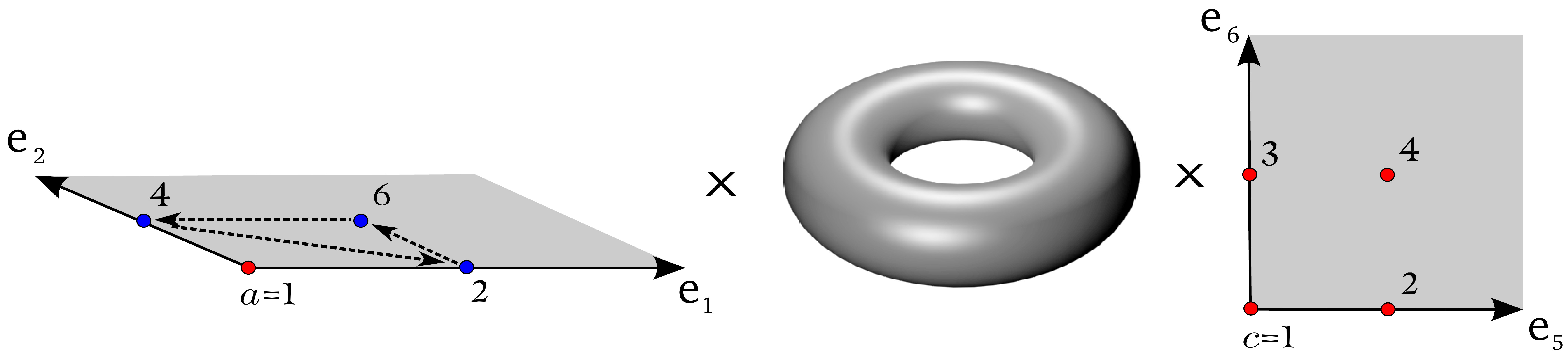}\\
\vspace{0.3cm}
\footnotesize{
\begin{tabular}{|c|c|c|c|c|}
\multicolumn{5}{c}{$T_3$}\\
\hline
\backslashbox{$a$}{$c$} & 1 & 2 & 3 & 4 \\\hline
1 & $0$ & $e_5$ & $e_6$ & $e_5+e_6$  \\
2 & \cellcolor[gray]{0.7} $e_1$ & \cellcolor[gray]{0.7} $e_1+e_5$ & \cellcolor[gray]{0.7} $e_1+e_6$ & \cellcolor[gray]{0.7} $e_1+e_5+e_6$ \\
\hline
\end{tabular}}
\end{center}
\caption{Fixed tori of the $T_3$ sector. The fixed points $a=2,4,6$ are identified upon $\theta$ and $\theta^2$.}
\label{tab:T3}
\end{table}

Consider now the \eexee\ heterotic string compactified on this orbifold background\footnote{For a more detailed discussion on this issues see e.g. \cite{Bailin:1999nk,Choibook,Vaudrevange:2008sm,RamosSanchez:2008tn}}. The ten bosonic fields are split into the standard four space-time coordinates plus the three complexified ones we have previously discussed. In the bosonic picture, sixteen extra left-moving coordinates give rise to the gauge theory after compactification. Orbifolding itself leads to an automatic breaking of the \eexee\ gauge group. Since the Bianchi identities must be satisfied, contemplating a space with curvature singularities forces us to account for such effects by modifying the gauge structure. This is usually done by embedding the orbifold identifications into the gauge space. Here we concentrate on translational embeddings only. Consider the map
\begin{equation}
g=(\theta^k,n_\alpha e_\alpha)\hookrightarrow V_g=kV+n_\alpha W_\alpha\, ,
\end{equation}
with the shift $V$ and the Wilson lines $W_\alpha$ being sixteen dimensional vectors subjected to certain constraints resulting from consistency with the space group multiplication and with modular transformations. The \zsix\ compactification requires a shift of order six (i.e. $6V$ is a vector of the \eexee\ lattice) and allows for three Wilson lines. The lattice vectors $e_3$ and $e_4$ support an order three Wilson line $W_3$, while $e_5$ and $e_6$ allow for $W_2$ and $W_2^\prime$ both of order two. Further requirements on these vectors can be found in refs. \cite{Dixon:1986jc,Katsuki1989}. Among those lattice shifts which satisfy the modular invariance constraints, those related by automorphisms and lattice vectors\footnote{In more explicit terms, given any two allowed shifts $V, V^\prime$, if one can find an automorphism $\sigma$ of the lattice such that $V=\sigma V^\prime+6\Lambda$, the physical theories arising from each of them are equivalent. Here $\Lambda$ is a vector in the gauge lattice. If the shifts are related only by $V=\sigma V^\prime+\Lambda$, the twisted spectra can differ but the untwisted sectors of both models are identical.} lead to models which are physically equivalent. The set of all inequivalent shift embeddings for \zsix\ can be found in ref. \cite{Katsuki1989}.\\
\subsection{Gauge Embeddings and Matter Spectra}
Let us now discuss the spectrum of physical states. They correspond to tensor products of the left and right moving parts together with an additional piece containing the information about the conjugacy class for the constructing element of the string. We concentrate on the massless spectrum: In the untwisted sector one finds the $4D$ supergravity multiplet, the geometric moduli, some matter fields and the gauge bosons. With respect to the former ones one should point out that the breaking induced by the embedding leaves the sixteen generators of the \eexee\ Cartan subalgebra untouched. The roots which survive are those satisfying the conditions
\begin{equation}
 p\cdot V=0\mmod 1 \, , \quad  p\cdot W_\alpha=0\mmod 1 \,.
\label{eq:Phys}
\end{equation}
The effect of the right moving fermionic fields can be parameterized -upon bosonization- in terms of weights $q$ from the spinor or vector lattice of \SO{8}. The boundary conditions on a twisted state of the $T_k$ sector make these weights to get shifted by $q_{\rm sh}=q+kv$. From the side of the left movers one is equipped with the oscillator modes $\alpha^i_{-w^i}$, $\alpha^{i*}_{-w^{i*}}$ ($w^i=v^i\mmod 1$, $w^{i*}=1-v^i\mmod 1$ with $0\leq w^i, w^{i*} < 1 $), as well as shifted weights in the gauge lattice $p_{\rm sh}=p+V_g$, if $g$ is the constructing element of the string. The space group part of the state we denote as $|g\rangle$, but one has to bear in mind that it does not only depend on the constructing element but also on the conjugacy class associated to it. A state of the form
\begin{equation}
 |\phi\rangle\sim |q_{\rm sh}\rangle \otimes \left(\prod_{i=1}^3 (\alpha^i_{-w^i})^{N^i}(\alpha^{i*}_{-w^{i*}})^{N^{i*}}|p_{\rm sh}\rangle \right)\otimes |g\rangle
\label{eq:phys}
\end{equation}
will be part of the massless spectrum if the level matching condition is satisfied
\begin{equation}
\frac{q_{\rm sh}^2}{2}-\frac{1}{2}+\delta_c=\frac{p_{\rm sh}^2}{2}+\tilde{N}-1+\delta_c=0
\label{eq:levelm}
\end{equation}
with $\delta_c$ a shift in the normal ordering constant and $\tilde{N}=w^iN^i+w^{i*}N^{i*}$ being the number operator. In addition to that, one has to ensure that the state is invariant under the space group. Such a condition results in a set of projection conditions
\begin{equation}
p_{sh}\cdot V_h-R\cdot v_h=0\mmod 1\,.
\label{proy}
\end{equation}
for any element $h\in S$ commuting with $g$.\\

When no Wilson lines are switched on, all fixed points are degenerate in the sense that all are populated by the same matter. This is not the case for the fixed tori, since there are some \emph{special} tori which suffer of additional identifications under the point group. For them, the projection conditions are relaxed and the gauge symmetry is then enhanced implying that matter has to come in higher representations. Furthermore, it can be shown that at such tori, SUSY is also enhanced to $\mathcal{N}=2$, so that matter at those tori is vector-like and hence likely to be decoupled from the low energy theory. The Wilson lines are a freedom to further break the gauge group one obtains with a given shift $V$: They have to be chosen in such a manner that the breaking to the standard model gauge group (times extra factors) is accomplished, while giving rise to three families of quarks and leptons plus a Higgs pair (and maybe some additional vector-like exotics).\\\\
The starting point of the Mini--Landscape searches were shift embeddings leading to an \es\ or \SO{10} factor \cite{Nilles:2008gq}, in the spirit of identifying a GUT breaking such as
\begin{equation}
 SU(3)\times SU(2)\times U(1)_Y \subset SU(5) \subset SO(10)\text{ or }E_6\, .
\end{equation}
One can compute the spectrum at the GUT level and find $\mathbf{27}$- or $\mathbf{16}$-plets localized at the different sectors. Since each multiplet provides a complete family, one has to bear in mind how Wilson lines split the degeneracy of the fixed points, so that in the end one can have a three family structure. Thanks to these observations two \SO{10} shifts
\begin{align}
 V_{SO(10)}^{1} &=\left(\frac13 , \frac12, \frac12,0,0,0,0,0\right)\left(\frac13,0,0,0,0,0,0,0\right)\,, \label{eq:so1}\\
V_{SO(10)}^{2} &=\left(\frac13 , \frac13, \frac13,0,0,0,0,0\right)\left(\frac16,\frac16,0,0,0,0,0,0\right)\, \label{eq:so2}
\end{align}
and two \es\ shifts
\begin{align}
 V_{E_6}^{1} &=\left(\frac12 , \frac13, \frac16,0,0,0,0,0\right)\left(0,0,0,0,0,0,0,0\right)\,, \label{eq:e1}\\
V_{E_6}^{2} &=\left(\frac23 , \frac13, \frac13,0,0,0,0,0\right)\left(\frac16,\frac16 \,
,0,0,0,0,0,0\right)\,. \label{eq:e2}
\end{align}
were shown to be good starting points for the heterotic road to the MSSM \cite{Lebedev:2007hv}. \\

It was observed that when both order two Wilson lines are switched on, there are three fixed points where the \SO{10} or \es\ factor remains unbroken, such that if one has a model with the right matter at $T_1$ ($T_5$), the fixed points we have just mentioned will support a three complete family structure. This approach is unfortunately unsuccessful since the Wilson lines do not suffice for the desired breaking of the gauge group in one $E_8$ factor \cite{Buchmuller:2006ik}.\\

Another alternative is to take one Wilson line of order three and one of order two, so that instead of three, one has two fixed points in the first (fifth) twisted sector where the states furnish complete representations of the unified group. The third family gets completed by pieces coming from the untwisted and twisted sectors. This strategy was more successful and out of all possible models about one percent of them were found to have the SM spectrum plus vector like exotics, in addition to  a non anomalous $U(1)_Y$ with the standard \SU{5} normalization.\\

The previous results are a good sign of the likelihood of having a stringy MSSM in this framework. Nevertheless one has to make sure that not only the spectrum but the interactions are in a good shape. From the orbifold CFT \cite{Dixon:1986qv,Hamidi:1986vh} one is equipped with a set of selection rules which allow to determine which couplings are non-vanishing. Such selection rules permit an interpretation as discrete $R$ and non-$R$ symmetries from the field theory point of view. Some further permutation symmetries between fixed points (subject to be broken by the WL) permit to combine the states into irreducible representations of some discrete non-Abelian flavor group (for a more detailed discussion on these issues see refs. \cite{Font:1988mm,Kobayashi:2004ya,Kobayashi:2006wq}).\\

A common feature of the models is the presence of an anomalous $U(1)$ factor. The FI term it induces has to be canceled in a supersymmetric way by VEVs of singlet fields. By assigning VEVs to some SM singlets the gauge group is broken further down to $SU(3)\times SU(2)\times U(1)_Y\times G_{\rm Hidden}$, where $G_{\rm Hidden}$ denotes a hidden sector which may be of use for SUSY breaking purposes. The above mentioned VEVs also lead to mass terms for the exotics, induce non-trivial Yukawa couplings and can give rise to an effective $\mu$-term.\\

To get some more profound insights on this analysis, promising models were further required to support a renormalizable (trilinear) top-Yukawa coupling. This is motivated by the observation of the high mass of the top-quark compared to all other SM particles. Couplings were checked up to order eight and at that level it was expected that all exotics could be decoupled. A physical motivation for this is, that such particles should remain far beyond observation. Surprisingly, these two requirements are met by almost all promising models. \\
\subsection{Origin of the Top-Yukawa Coupling}
Let us elaborate a bit more on the trilinear top-Yukawa coupling. The constraints imposed by the CFT only allow for the following coupling candidates
\begin{equation*}
 UUU \, , \quad T_2T_4U \, , \quad T_3T_3 U \, , \quad T_5T_4T_3 \, , \quad T_5T_5T_2 \, .
\end{equation*}
Among them, those involving only twisted sectors require the fields to be at the same fixed point in order not to be instantonically suppressed. In principle, one needs to specify the WLs in order to find out which trilinear couplings are allowed. However, some of these couplings may descend from the underlying GUT and can thus be checked at the stage in which all WLs are switched off. This is the case for the purely untwisted coupling:
Here one has to check the untwisted fields for the possibility of a coupling such as $\mathbf{16}\cdot\mathbf{16}\cdot\mathbf{10}$ (for models with \SO{10}) or $(\mathbf{27})^3$ (in \es). If this is the case one simply has to look for Wilson lines which leave the relevant pieces for the top-Yukawa unprojected. It is worth to remark that such untwisted coupling occurs relatively often. We devote the remainder of this section to present some specific examples with special focus on the origin of the renormalizable coupling.\\\\
Consider a shift equivalent to the one presented in \eqref{eq:so1},
\begin{equation}
V^1_{SO(10)} = \left(\frac13 , -\frac12, -\frac12,0,0,0,0,0\right)\left(\frac12, -\frac16, -\frac12,
 -\frac12,-\frac12,-\frac12,-\frac12,\frac12\right)\,.
\label{b1}
\end{equation}
It leads to the gauge group $[SO(10) \times SU(2) \times SU(2) \times U(1)]\times[ SO(14) \times U(1)]$, the squared brackets are used to distinguish the \ee\ factor giving rise to these groups. The spectrum of left-chiral superfields is presented in table \ref{tab:spectrumso}.\\
\begin{table}[h!]
\centering
\renewcommand{\arraystretch}{1.4}
\begin{center}
\footnotesize{
\begin{tabular}{|c|c|c|c|c|}\hline
$U$ & $T_2$ & $T_3$ & $T_4$ & $T_5$ \\ \hline
$(\mathbf{10},\mathbf{2},\mathbf{2},\mathbf{1})_{0,0}$ & $\textcolor{red}{3}(\mathbf{1},\mathbf{2},\mathbf{2},\mathbf{1})_{-\frac{28}{3}, -\frac{2}{3}}$ & $4+\textcolor{red}{4}(\mathbf{1},\mathbf{1},\mathbf{2},\mathbf{1})_{12,-2}$ & $3+\textcolor{red}{3}(\mathbf{1},\mathbf{2},\mathbf{2},\mathbf{1})_{\frac{28}{3}, \frac{2}{3}}$  & 12 $(\overline{\mathbf{16}},\mathbf{1},\mathbf{1},\mathbf{1})_{\frac{14}{3},\frac13}$ \\
$(\mathbf{1},\mathbf{2},\mathbf{2},\mathbf{1})_{4,6}$& $3+\textcolor{red}{3}(\mathbf{10},\mathbf{1},\mathbf{1},\mathbf{1})_{-\frac{28}{3}, -\frac{2}{3}}$ & $\textcolor{red}{4}(\mathbf{1},\mathbf{1},\mathbf{2},\mathbf{14})_{0,0}$ & $\textcolor{red}{3}(\mathbf{1},\mathbf{1},\mathbf{1},\mathbf{14})_{-\frac{20}{3}, -\frac{10}{3}}$ & 12$(\mathbf{1},\mathbf{1},\mathbf{2},\mathbf{1})_{\frac{20}{3}, \frac{10}{3}}$\\
$(\overline{\mathbf{16}},\mathbf{1},\mathbf{2},\mathbf{1})_{-2,-3}$ & $3+\textcolor{red}{3}(\mathbf{1},\mathbf{1},\mathbf{1},\mathbf{14})_{\frac{20}{3}, \frac{10}{3}}$ & $\textcolor{red}{4}(\mathbf{1},\mathbf{1},\mathbf{2},\mathbf{1})_{-12, 2}$ &$ \textcolor{red}{3}(\mathbf{10},\mathbf{1},\mathbf{1},\mathbf{1} )_{\frac{28}{3}, \frac{2}{3}} $& $24(\mathbf{1},\mathbf{2},\mathbf{1},\mathbf{1})_{\frac83 ,-\frac83 }$\\
$(\mathbf{1},\mathbf{1},\mathbf{1},\mathbf{64})_{6,-1}$ & $\textcolor{red}{9} (\mathbf{1},\mathbf{1},\mathbf{1},\mathbf{1})_{-\frac{16}{3},\frac{16}{3}} $ & & $\textcolor{red}{9}(\mathbf{1},\mathbf{1},\mathbf{1},\mathbf{1})_{\frac{16}{3},-\frac{16}{3}} $  & \\
$(\overline{\mathbf{16}},\mathbf{2},\mathbf{1},\mathbf{1})_{2,3}$ &  & &  & \\
$(\mathbf{10},\mathbf{1},\mathbf{1},\mathbf{1})_{-4,-6}$ & & & & \\
$(\mathbf{1},\mathbf{1},\mathbf{1},\mathbf{14})_{12,-2}$ & & & & \\ \hline
\end{tabular}}
\end{center}
\caption{Massless spectrum of left-chiral superfields for the shift \eqref{b1}. The multiplicities are shown left of the parenthesis, with the special tori highlighted in red. Bold numbers label the representations of $SO(10) \times SU(2) \times SU(2)\times SO(14)$ under which the different fields transform. The subindices are the corresponding $U(1)$ charges. Note that the first twisted sector does not contain left-chiral fields giving rise only to the CPT conjugates of the states from $T_5$.}
\label{tab:spectrumso}
\end{table}

\noindent From this table we can see that $\overline{\mathbf{16}}$-plets arise from the $T_5$ sector. Switching on $W_2$ and $W_3$ reduces the multipicity of these states from twelve to two. They will give rise to the light families. The multiplicity of two is a consequence of a surviving $D_4$ family symmetry. 
From the untwisted sector, the following states are of capital interest for us
\begin{equation*}
\Phi_1=(\overline{\mathbf{16}},\mathbf{1},\mathbf{2},\mathbf{1})_{-2,-3} \quad \Phi_2=(\overline{\mathbf{16}},\mathbf{2},\mathbf{1},\mathbf{1})_{2,3} \quad \Phi_3=(\mathbf{10},\mathbf{2},\mathbf{2},\mathbf{1})_{0,0}\,.
\end{equation*}
Their corresponding $R$-charges are $(0,-1,0,0)$, $(0,0,-1,0)$ and $(0,0,0,-1)$ respectively. This means that the coupling $\Phi_1\Phi_2\Phi_3$ is allowed and happens to be the only possible renormalizable coupling between untwisted fields. Note also that the piece $\Phi_3\Phi_3$ is a neutral monomial under all selection rules. All one needs is to find a configuration of Wilson lines which does not project the relevant pieces. To show that this is possible, consider for instance the Wilson lines
\begin{align*}
W_2 =& \left(0, -\frac12 ,-\frac12, -\frac12, \frac12 , 0, 0, 0\right)\left(4, -3 , -\frac72 , -4, -3, -\frac72,-\frac92,\frac72\right)\, \text{ and} \\
W_3 =&\left(-\frac12 , -\frac12,\frac16,\frac16,\frac16,\frac16,\frac16,\frac16\right)\left(\frac13 ,0 , 0, \frac23, 0,\frac53,-2,0\right)\, .
\end{align*}
These backgrounds lead to the breaking
\begin{align*}
SO(10) \times SU(2) \times SU(2) \times SO(14) \times U(1)^2 \rightarrow SU(3) \times SU(2) \times SU(4) \times SU(2) \times U(1)^9\,.
\end{align*}
The $U(1)$'s can be rotated in such way that one of them corresponds to the standard $SU(5)$ hypercharge which, in fact, is non anomalous. Concerning the untwisted fields of our interest, one can identify the splitting pattern
\begin{align*}
(\mathbf{10},\mathbf{2},\mathbf{2},\mathbf{1})_{0,0}  & \rightarrow  \quad (\mathbf{1},\mathbf{2},\mathbf{1}, \mathbf{1})_{1/2,...} + (\mathbf{1},\mathbf{2},\mathbf{1}, \mathbf{1})_{-1/2,....} \, ,\\
(\overline{\mathbf{16}}, \mathbf{1},\mathbf{2}, \mathbf{1})_{-2,-3} & \rightarrow  \quad (\overline{\mathbf{3}},\mathbf{1},\mathbf{1}, \mathbf{1})_{-2/3,....} + (\mathbf{1},\mathbf{1},\mathbf{1},\mathbf{1})_{0,...}+(\mathbf{1},\mathbf{1},\mathbf{1},\mathbf{1})_{0,...} \, ,\\
(\overline{\mathbf{16}},\mathbf{2},\mathbf{1},\mathbf{1})_{2,3} & \rightarrow  \quad (\mathbf{3}, \mathbf{2},\mathbf{1},\mathbf{1})_{1/6,...} + (\mathbf{1}, \mathbf{1},\mathbf{1},\mathbf{1})_{0,....} \, ,
\end{align*}
so that one ends up having a top-Yukawa coupling of order one. The complete spectrum upon WLs and suitable VEV configurations can be found in ref. \cite{Lebedev:2007hv}. This detailed example was introduced to describe how one can easily determine the presence of renormalizable couplings in the untwisted sector. When looking at the other promising shifts, it is found that $V_{E_6}^2$ is the only one which does not allow for an untwisted coupling. The model building is thus more difficult in this model since the trilinear coupling is much more sensitive to the Wilson lines\footnote{Out of 1700 alternatives only two models end up being of physical relevance.}. The neutral Higgs monomial exists in all models where the purely untwisted trilinear coupling is possible.


\section{Lessons from the \zsix\ Mini--Landscape}
\label{sec:LessonsfromZ6}

The constructions of the \zsix\ orbifold allow for hundreds of models
with a realistic MSSM structure representing a ``fertile patch'' of the
heterotic landscape. It is therefore interesting to analyze why this
patch is so fertile and what makes this model building approach so
successful. We shall concentrate our discussion on the benchmark models
of \cite{Lebedev:2007hv} as representatives for this class of models.

Key properties of the low energy effective theory depend on the geography
of fields in extra dimensions. In the heterotic braneworld we can have
3 classes of fields:

\begin{itemize}

\item fields in the untwisted sector, the 10-dimensional bulk

\item fields in the $T_1$ ($T_5$) twisted sector at fixed points
in extra dimensions (representing ``3-branes'')

\item fields in the $T_2$ ($T_4$) and $T_3$-sectors at fixed
tori in extra dimensions (representing ``5-branes'')

\end{itemize}

At these various locations we also have a very specific gauge group
structure. While in $4D$ we have the gauge group
$SU(3)\times SU(2)\times U(1)$, there
could be ``enhanced'' gauge symmetries (like $SO(10)$ or $SU(6)$) at some
fixed points or fixed tori. Fields located there
come in representations of the enhanced gauge group. This allows the
coexistence of complete and split multiplets of the underlying grand
unified group. In addition the different sectors obey various degrees
of supersymmetry. Sectors with fixed points have just $\mathcal{N}=1$ supersymmetry
while fixed tori enjoy remnants of an underlying $\mathcal{N}=2$ and the untwisted
sector an $\mathcal{N}=4$ supersymmetry. All these properties are important for
the nature of the 4-dimensional effective theory. We shall now
analyze the geometrical properties of the successful models of
the Mini--Landscape.

\subsection{The Higgs system}
\label{sec:higgs}

The MSSM contains one pair of Higgs-doublet superfields $H_u$ and $H_d$.
This is a vector like pair and we have to face the so-called $\mu$-problem. In string constructions we usually have a larger number of Higgs pairs
and we have to answer the questions why all of them except one pair
become heavy. For a large class of models in the Mini--Landscape we find
a situation where the $\mu$-problem is solved in a miraculous way \cite{Lebedev:2007hv,Kappl:2008ie}:
exactly one pair of Higgs doublets remains massless ($\mu=0$). These
fields reside in the untwisted sector and they are neutral under all
selection rules. This implies that if a term $\mu H_u H_d$ is forbidden
in the superpotential $(\mathcal{W})$, the same applies to a constant term in $\mathcal{W}$.
$\mu$ and $\langle \mathcal{W}\rangle$ vanish in the supersymmetric vacuum (a solution discussed
earlier in ref \cite{Casas:1992mk} in the framework of field theoretic models).
This avoids deep supersymmetric AdS-vacua.

Higgs fields in the untwisted sector are directly related to gauge
fields in extra dimensions and reveal so-called gauge-Higgs
unification. These Higgs fields represent continuous Wilson lines
as discussed in ref. \cite{Ibanez:1987xa,Forste:2005rs,Hebecker:2012qp} and allow a smooth breakdown of
the electroweak symmetry.

This is a remarkable property and one might like to understand the
origin of this behaviour. It can be shown \cite{Lebedev:2007hv,Kappl:2008ie,Lee:2011dya} that it is the result
of an underlying (discrete) $R$-symmetry and its action
on fields from the untwisted sector. Such $R$-symmetries are remnants
of the extra dimensional Lorentz group $SO(6) \subset SO(9,1)$.

This is the first lesson from the Mini--Landscape. The Higgs pair
$H_u$ and $H_d$ should live in the bulk (untwisted sector). This
allows a solution of  the $\mu$-problem in supersymmetric Minkowski
space via an $R$-symmetry (originating from the higher dimensional
Lorentz group).

\subsection{The top-quark}

The mass of the top-quark is of order of the weak scale and we thus
expect its Yukawa coupling to be of the order of the gauge coupling,
exhibiting gauge-top unification \cite{Hosteins:2009xk}. In string theory these
couplings are given directly by the string coupling and we expect
the top-quark Yukawa coupling at the trilinear level in the
superpotential. Given the fact that $H_u$ is a field in the
untwisted sector there remain only few allowed couplings as we have seen
in the previous section.

In the Mini--Landscape we find that both the left-handed and
right-handed top-multiplet have to be in the untwisted sector
(bulk) to guarantee a sufficiently large Yukawa coupling. The
location of the other members of the third family is rather model
dependent, they are distributed over various sectors. Very often
the top-quark Yukawa coupling is the only trilinear Yukawa
coupling in the model.

This is the second lesson from the Mini--Landscape: left- and
right-handed top-quark multiplets should be bulk fields
leading to gauge-top unification.

\subsection{The first two families of quarks and leptons}

As explained in the last chapter we aimed at a grand unified picture
with families in the 16-dimensional spinor
representation of $SO(10)$. This did not
work out for 3 localized families. For the first two families,
however, it is  possible. The models show that these families
live at fixed points (3-branes) of the $T_5$-sector, more specifically
 at the points $a=b=c=1$ and $a=b=1$, $c=3$ in table \ref{tab:T1}. Since the
Higgs bosons $H_u$, $H_d$ live in the untwisted sector there are no
allowed trilinear Yukawa couplings and thus quark-and lepton
masses are suppressed. The specific location of the two families
at the points $c=1$ and $c=3$ in the third torus gives rise to a
$D_4$ family symmetry \cite{Ko:2007dz,Kobayashi:2006wq} and thus avoids the problem of flavour
changing neutral currents. This is another example of a
discrete symmetry which are rather common in the Mini--Landscape
and exhibit the rich symmetry structure of a successful MSSM model.

Thus we have a third lesson from the Mini--Landscape:
the first two families are located
at fixed points in extra dimension and exhibit a ($D_4$) family symmetry.
Due to the absence of trilinear Yukawa couplings the masses of
these quarks and leptons are suppressed.

\subsection{The pattern of supersymmetry breakdown}

This discussion is a bit more delicate, as it has to address the
question of moduli stabilization. A source for SUSY breakdown is
gaugino condensation in the hidden sector \cite{Nilles:1982ik,Ferrara:1982qs}. As discussed in
ref \cite{Lebedev:2006tr} we can determine the gauge groups in the hidden sector and
this leads to an acceptable gravitino mass in the (multi) TeV-range
(provided the dilaton is fixed at a realistic grand unified gauge
coupling). If we assume a stabilization of moduli in the spirit
discussed in \cite{Kappl:2010yu,Anderson:2011cza}, we would then remain with a run-away dilaton
and a positive vacuum energy. It can be shown that the adjustment
 of the vacuum energy with a matter superpotential (downlifting
the vacuum energy)
one can fix the dilaton as well. The resulting picture \cite{Lowen:2008fm} is
reminiscent for a scheme known as mirage mediation\footnote{The term ``mirage mediation'' was first introduced in ref. \cite{LoaizaBrito:2005fa}.} \cite{Choi:2004sx,Choi:2005ge}, at least
for gaugino masses and A-parameters. Scalar masses are more model
dependent and could be as large as the gravitino mass $m_{3/2}$ \cite{Lebedev:2006qq}
while gaugino masses are suppressed by a factor
$\log ({M_{\rm Planck}/{m_{3/2}}})$. The
mirage scheme gives rise to a compressed pattern of gaugino
masses \cite{Choi:2007ka}.

Scalar masses depend strongly on the location of fields in extra
dimension. Fields located at fixed points only feel $\mathcal{N}=1$ supersymmetry
and we expect masses of order $m_{3/2}$ for the first two families of
quarks and leptons. In contrast, fields on fixed tori feel remnants
of $\mathcal{N}=2$  and bulk fields remnants of $\mathcal{N}=4$ (at tree level). Thus Higgs
bosons
and scalar partners of the top-quark feel more protection and are thus
lighter by a factor of order $\log ({M_{\rm Planck}/{m_{3/2}}})$ compared to
$m_{3/2}$. We thus expect $m_{3/2}$
and scalar masses of the first two families
in the multi-TeV range while stops and Higgs bosons are in the TeV range.
This is a result of the location of fields in the
extra dimensions and provides the fourth lesson of the Mini--Landscape \cite{Krippendorf:2012ir}.

\subsection{The lessons}

Geometric properties of models from the heterotic Mini--Landscape
provide us with a set of lessons for successful model building. We can extract some generic features that might be of more
general validity. These are

\begin{itemize}

\item the Higgs bosons $H_u$ and $H_d$ are bulk fields

\item the top-quark lives in the bulk as well

\item the first two families are located at points in extra dimensions
and exhibit family symmetries

\item there is a special pattern of soft SUSY breaking terms resulting
in a mirage picture for gauginos, A-parameters
and remnants of $\mathcal{N}=4$ supersymmetry for the bulk fields.

\end{itemize}

It seems that these properties are crucial for a successful
MSSM construction. Key ingredients are the solution of the
$\mu$-problem and the quest for gauge-top unification. Therefore
such properties might very well extend beyond the \zsix\ model and
even beyond the heterotic constructions. So let us now explore
other models to further analyze this picture.

\section{Construction of $\boldsymbol{\mathbb{Z}_2 \times \mathbb{Z}_4}$ Orbifold}
\label{sec:Z2Z4Geometry}
The $\mathbbm{Z}_2\times\mathbbm{Z}_4$ orbifold of our interest results from dividing $\mathbbm{C}^3$ by the factorizable lattice spanned by the roots of $SU(2)^2\times SO(4)\times SO(4)$. Out of the resulting torus, we mod out the point group generated by the following twist vectors
\begin{align}
\textstyle v_2=(0,\frac{1}{2},-\frac{1}{2},0) \, , \quad v_4=(0,0,\frac{1}{4},-\frac{1}{4}) \, ,
\end{align}
which clearly correspond to isometries of the lattice and comply with the conditions for $\mathcal{N}=1$ SUSY in 4D. The $v_2$ twist indicates that the $\mathbbm{Z}_2$ generator acts as a simultaneous reflection on the first two planes, while leaving the third invariant. Similarly, the twist $v_4$ implies that the $\mathbbm{Z}_4$ generator rotates the second and third planes by $\pi/2$ counter and clockwise, respectively.
\subsection{Fixed Point Structure}
Here we start discussing the geometry of the orbifold, but many stringy properties of the fixed point/tori are left to be studied in the upcoming sections, where the appropriate machinery is constructed. Lets denote by $\theta$ and $\omega$ the generators of the $\mathbbm{Z}_2$ and $\mathbbm{Z}_4$ factors, respectively. The fixed points/tori of $\theta^{k_1}\omega^{k_2}$ are said to belong to the twisted sector $T_{(k_1,k_2)}$. Each particular sector is studied separately in tables \ref{T01} to \ref{T12}. For the sectors $T_{(0,1)}$, $T_{(0,2)}$ and $T_{(0,3)}$ one finds fixed tori due to the trivial action of $\omega$ on the first complex plane. A similar situation will happen for the case of $T_{(1,0)}$ and $T_{(1,2)}$ where, respectively the third and second planes are invariant. The only sectors where one finds fixed points are $T_{(1,1)}$ and $T_{(1,3)}$.  For each twisted sector, the fixed points/tori on the fundamental domain are a maximal set of representatives for the conjugacy classes of the space group (which act non freely on $\mathbbm{C}^3$). For model building purposes it is necessary to find the generating elements of each inequivalent fixed point, in order to determine the right gauge shift and its corresponding set of commuting elements. The fixed points have been labeled with indices $a$, $b$ and $c$ denoting the location on each complex plane. The twisted sectors $T_{(0,1)}$ and $T_{(0,3)}$ as well as $T_{(1,1)}$ and $T_{(1,3)}$ are the inverse of each other and have the same fixed points/tori but the generating elements differ depending on which sector one is considering (see tables \ref{T01} and \ref{T11}). The remaining sectors are self inverse and are generated by point group elements of order two, so that in general, one has further identifications induced by the $\mathbbm{Z}_4$ generator. Similar as in the \zsix\ case, those fixed tori which do not have a unique representative within the fundamental domain we call \emph{special}. One can show that there is no element of the form $(\theta,\lambda)\in S$ which commutes with the generating elements of those special fixed tori. This feature will again lead to an enhancement of the local gauge symmetry, but in contrast to the \zsix\ case, the presence of the additional $\mathbb{Z}_2$ twist forbids the local enhancement of the supersymmetries. \\\\
Note that in the sectors $T_{(1,0)}$ and $T_{(1,2)}$ the identifications under the $\mathbbm{Z}_4$ generator will lead to only three inequivalent fixed points in the $SO(4)$ plane. This is the same amount of fixed points one finds for the $SU(3)$ plane of the \zsix\ orbifold.\\\\
Our choice of geometry allows the following set of inequivalent Wilson lines:
\begin{enumerate}[(i)]
\item In the first complex plane one has two Wilson lines $W_1$ and $W_2$ at disposal, because the point group action does not relate the lattice vectors $e_1$ and $e_2$. These lines are of order two, since \[\theta e_\alpha+e_\alpha=0 \, ,\quad\alpha=1,2\, .\]
\item In the second and third planes we have the identifications:
\begin{equation}
\omega e_3=e_4 \, ,\quad \omega^3 e_5= e_6\, ,
\end{equation}
so that there is only one inequivalent Wilson line per complex plane\footnote{This equivalence is up to lattice vectors from the gauge lattice, which we omit for simplicity.}. These ones we will denote by $W_3$ and $W_4$. Due to the conditions
\begin{equation}
\omega^2 e_3+e_3=0 \, ,\quad\text{ and }\quad\omega^2e_5+e_5=0 \, ,
\end{equation}
these Wilson lines are of order two as well.
\end{enumerate}
\begin{table}[h!]
\centering
\renewcommand{\arraystretch}{1.4}
\begin{center}
\includegraphics[height=30mm,width=100mm]{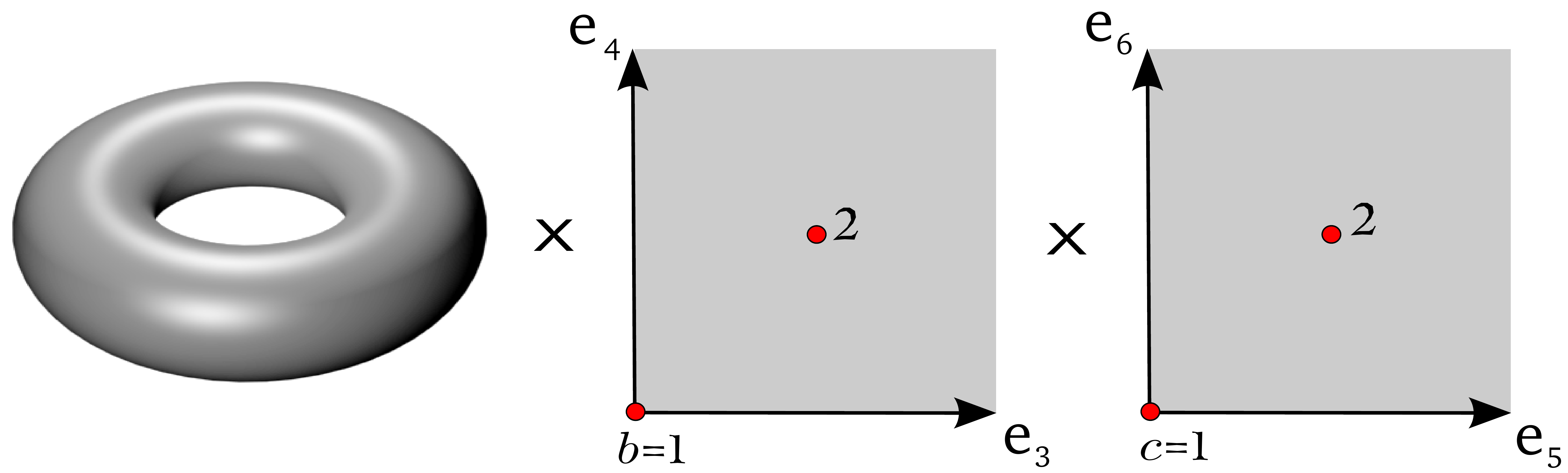}\\
\vspace{0.3cm}
\footnotesize{
\begin{tabular}{|c|c|c|p{1cm}|c|c|c|}
\multicolumn{3}{c}{$T_{(0,1)}$} & \multicolumn{1}{c}{} & \multicolumn{3}{c}{$T_{(0,3)}$}\\\cline{1-3}\cline{5-7}
\backslashbox{$b$}{$c$} & 1 & 2 & & \backslashbox{$b$}{$c$} & 1 & 2\\ \cline{1-3}\cline{5-7}
1 & $0$ & $e_6$ & & 1 & $0$ & $e_5$ \\
2 & $e_3$ & $e_3+e_6$ & & 2 & $e_4$ & $e_4+e_5$\\\cline{1-3}\cline{5-7}
\end{tabular}}
\end{center}
\caption{Fixed tori of the $T_{(0,1)}$ and $T_{(0,3)}$ sectors. The tables below help to deduce the generating element of each fixed torus. Consider a fixed torus located at the position $b$ and $c$ in the last two planes, such fixed torus is generated by a space group element $(\omega,\lambda_{bc})$ if the fixed torus belongs to the $T_{(1,0)}$ sector, or $(\omega^3,\lambda^\prime_{bc})$ for $T_{(0,3)}$. The lattice vectors $\lambda_{bc}$ and $\lambda^\prime_{bc}$ can be found in the $bc$-th entry of the left and right tables below the picture.}
\label{T01}
\end{table}
\begin{table}[h!]
\centering
\renewcommand{\arraystretch}{1.4}
\begin{center}
\includegraphics[height=30mm,width=100mm]{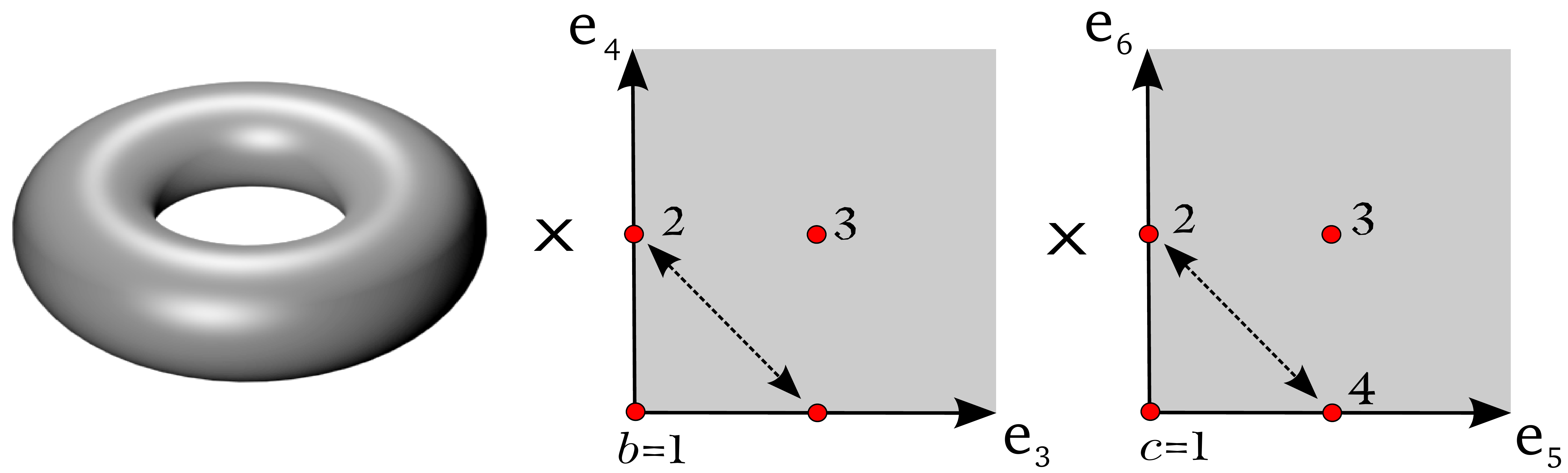}\\
\vspace{0.3cm}
\footnotesize{
\begin{tabular}{|c|c|c|c|c|}
\multicolumn{5}{c}{$T_{(0,2)}$}\\
\hline
\backslashbox{$b$}{$c$} & 1 & 2 & 3 & 4\\\hline
1 & $0$ & $e_5+e_6$ & \cellcolor[gray]{0.7} $e_6$ &  \\
2 &  $e_3+e_4$ & $e_3+e_4+e_5+e_6$ & \cellcolor[gray]{0.7} $e_3+e_4+e_5+e_6$  &  \\
3 & \cellcolor[gray]{0.7}  $e_4$ & \cellcolor[gray]{0.7} $e_4+e_5+e_6$ & \cellcolor[gray]{0.7} $e_4+e_6$ & \cellcolor[gray]{0.7} $e_4+e_5$ \\
\hline
\end{tabular}}
\end{center}
\caption{Fixed tori of $T_{(0,2)}$. The arrows in the last two planes have to be understood as identifications acting simultaneously such that they reproduce the effects of the $\mathbbm{Z}_4$ generator of the point group. Similarly as in table \ref{T01}, the generating elements can be found below the picture. Those entries which are left blank do not correspond to additional inequivalences. Shaded cells have been put to denote special fixed tori.}
\label{T02}
\end{table}
\begin{table}[h!]
\centering
\renewcommand{\arraystretch}{1.4}
\begin{center}
\includegraphics[height=30mm,width=100mm]{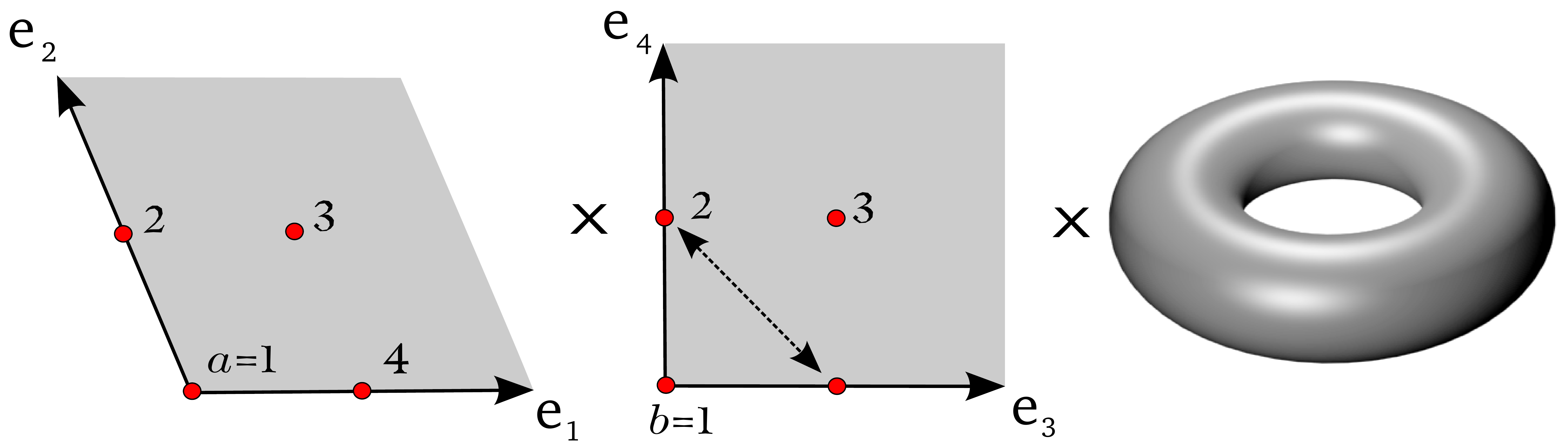}\\
\vspace{0.3cm}
\footnotesize{
\begin{tabular}{|c|c|c|c|}
\multicolumn{4}{c}{$T_{(1,0)}$}\\
\hline
\backslashbox{$a$}{$b$} & 1 & 2 & 3 \\\hline
1 & $0$ & $e_3+e_4$ & \cellcolor[gray]{0.7} $e_4$    \\
2 &  $e_2$ & $e_2+e_3+e_4$ & \cellcolor[gray]{0.7}  $e_2+e_4$   \\
3 &  $e_1+e_2$ &  $e_1+e_2+e_3+e_4$ & \cellcolor[gray]{0.7} $e_1+e_2+e_4$    \\
4 &  $e_1$ &  $e_1+e_3+e_4$ & \cellcolor[gray]{0.7} $e_1+e_4$   \\\hline
\end{tabular}}
\end{center}
\caption{Fixed tori of the $T_{(1,0)}$ sector. This sector is associated to the $\mathbbm{Z}_2$ generator of the point group which happens to act trivially on the third plane. For this reason, the fixed tori are identified only up to rotations by $\pi/2$ on the second plane. The special fixed tori are associated to the shaded cells in the table.}
\label{T10}
\end{table}
\begin{table}[h!]
\centering
\renewcommand{\arraystretch}{1.4}
\begin{center}
\includegraphics[height=30mm,width=100mm]{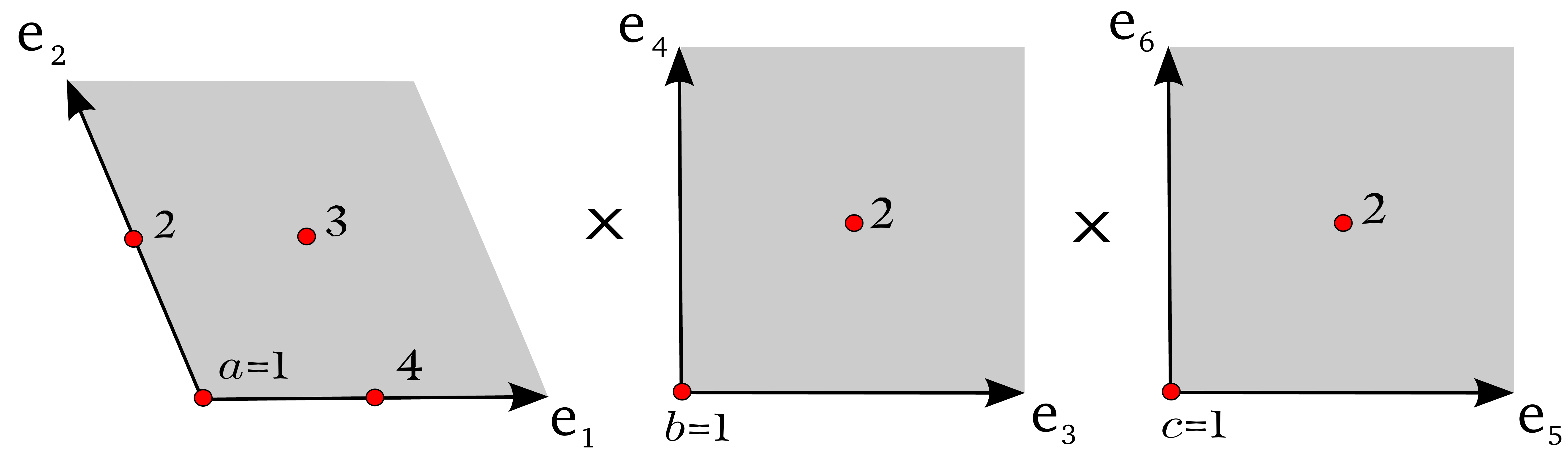}\\
\vspace{0.3cm}
\footnotesize{
\begin{tabular}{|c|c|c|c|c|}
\multicolumn{5}{c}{$T_{(1,1)}$}\\
\hline
\backslashbox{$a$}{$bc$} & 11 & 12 & 21 & 22 \\\hline
1 & $0$ & $e_6$ & $e_4$ & $e_4+e_6$  \\
2 &  $e_2$ &  $e_2+e_6$ & $e_2+e_4$ & $e_2+e_4+e_6$ \\
3 &  $e_1+e_2$ & $e_1+e_2+e_6$ & $e_1+e_2+e_4$ & $e_1+e_2+e_4+e_6$   \\
4 &  $e_1$ & $e_1+e_6$ & $e_1+e_4$ & $e_1+e_4+e_6$  \\\hline
\multicolumn{5}{c}{}\\
\multicolumn{5}{c}{$T_{(1,3)}$}\\
\hline
\backslashbox{$a$}{$bc$} & 11 & 12 & 21 & 22 \\\hline
1 & $0$ & $e_5$ & $e_3$ & $e_3+e_5$  \\
2 &  $e_2$ &  $e_2+e_5$ & $e_2+e_3$ & $e_2+e_3+e_5$ \\
3 &  $e_1+e_2$ & $e_1+e_2+e_5$ & $e_1+e_2+e_3$ & $e_1+e_2+e_3+e_5$   \\
4 &  $e_1$ & $e_1+e_5$ & $e_1+e_3$ & $e_1+e_3+e_5$  \\\hline
\end{tabular}}
\end{center}
\caption{Fixed points of the sectors $T_{(1,1)}$ and $T_{(1,3)}$. }
\label{T11}
\end{table}
\begin{table}[h!]
\centering
\renewcommand{\arraystretch}{1.4}
\begin{center}
\includegraphics[height=30mm,width=100mm]{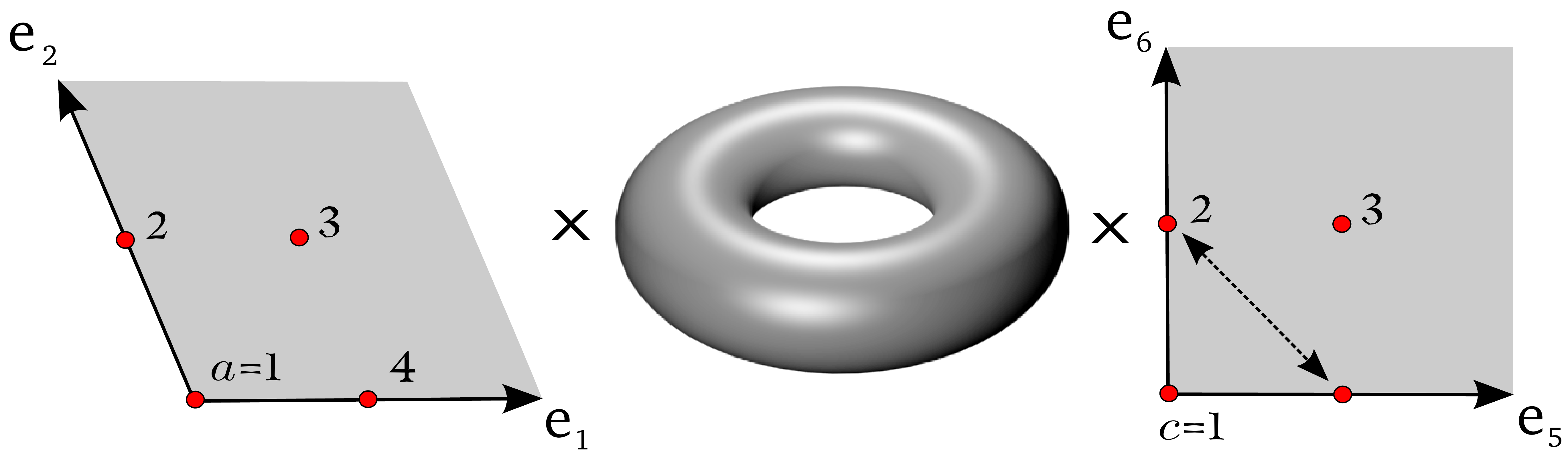}\\
\vspace{0.3cm}
\footnotesize{
\begin{tabular}{|c|c|c|c|}
\multicolumn{4}{c}{$T_{(1,2)}$}\\
\hline
\backslashbox{$a$}{$c$} & 1 & 2 & 3 \\\hline
1 & $0$ & $e_5+e_6$ & \cellcolor[gray]{0.7} $e_6$    \\
2 &  $e_2$ & $e_2+e_5+e_6$ & \cellcolor[gray]{0.7}  $e_2+e_6$   \\
3 &  $e_1+e_2$ &  $e_1+e_2+e_5+e_6$ & \cellcolor[gray]{0.7} $e_1+e_2+e_6$    \\
4 &  $e_1$ &  $e_1+e_5+e_6$ & \cellcolor[gray]{0.7} $e_1+e_6$   \\\hline
\end{tabular}}
\end{center}
\caption{Fixed tori of $T_{(1,2)}$. The picture is very similar to that one observes in $T_{(1,0)}$}
\label{T12}
\end{table}
\subsection{Gauge Embeddings}
\label{embed}
In order to consider all physically inequivalent models we can obtain from the $\mathbbm{Z}_2\times\mathbbm{Z}_4$ orbifold, it is necessary to establish all possible realizations of the gauge embedding which are consistent with the constraints derived from a modular invariant partition function. Physically speaking, two models are \emph{equivalent} if they share the same field content and the same interactions. For models which are free of Wilson lines, the embedding requires only the specification of two shifts describing the action of the $\mathbbm{Z}_2$ and $\mathbbm{Z}_4$ subgroups of the point group.\\\\
Consistency of the embedding requires $2V_2$ and $4V_4$ to belong to the gauge lattice. Modular invariance requires of the following conditions to be satisfied \cite{Ploger:2007iq}: \newpage
\begin{align}
2\left(V_2^2-\frac{1}{2}\right)=0\mod 2 \, ,\label{1mod}\\
4\left(V_4^2-\frac{1}{8}\right)=0\mod 2 \, ,\label{2mod}\\
2\left(V_2\cdot V_4 + \frac{1}{8}\right)=0\mod 2 \, .\label{3mod}
\end{align}
Embeddings related by lattice automorphisms will lead to equivalent spectra. For each inequivalent set of shifts, one finds two inequivalent (\emph{brother}) models upon addition of lattice shifts. We therefore developed an algorithm to find all those inequivalent embeddings and the details are deferred to appendix \ref{app:embeddings}. We found 144 modular invariant embeddings and their brother model which realize all possible breaking structures expected for point groups of order eight \cite{Carter70}. We find that the shifts can break one $E_8$ factor down to $E_6\times U(1)^2$, $E_6\times SU(2)\times U(1)$, $ SO(10)\times SU(4)$, $SO(10) \times SU(2)^2\times U(1)$, $SO(10)\times U(1)^3$, $SU(5)\times SU(3)\times U(1)^2$, $SU(4)^2\times SU(2)\times U(1)$ and $SU(4)\times SU(2)^2\times U(1)$. From them we can see the diversity of gauge groups that can be reached with this orbifold. Note that these groups can support GUT schemes ranging from those based on $E_6$, \SO{10}, Pati-Salam or SU(5). A list of the shift embeddings as well as their gauge group can be found in appendix \ref{app:Shiftembbedings}.

\section{Geometrical Features and Interpretation}
\label{sec:Symmetries}
Equipped with the class of inequivalent models we have identified so far, how many of them are of physical relevance and how can one engineer a strategy to find interesting models. We attempt to keep our discussion in general terms while avoiding any explicit construction. Instead, we combine the mentioned results with our knowledge of the orbifold geometry and the role it plays in physical theories. With this we want to arrive at a refined selection of suitable candidates. The general features of these models will be discussed in this section. A more detailed analysis will be given in a future publication.
\subsection{Gauge Topography}
Among the valid embeddings we see that a vast amount of them have a gauge group containing an \SO{10} or \es\ factor. Our intention is to find out which of them allow for a set of Wilson lines which modifies the physical theory in a phenomenologically meaningful way. Having all shift embeddings we computed the spectra. The results can be found in appendix \ref{app:mattercontent}, where we show the sectors at which \st- or \ts-plets arise. At the GUT level, the multiplicities for the states in $T_{(0,1)}$, $T_{(0,3)}$, $T_{(1,1)}$ and $T_{(1,3)}$ are the same as the number of fixed points/tori in each sector. The same occurs for matter sitting at the ordinary tori of $T_{(0,2)}$, $T_{(1,0)}$ and $T_{(1,2)}$. At the special singularities, the gauge group is enhanced. Hence the special fixed tori contain the same matter as the ordinary ones plus extra states needed to complete representations of the enhanced gauge group. That is the reason for which from now on, special and ordinary tori are discussed in a separate fashion.\\

One can determine how the Wilson lines affect the mass equation and the projectors at each of the singularities. When the WL configuration affects neither of those, the fixed point/torus is \emph{protected} in the sense that the states at the GUT level remain intact. There may also be some \emph{split} singularities, where the mass equation remains unchanged but the the Wilson lines project out some parts of the GUT multiplet. All other fixed points/tori are regarded as \emph{unshielded} because the matter which appears there is extremely sensitive to the specific choice of the WLs.\\
It is clear that having a spectrum at the GUT level only helps to make statements about protected and split singularities. Nevertheless, by insisting on a picture in which (some) families are complete GUT multiplets one can keep track of the splitting of the degeneracies to find out what are the relevant multiplicities under a given configuration. The results of our analysis are depicted in table \ref{degeneraciesZ2Z4}. As pointed out in section \ref{sec:Z2Z4Geometry} one can have four Wilson lines of order two. The combinatorics leads to sixteen configurations, each corresponding to a column. For each configuration we computed the embedding of the generators and their centralizers. Then we look at all those fixed points sharing the same corrections to the mass equation and the same projectors. Each fixed point/tori corresponds to a box in the table. Here the first distinction is made between the ordinary and special singularities: If two special tori have boxes marked with the same color within the same column and sector, they will have the same matter content under the WL configuration the column corresponds to. The same applies to ordinary singularities.\\

Table \ref{degeneraciesZ2Z4} illustrates how the orbifold topography gets affected by the backgrounds and serves to determine the multiplicities in which a state can appear. In table \ref{protected} the points are displayed in the same manner as before, but now, the green boxes represent protected fixed points, blue ones are split and the remainder are unshielded. To explain how to interpret our results consider for instance the configuration number 2, in which only $W_1$ is non trivial. Note that the first complex plane is a fixed torus of $T_{(0,1)}$ and $T_{(0,2)}$. This means that the constructing elements for the fixed tori at those sectors are independent of $e_1$, so that the mass equation at this sectors does not suffer from any modification. However, the presence of the twisted torus implies that any constructing element commutes with $(\mathbbm{1},e_1)$ so that $W_1$ projects out states. This is the reason why all fixed points from the sectors mentioned above appear blue in the second column. The remaining sectors contain some fixed points/tori which are only protected if they are located along the vertical axis of the first plane. For instance, this can be observed in table \ref{T10} where the $T_{(1,0)}$ sector has four ordinary and two special fixed points unaffected and thus highlighted in green.\\

One can now use table \ref{protected} to deduce which are the appealing backgrounds for complete families out of \SO{10} or \es\ , sitting at the protected or split fixed points. One has to recall that no complete families can arise from \st-plets at split singularities since the Wilson lines project out pieces which are needed to complete a family. This situation is a bit more subtle in the case of \es\ because the \ts's can appear at points where only one WL acts. If such a Wilson line is adequately chosen one can retain a whole family at this particular location. This shows that \es\ models are somehow harder to deal with and the distinction between protected and split singularities does not apply in the way it was meant to. With a bit more of focus on \SO{10} models where our argumentation makes perfect sense, let us stress some features of the Wilson line configurations. First of all, one Wilson line is not enough to break an \SO{10} factor down to the SM gauge group, for this reason configurations 1-5 are ruled out. Two WLs are enough to achieve the breaking and one can see from the table that configuration 6 permits a three family configuration provided the presence of GUT multiplets at the ordinary tori of $T_{(1,2)}$. There is one \SO{10} model, denoted by $67_1$ in table \ref{matterso10} of appendix \ref{app:mattercontent} that provides this distribution.\\

Configuration 11 could also work out if one finds a model with matter at $T_{(0,1)}, T_{(0,2)}$ and $T_{(0,3)}$. Only at the level of \es\ one can find a model (number $3_1$ in \ref{mattere6} of appendix \ref{app:mattercontent}) which fulfills these requirements. This possibility is ruled out because two WLs do not suffice for the breaking of \es. The analysis suggests that the only alternative to have three complete families is based on model $67_1$. It can be shown that there are WL configurations which allow for a model with the right spectrum and a non anomalous hypercharge. The main drawback of this model is that the families are almost identical\footnote{Since the \st-plets belong to the same twisted sector, the gauge momentum and the $R$-charge are identical, they are only distinguished by their transformation under the flavor group.}, for this reason, the presence of renormalizable couplings with an untwisted Higgs will benefit all the families, making it very difficult to retain the picture in which the top can be made heavy while the other particles remain light.
\begin{table}[H]
\centering
\renewcommand{\arraystretch}{1.15}
\begin{center}
\tiny{
\begin{tabular}{|c|r|c|c|c|c|c|c|c|c|c|c|c|c|c|c|c|c|}
\cline{2-18}
\multicolumn{1}{c|}{} & Config. & 1 & 2 & 3 & 4 & 5 & 6 & 7 & 8 & 9 & 10 & 11 & 12 & 13 & 14 & 15 & 16 \\\cline{2-18}
\multicolumn{1}{c|}{} & \multicolumn{1}{|c|}{$W_1$} & \multicolumn{1}{c}{} & \multicolumn{1}{c}{$\checkmark$} & \multicolumn{1}{c}{} & \multicolumn{1}{c}{} & \multicolumn{1}{c}{} & \multicolumn{1}{c}{$\checkmark$} & \multicolumn{1}{c}{$\checkmark$} & \multicolumn{1}{c}{$\checkmark$} & \multicolumn{1}{c}{} & \multicolumn{1}{c}{} & \multicolumn{1}{c}{} & \multicolumn{1}{c}{$\checkmark$} & \multicolumn{1}{c}{$\checkmark$} & \multicolumn{1}{c}{$\checkmark$} & \multicolumn{1}{c}{} & \multicolumn{1}{c|}{$\checkmark$} \\\cline{2-2}
\multicolumn{1}{c|}{} & \multicolumn{1}{|c|}{$W_2$} & \multicolumn{1}{c}{} & \multicolumn{1}{c}{} & \multicolumn{1}{c}{$\checkmark$} & \multicolumn{1}{c}{} & \multicolumn{1}{c}{} & \multicolumn{1}{c}{$\checkmark$} & \multicolumn{1}{c}{} & \multicolumn{1}{c}{} & \multicolumn{1}{c}{$\checkmark$} & \multicolumn{1}{c}{$\checkmark$} & \multicolumn{1}{c}{} & \multicolumn{1}{c}{$\checkmark$} & \multicolumn{1}{c}{$\checkmark$} & \multicolumn{1}{c}{} & \multicolumn{1}{c}{$\checkmark$} & \multicolumn{1}{c|}{$\checkmark$} \\\cline{2-2}
\multicolumn{1}{c|}{} & \multicolumn{1}{|c|}{$W_3$} & \multicolumn{1}{c}{} & \multicolumn{1}{c}{} & \multicolumn{1}{c}{} & \multicolumn{1}{c}{$\checkmark$} & \multicolumn{1}{c}{} & \multicolumn{1}{c}{} & \multicolumn{1}{c}{$\checkmark$} & \multicolumn{1}{c}{} & \multicolumn{1}{c}{$\checkmark$} & \multicolumn{1}{c}{} & \multicolumn{1}{c}{$\checkmark$} & \multicolumn{1}{c}{$\checkmark$} & \multicolumn{1}{c}{} & \multicolumn{1}{c}{$\checkmark$} & \multicolumn{1}{c}{$\checkmark$} & \multicolumn{1}{c|}{$\checkmark$} \\\cline{2-2}
\multicolumn{1}{c|}{} & \multicolumn{1}{|c|}{$W_5$} & \multicolumn{1}{c}{} & \multicolumn{1}{c}{} & \multicolumn{1}{c}{} & \multicolumn{1}{c}{} & \multicolumn{1}{c}{$\checkmark$} & \multicolumn{1}{c}{} & \multicolumn{1}{c}{} & \multicolumn{1}{c}{$\checkmark$} & \multicolumn{1}{c}{} & \multicolumn{1}{c}{$\checkmark$} & \multicolumn{1}{c}{$\checkmark$} & \multicolumn{1}{c}{} & \multicolumn{1}{c}{$\checkmark$} & \multicolumn{1}{c}{$\checkmark$} & \multicolumn{1}{c}{$\checkmark$} & \multicolumn{1}{c|}{$\checkmark$} \\\cline{2-18}
\multicolumn{18}{c}{}\\\hline
\multirow {4}*{$T_{(0,1)}$, $T_{(0,3)}$} & $bc=11$ & \cellcolor{green} & \cellcolor{green} & \cellcolor{green} & \cellcolor{green} & \cellcolor{green} & \cellcolor{green} & \cellcolor{green} & \cellcolor{green} & \cellcolor{green} & \cellcolor{green} & \cellcolor{green} & \cellcolor{green} & \cellcolor{green} & \cellcolor{green} & \cellcolor{green} & \cellcolor{green} \\\cline{2-18}
& $12$& \cellcolor{green} & \cellcolor{green} & \cellcolor{green} & \cellcolor{green} & \cellcolor{red} & \cellcolor{green} & \cellcolor{green} & \cellcolor{red} & \cellcolor{green} & \cellcolor{red} & \cellcolor{red} & \cellcolor{green} & \cellcolor{red} & \cellcolor{red} & \cellcolor{red} & \cellcolor{red} \\\cline{2-18}
& $21$& \cellcolor{green} & \cellcolor{green} & \cellcolor{green} & \cellcolor{red} & \cellcolor{green} & \cellcolor{green} & \cellcolor{red} & \cellcolor{green} & \cellcolor{red} & \cellcolor{green} & \cellcolor{blue} & \cellcolor{red} & \cellcolor{green} & \cellcolor{blue} & \cellcolor{blue} & \cellcolor{blue} \\\cline{2-18}
& $22$& \cellcolor{green} & \cellcolor{green} & \cellcolor{green} & \cellcolor{red} & \cellcolor{red} & \cellcolor{green} & \cellcolor{red} & \cellcolor{red} & \cellcolor{red} & \cellcolor{red} & \cellcolor{orange} & \cellcolor{red} & \cellcolor{red} & \cellcolor{orange} & \cellcolor{orange} & \cellcolor{orange} \\\hline
\multicolumn{18}{c}{}\\\hline
\multirow {11}*{$T_{(0,2)}$} & $bc=11$ & \cellcolor{green} & \cellcolor{green} & \cellcolor{green} & \cellcolor{green} & \cellcolor{green} & \cellcolor{green} & \cellcolor{green} & \cellcolor{green} & \cellcolor{green} & \cellcolor{green} & \cellcolor{green} & \cellcolor{green} & \cellcolor{green} & \cellcolor{green} & \cellcolor{green} & \cellcolor{green} \\\cline{2-18}
& $12$& \cellcolor{green} & \cellcolor{green} & \cellcolor{green} & \cellcolor{green} & \cellcolor{red} & \cellcolor{green} & \cellcolor{green} & \cellcolor{red} & \cellcolor{green} & \cellcolor{red} & \cellcolor{red} & \cellcolor{green} & \cellcolor{red} & \cellcolor{red} & \cellcolor{red} & \cellcolor{red} \\\cline{2-18}
& $21$& \cellcolor{green} & \cellcolor{green} & \cellcolor{green} & \cellcolor{red} & \cellcolor{green} & \cellcolor{green} & \cellcolor{red} & \cellcolor{green} & \cellcolor{red} & \cellcolor{green} & \cellcolor{blue} & \cellcolor{red} & \cellcolor{green} & \cellcolor{blue} & \cellcolor{blue} & \cellcolor{blue} \\\cline{2-18}
& $22$& \cellcolor{green} & \cellcolor{green} & \cellcolor{green} & \cellcolor{red} & \cellcolor{red} & \cellcolor{green} & \cellcolor{red} & \cellcolor{red} & \cellcolor{red} & \cellcolor{red} & \cellcolor{orange} & \cellcolor{red} & \cellcolor{red} & \cellcolor{orange} & \cellcolor{orange} & \cellcolor{orange}\\\cline{2-18}
& \multicolumn{17}{|c}{}\\\cline{2-18}
&\cellcolor[gray]{0.7} $13$& \cellcolor{green} & \cellcolor{green} & \cellcolor{green} & \cellcolor{green} & \cellcolor{green} & \cellcolor{green} & \cellcolor{green} & \cellcolor{green} & \cellcolor{green} & \cellcolor{green} & \cellcolor{green} & \cellcolor{green} & \cellcolor{green} & \cellcolor{green} & \cellcolor{green} & \cellcolor{green} \\\cline{2-18}
&\cellcolor[gray]{0.7} $31$& \cellcolor{green} & \cellcolor{green} & \cellcolor{green} & \cellcolor{red} & \cellcolor{red} & \cellcolor{green} & \cellcolor{red} & \cellcolor{red} & \cellcolor{red} & \cellcolor{red} & \cellcolor{red} & \cellcolor{red} & \cellcolor{red} & \cellcolor{red} & \cellcolor{red} & \cellcolor{red} \\\cline{2-18}
&\cellcolor[gray]{0.7} $33$& \cellcolor{green} & \cellcolor{green} & \cellcolor{green} & \cellcolor{red} & \cellcolor{green} & \cellcolor{green} & \cellcolor{red} & \cellcolor{green} & \cellcolor{red} & \cellcolor{green} & \cellcolor{blue} & \cellcolor{red} & \cellcolor{green} & \cellcolor{blue} & \cellcolor{blue} & \cellcolor{blue} \\\cline{2-18}
&\cellcolor[gray]{0.7} $34$& \cellcolor{green} & \cellcolor{green} & \cellcolor{green} & \cellcolor{red} & \cellcolor{green} & \cellcolor{green} & \cellcolor{red} & \cellcolor{green} & \cellcolor{red} & \cellcolor{green} & \cellcolor{blue} & \cellcolor{red} & \cellcolor{green} & \cellcolor{blue} & \cellcolor{blue} & \cellcolor{blue} \\\cline{2-18}
&\cellcolor[gray]{0.7} $32$& \cellcolor{green} & \cellcolor{green} & \cellcolor{green} & \cellcolor{red} & \cellcolor{red} & \cellcolor{green} & \cellcolor{red} & \cellcolor{red} & \cellcolor{red} & \cellcolor{red} & \cellcolor{red} & \cellcolor{red} & \cellcolor{red} & \cellcolor{red} & \cellcolor{red} & \cellcolor{red} \\\cline{2-18}
&\cellcolor[gray]{0.7} $23$& \cellcolor{green} & \cellcolor{green} & \cellcolor{green} & \cellcolor{green} & \cellcolor{green} & \cellcolor{green} & \cellcolor{green} & \cellcolor{green} & \cellcolor{green} & \cellcolor{green} & \cellcolor{green} & \cellcolor{green} & \cellcolor{green} & \cellcolor{green} & \cellcolor{green} & \cellcolor{green} \\\hline
\multicolumn{18}{c}{}\\\hline
\multirow {13}*{$T_{(1,0)}$} & $ab=11$ & \cellcolor{green} & \cellcolor{green} & \cellcolor{green} & \cellcolor{green} & \cellcolor{green} & \cellcolor{green} & \cellcolor{green} & \cellcolor{green} & \cellcolor{green} & \cellcolor{green} & \cellcolor{green} & \cellcolor{green} & \cellcolor{green} & \cellcolor{green} & \cellcolor{green} & \cellcolor{green} \\\cline{2-18}
& $12$& \cellcolor{green} & \cellcolor{green} & \cellcolor{green} & \cellcolor{red} & \cellcolor{green} & \cellcolor{green} & \cellcolor{red} & \cellcolor{green} & \cellcolor{red} & \cellcolor{green} & \cellcolor{red} & \cellcolor{red} & \cellcolor{green} & \cellcolor{red} & \cellcolor{red} & \cellcolor{red} \\\cline{2-18}
& $21$& \cellcolor{green} & \cellcolor{green} & \cellcolor{red} & \cellcolor{green} & \cellcolor{green} & \cellcolor{red} & \cellcolor{green} & \cellcolor{green} & \cellcolor{blue} & \cellcolor{red} & \cellcolor{green} & \cellcolor{blue} & \cellcolor{red} & \cellcolor{green} & \cellcolor{blue} & \cellcolor{blue} \\\cline{2-18}
& $22$& \cellcolor{green} & \cellcolor{green} & \cellcolor{red} & \cellcolor{red} & \cellcolor{green} & \cellcolor{red} & \cellcolor{red} & \cellcolor{green} & \cellcolor{orange} & \cellcolor{red} & \cellcolor{red} & \cellcolor{orange} & \cellcolor{red} & \cellcolor{red} & \cellcolor{orange} & \cellcolor{orange} \\\cline{2-18}
& $41$& \cellcolor{green} & \cellcolor{red} & \cellcolor{green} & \cellcolor{green} & \cellcolor{green} & \cellcolor{blue} & \cellcolor{blue} & \cellcolor{red} & \cellcolor{green} & \cellcolor{green} & \cellcolor{green} & \cellcolor{violet} & \cellcolor{blue} & \cellcolor{blue} & \cellcolor{green} & \cellcolor{violet} \\\cline{2-18}
& $42$& \cellcolor{green} & \cellcolor{red} & \cellcolor{green} & \cellcolor{red} & \cellcolor{green} & \cellcolor{blue} & \cellcolor{orange} & \cellcolor{red} & \cellcolor{red} & \cellcolor{green} & \cellcolor{red} & \cellcolor{yellow} & \cellcolor{blue} & \cellcolor{orange} & \cellcolor{red} & \cellcolor{yellow} \\\cline{2-18}
& $31$& \cellcolor{green} & \cellcolor{red} & \cellcolor{red} & \cellcolor{green} & \cellcolor{green} & \cellcolor{orange} & \cellcolor{blue} & \cellcolor{red} & \cellcolor{blue} & \cellcolor{red} & \cellcolor{green} & \cellcolor{cyan} & \cellcolor{orange} & \cellcolor{blue} & \cellcolor{blue} & \cellcolor{cyan} \\\cline{2-18}
& $32$& \cellcolor{green} & \cellcolor{red} & \cellcolor{red} & \cellcolor{red} & \cellcolor{green} & \cellcolor{orange} & \cellcolor{orange} & \cellcolor{red} & \cellcolor{orange} & \cellcolor{red} & \cellcolor{red} & \cellcolor{teal} & \cellcolor{orange} & \cellcolor{orange} & \cellcolor{orange} & \cellcolor{teal} \\\cline{2-18}
& \multicolumn{17}{|c}{}\\\cline{2-18}
&\cellcolor[gray]{0.7} $13$& \cellcolor{green} & \cellcolor{green} & \cellcolor{green} & \cellcolor{green} & \cellcolor{green} & \cellcolor{green} & \cellcolor{green} & \cellcolor{green} & \cellcolor{green} & \cellcolor{green} & \cellcolor{green} & \cellcolor{green} & \cellcolor{green} & \cellcolor{green} & \cellcolor{green} & \cellcolor{green} \\\cline{2-18}
&\cellcolor[gray]{0.7} $23$& \cellcolor{green} & \cellcolor{green} & \cellcolor{red} & \cellcolor{green} & \cellcolor{green} & \cellcolor{red} & \cellcolor{green} & \cellcolor{green} & \cellcolor{red} & \cellcolor{red} & \cellcolor{green} & \cellcolor{red} & \cellcolor{red} & \cellcolor{green} & \cellcolor{red} & \cellcolor{red} \\\cline{2-18}
&\cellcolor[gray]{0.7} $43$& \cellcolor{green} & \cellcolor{red} & \cellcolor{green} & \cellcolor{green} & \cellcolor{green} & \cellcolor{blue} & \cellcolor{red} & \cellcolor{red} & \cellcolor{green} & \cellcolor{green} & \cellcolor{green} & \cellcolor{blue} & \cellcolor{blue} & \cellcolor{red} & \cellcolor{green} & \cellcolor{blue} \\\cline{2-18}
&\cellcolor[gray]{0.7} $33$& \cellcolor{green} & \cellcolor{red} & \cellcolor{red} & \cellcolor{green} & \cellcolor{green} & \cellcolor{orange} & \cellcolor{red} & \cellcolor{red} & \cellcolor{red} & \cellcolor{red} & \cellcolor{green} & \cellcolor{orange} & \cellcolor{orange} & \cellcolor{red} & \cellcolor{red} & \cellcolor{orange} \\\hline
\multicolumn{18}{c}{}\\\hline
\multirow {16}*{$T_{(1,1)}$, $T_{(1,3)}$} & $abc=111$& \cellcolor{green} & \cellcolor{green} & \cellcolor{green} & \cellcolor{green} & \cellcolor{green} & \cellcolor{green} & \cellcolor{green} & \cellcolor{green} & \cellcolor{green} & \cellcolor{green} & \cellcolor{green} & \cellcolor{green} & \cellcolor{green} & \cellcolor{green} & \cellcolor{green} & \cellcolor{green} \\\cline{2-18}
& $112$& \cellcolor{green} & \cellcolor{green} & \cellcolor{green} & \cellcolor{green} & \cellcolor{red} & \cellcolor{green} & \cellcolor{green} & \cellcolor{red} & \cellcolor{green} & \cellcolor{red} & \cellcolor{red} & \cellcolor{green} & \cellcolor{red} & \cellcolor{red} & \cellcolor{red} & \cellcolor{red} \\\cline{2-18}
& $121$& \cellcolor{green} & \cellcolor{green} & \cellcolor{green} & \cellcolor{red} & \cellcolor{green} & \cellcolor{green} & \cellcolor{red} & \cellcolor{green} & \cellcolor{red} & \cellcolor{green} & \cellcolor{blue} & \cellcolor{red} & \cellcolor{green} & \cellcolor{blue} & \cellcolor{blue} & \cellcolor{blue} \\\cline{2-18}
& $122$& \cellcolor{green} & \cellcolor{green} & \cellcolor{green} & \cellcolor{red} & \cellcolor{red} & \cellcolor{green} & \cellcolor{red} & \cellcolor{red} & \cellcolor{red} & \cellcolor{red} & \cellcolor{orange} & \cellcolor{red} & \cellcolor{red} & \cellcolor{orange} & \cellcolor{orange} & \cellcolor{orange} \\\cline{2-18}
& $211$& \cellcolor{green} & \cellcolor{green} & \cellcolor{red} & \cellcolor{green} & \cellcolor{green} & \cellcolor{red} & \cellcolor{green} & \cellcolor{green} & \cellcolor{blue} & \cellcolor{blue} & \cellcolor{green} & \cellcolor{blue} & \cellcolor{blue} & \cellcolor{green} & \cellcolor{violet} & \cellcolor{violet} \\\cline{2-18}
& $212$& \cellcolor{green} & \cellcolor{green} & \cellcolor{red} & \cellcolor{green} & \cellcolor{red} & \cellcolor{red} & \cellcolor{green} & \cellcolor{red} & \cellcolor{blue} & \cellcolor{orange} & \cellcolor{red} & \cellcolor{blue} & \cellcolor{orange} & \cellcolor{red} & \cellcolor{yellow} & \cellcolor{yellow} \\\cline{2-18}
& $221$& \cellcolor{green} & \cellcolor{green} & \cellcolor{red} & \cellcolor{red} & \cellcolor{green} & \cellcolor{red} & \cellcolor{red} & \cellcolor{green} & \cellcolor{orange} & \cellcolor{blue} & \cellcolor{blue} & \cellcolor{orange} & \cellcolor{blue} & \cellcolor{blue} & \cellcolor{cyan} & \cellcolor{cyan} \\\cline{2-18}
& $222$& \cellcolor{green} & \cellcolor{green} & \cellcolor{red} & \cellcolor{red} & \cellcolor{red} & \cellcolor{red} & \cellcolor{red} & \cellcolor{red} & \cellcolor{orange} & \cellcolor{orange} & \cellcolor{orange} & \cellcolor{orange} & \cellcolor{orange} & \cellcolor{orange} & \cellcolor{teal} & \cellcolor{teal} \\\cline{2-18}
& $411$& \cellcolor{green} & \cellcolor{red} & \cellcolor{green} & \cellcolor{green} & \cellcolor{green} & \cellcolor{blue} & \cellcolor{blue} & \cellcolor{blue} & \cellcolor{green} & \cellcolor{green} & \cellcolor{green} & \cellcolor{violet} & \cellcolor{violet} & \cellcolor{violet} & \cellcolor{green} & \cellcolor{magenta} \\\cline{2-18}
& $412$& \cellcolor{green} & \cellcolor{red} & \cellcolor{green} & \cellcolor{green} & \cellcolor{red} & \cellcolor{blue} & \cellcolor{blue} & \cellcolor{orange} & \cellcolor{green} & \cellcolor{red} & \cellcolor{red} & \cellcolor{violet} & \cellcolor{yellow} & \cellcolor{yellow} & \cellcolor{red} & \cellcolor{pink} \\\cline{2-18}
& $421$& \cellcolor{green} & \cellcolor{red} & \cellcolor{green} & \cellcolor{red} & \cellcolor{green} & \cellcolor{blue} & \cellcolor{orange} & \cellcolor{blue} & \cellcolor{red} & \cellcolor{green} & \cellcolor{blue} & \cellcolor{yellow} & \cellcolor{violet} & \cellcolor{cyan} & \cellcolor{blue} & \cellcolor{brown} \\\cline{2-18}
& $422$& \cellcolor{green} & \cellcolor{red} & \cellcolor{green} & \cellcolor{red} & \cellcolor{red} & \cellcolor{blue} & \cellcolor{orange} & \cellcolor{orange} & \cellcolor{red} & \cellcolor{red} & \cellcolor{orange} & \cellcolor{yellow} & \cellcolor{yellow} & \cellcolor{teal} & \cellcolor{orange} & \cellcolor{olive} \\\cline{2-18}
& $311$& \cellcolor{green} & \cellcolor{red} & \cellcolor{red} & \cellcolor{green} & \cellcolor{green} & \cellcolor{orange} & \cellcolor{blue} & \cellcolor{blue} & \cellcolor{blue} & \cellcolor{blue} & \cellcolor{green} & \cellcolor{cyan} & \cellcolor{cyan} & \cellcolor{violet} & \cellcolor{violet} & \cellcolor{purple} \\\cline{2-18}
& $312$& \cellcolor{green} & \cellcolor{red} & \cellcolor{red} & \cellcolor{green} & \cellcolor{red} & \cellcolor{orange} & \cellcolor{blue} & \cellcolor{orange} & \cellcolor{blue} & \cellcolor{orange} & \cellcolor{red} & \cellcolor{cyan} & \cellcolor{teal} & \cellcolor{yellow} & \cellcolor{yellow} & \cellcolor{black} \\\cline{2-18}
& $321$& \cellcolor{green} & \cellcolor{red} & \cellcolor{red} & \cellcolor{red} & \cellcolor{green} & \cellcolor{orange} & \cellcolor{orange} & \cellcolor{blue} & \cellcolor{orange} & \cellcolor{blue} & \cellcolor{blue} & \cellcolor{teal} & \cellcolor{cyan} & \cellcolor{cyan} & \cellcolor{cyan} & \cellcolor{white} \\\cline{2-18}
& $322$& \cellcolor{green} & \cellcolor{red} & \cellcolor{red} & \cellcolor{red} & \cellcolor{red} & \cellcolor{orange} & \cellcolor{orange} & \cellcolor{orange} & \cellcolor{orange} & \cellcolor{orange} & \cellcolor{orange} & \cellcolor{teal} & \cellcolor{teal} & \cellcolor{teal} & \cellcolor{teal} & \cellcolor{darkgray} \\\hline
\multicolumn{18}{c}{}\\\hline
\multirow {13}*{$T_{(1,2)}$} & $ac=11$ & \cellcolor{green} & \cellcolor{green} & \cellcolor{green} & \cellcolor{green} & \cellcolor{green} & \cellcolor{green} & \cellcolor{green} & \cellcolor{green} & \cellcolor{green} & \cellcolor{green} & \cellcolor{green} & \cellcolor{green} & \cellcolor{green} & \cellcolor{green} & \cellcolor{green} & \cellcolor{green} \\\cline{2-18}
& $12$& \cellcolor{green} & \cellcolor{green} & \cellcolor{green} & \cellcolor{green} & \cellcolor{red} & \cellcolor{green} & \cellcolor{green} & \cellcolor{red} & \cellcolor{green} & \cellcolor{red} & \cellcolor{red} & \cellcolor{green} & \cellcolor{red} & \cellcolor{red} & \cellcolor{red} & \cellcolor{red} \\\cline{2-18}
& $21$& \cellcolor{green} & \cellcolor{green} & \cellcolor{red} & \cellcolor{green} & \cellcolor{green} & \cellcolor{red} & \cellcolor{green} & \cellcolor{green} & \cellcolor{red} & \cellcolor{blue} & \cellcolor{green} & \cellcolor{red} & \cellcolor{blue} & \cellcolor{green} & \cellcolor{blue} & \cellcolor{blue} \\\cline{2-18}
& $22$& \cellcolor{green} & \cellcolor{green} & \cellcolor{red} & \cellcolor{green} & \cellcolor{red} & \cellcolor{red} & \cellcolor{green} & \cellcolor{red} & \cellcolor{red} & \cellcolor{orange} & \cellcolor{red} & \cellcolor{red} & \cellcolor{orange} & \cellcolor{red} & \cellcolor{orange} & \cellcolor{orange} \\\cline{2-18}
& $41$& \cellcolor{green} & \cellcolor{red} & \cellcolor{green} & \cellcolor{green} & \cellcolor{green} & \cellcolor{blue} & \cellcolor{red} & \cellcolor{blue} & \cellcolor{green} & \cellcolor{green} & \cellcolor{green} & \cellcolor{blue} & \cellcolor{violet} & \cellcolor{blue} & \cellcolor{green} & \cellcolor{violet} \\\cline{2-18}
& $42$& \cellcolor{green} & \cellcolor{red} & \cellcolor{green} & \cellcolor{green} & \cellcolor{red} & \cellcolor{blue} & \cellcolor{red} & \cellcolor{orange} & \cellcolor{green} & \cellcolor{red} & \cellcolor{red} & \cellcolor{blue} & \cellcolor{yellow} & \cellcolor{orange} & \cellcolor{red} & \cellcolor{yellow} \\\cline{2-18}
& $31$& \cellcolor{green} & \cellcolor{red} & \cellcolor{red} & \cellcolor{green} & \cellcolor{green} & \cellcolor{orange} & \cellcolor{red} & \cellcolor{blue} & \cellcolor{red} & \cellcolor{blue} & \cellcolor{green} & \cellcolor{orange} & \cellcolor{cyan} & \cellcolor{blue} & \cellcolor{blue} & \cellcolor{cyan} \\\cline{2-18}
& $32$& \cellcolor{green} & \cellcolor{red} & \cellcolor{red} & \cellcolor{green} & \cellcolor{red} & \cellcolor{orange} & \cellcolor{red} & \cellcolor{orange} & \cellcolor{red} & \cellcolor{orange} & \cellcolor{red} & \cellcolor{orange} & \cellcolor{teal} & \cellcolor{orange} & \cellcolor{orange} & \cellcolor{teal} \\\cline{2-18}
& \multicolumn{17}{|c}{}\\\cline{2-18}
&\cellcolor[gray]{0.7} $13$& \cellcolor{green} & \cellcolor{green} & \cellcolor{green} & \cellcolor{green} & \cellcolor{green} & \cellcolor{green} & \cellcolor{green} & \cellcolor{green} & \cellcolor{green} & \cellcolor{green} & \cellcolor{green} & \cellcolor{green} & \cellcolor{green} & \cellcolor{green} & \cellcolor{green} & \cellcolor{green}  \\\cline{2-18}
&\cellcolor[gray]{0.7} $23$& \cellcolor{green} & \cellcolor{green} & \cellcolor{red} & \cellcolor{green} & \cellcolor{green} & \cellcolor{red} & \cellcolor{green} & \cellcolor{green} & \cellcolor{red} & \cellcolor{red} & \cellcolor{green} & \cellcolor{red} & \cellcolor{red} & \cellcolor{green} & \cellcolor{red} & \cellcolor{red} \\\cline{2-18}
&\cellcolor[gray]{0.7} $43$& \cellcolor{green} & \cellcolor{red} & \cellcolor{green} & \cellcolor{green} & \cellcolor{green} & \cellcolor{blue} & \cellcolor{red} & \cellcolor{red} & \cellcolor{green} & \cellcolor{green} & \cellcolor{green} & \cellcolor{blue} & \cellcolor{blue} & \cellcolor{red} & \cellcolor{green} & \cellcolor{blue} \\\cline{2-18}
&\cellcolor[gray]{0.7} $33$& \cellcolor{green} & \cellcolor{red} & \cellcolor{red} & \cellcolor{green} & \cellcolor{green} & \cellcolor{orange} & \cellcolor{red} & \cellcolor{red} & \cellcolor{red} & \cellcolor{red} & \cellcolor{green} & \cellcolor{orange} & \cellcolor{orange} & \cellcolor{red} & \cellcolor{red} & \cellcolor{orange} \\\hline
\end{tabular}}
\end{center}
\caption{Constraints on the physical states for different configurations of Wilson lines in $\mathbbm{Z}_2\times\mathbbm{Z}_4$. Within the same twisted sector, boxes with the same color correspond to singularities sharing the same set of physical states. For the case of $T_{(0,2)}$, $T_{(1,0)}$ and $T_{(1,2)}$ we have separated special from ordinary fixed tori since they have different projection conditions at the level in which only the point group is embedded.}
\label{degeneraciesZ2Z4}
\end{table}

\begin{table}[H]
\centering
\renewcommand{\arraystretch}{1.15}
\begin{center}
\tiny{
\begin{tabular}{|c|r|c|c|c|c|c|c|c|c|c|c|c|c|c|c|c|c|}
\cline{2-18}
\multicolumn{1}{c|}{} & Config. & 1 & 2 & 3 & 4 & 5 & 6 & 7 & 8 & 9 & 10 & 11 & 12 & 13 & 14 & 15 & 16 \\\cline{2-18}
\multicolumn{1}{c|}{} & \multicolumn{1}{|c|}{$W_1$} & \multicolumn{1}{c}{} & \multicolumn{1}{c}{$\checkmark$} & \multicolumn{1}{c}{} & \multicolumn{1}{c}{} & \multicolumn{1}{c}{} & \multicolumn{1}{c}{$\checkmark$} & \multicolumn{1}{c}{$\checkmark$} & \multicolumn{1}{c}{$\checkmark$} & \multicolumn{1}{c}{} & \multicolumn{1}{c}{} & \multicolumn{1}{c}{} & \multicolumn{1}{c}{$\checkmark$} & \multicolumn{1}{c}{$\checkmark$} & \multicolumn{1}{c}{$\checkmark$} & \multicolumn{1}{c}{} & \multicolumn{1}{c|}{$\checkmark$} \\\cline{2-2}
\multicolumn{1}{c|}{} & \multicolumn{1}{|c|}{$W_2$} & \multicolumn{1}{c}{} & \multicolumn{1}{c}{} & \multicolumn{1}{c}{$\checkmark$} & \multicolumn{1}{c}{} & \multicolumn{1}{c}{} & \multicolumn{1}{c}{$\checkmark$} & \multicolumn{1}{c}{} & \multicolumn{1}{c}{} & \multicolumn{1}{c}{$\checkmark$} & \multicolumn{1}{c}{$\checkmark$} & \multicolumn{1}{c}{} & \multicolumn{1}{c}{$\checkmark$} & \multicolumn{1}{c}{$\checkmark$} & \multicolumn{1}{c}{} & \multicolumn{1}{c}{$\checkmark$} & \multicolumn{1}{c|}{$\checkmark$} \\\cline{2-2}
\multicolumn{1}{c|}{} & \multicolumn{1}{|c|}{$W_3$} & \multicolumn{1}{c}{} & \multicolumn{1}{c}{} & \multicolumn{1}{c}{} & \multicolumn{1}{c}{$\checkmark$} & \multicolumn{1}{c}{} & \multicolumn{1}{c}{} & \multicolumn{1}{c}{$\checkmark$} & \multicolumn{1}{c}{} & \multicolumn{1}{c}{$\checkmark$} & \multicolumn{1}{c}{} & \multicolumn{1}{c}{$\checkmark$} & \multicolumn{1}{c}{$\checkmark$} & \multicolumn{1}{c}{} & \multicolumn{1}{c}{$\checkmark$} & \multicolumn{1}{c}{$\checkmark$} & \multicolumn{1}{c|}{$\checkmark$} \\\cline{2-2}
\multicolumn{1}{c|}{} & \multicolumn{1}{|c|}{$W_5$} & \multicolumn{1}{c}{} & \multicolumn{1}{c}{} & \multicolumn{1}{c}{} & \multicolumn{1}{c}{} & \multicolumn{1}{c}{$\checkmark$} & \multicolumn{1}{c}{} & \multicolumn{1}{c}{} & \multicolumn{1}{c}{$\checkmark$} & \multicolumn{1}{c}{} & \multicolumn{1}{c}{$\checkmark$} & \multicolumn{1}{c}{$\checkmark$} & \multicolumn{1}{c}{} & \multicolumn{1}{c}{$\checkmark$} & \multicolumn{1}{c}{$\checkmark$} & \multicolumn{1}{c}{$\checkmark$} & \multicolumn{1}{c|}{$\checkmark$} \\\cline{2-18}
\multicolumn{18}{c}{}\\\hline
\multirow {4}*{$T_{(0,1)}$, $T_{(0,3)}$} & $bc=11$ & \cellcolor{green} & \cellcolor{blue} & \cellcolor{blue} & \cellcolor{green} & \cellcolor{green} & \cellcolor{blue} & \cellcolor{blue} & \cellcolor{blue} & \cellcolor{blue} & \cellcolor{blue} & \cellcolor{green} & \cellcolor{blue} & \cellcolor{blue} & \cellcolor{blue} & \cellcolor{blue} & \cellcolor{blue}\\\cline{2-18}
& $12$& \cellcolor{green} & \cellcolor{blue} & \cellcolor{blue} & \cellcolor{green} &   & \cellcolor{blue} & \cellcolor{blue} &   & \cellcolor{blue} &   &   & \cellcolor{blue} &   &   &   &  \\\cline{2-18}
& $21$& \cellcolor{green} & \cellcolor{blue} & \cellcolor{blue} &   & \cellcolor{green} & \cellcolor{blue} &   & \cellcolor{blue} &   & \cellcolor{blue} &   &   & \cellcolor{blue} &   &   &  \\\cline{2-18}
& $22$& \cellcolor{green} & \cellcolor{blue} & \cellcolor{blue} &   &   & \cellcolor{blue} &   &   &   &   &   &   &   &   &   &  \\\hline
\multicolumn{18}{c}{}\\\hline
\multirow {11}*{$T_{(0,2)}$} & $bc=11$ & \cellcolor{green} & \cellcolor{blue} & \cellcolor{blue} & \cellcolor{green} & \cellcolor{green} & \cellcolor{blue} & \cellcolor{blue} & \cellcolor{blue} & \cellcolor{blue} & \cellcolor{blue} & \cellcolor{green} & \cellcolor{blue} & \cellcolor{blue} & \cellcolor{blue} & \cellcolor{blue} & \cellcolor{blue} \\\cline{2-18}
& $12$& \cellcolor{green} & \cellcolor{blue} & \cellcolor{blue} & \cellcolor{green} &   & \cellcolor{blue} & \cellcolor{blue} &   & \cellcolor{blue} &   &   & \cellcolor{blue} &   &   &   &  \\\cline{2-18}
& $21$& \cellcolor{green} & \cellcolor{blue} & \cellcolor{blue} &   & \cellcolor{green} & \cellcolor{blue} &   & \cellcolor{blue} &   & \cellcolor{blue} &   &   & \cellcolor{blue} &   &   &  \\\cline{2-18}
& $22$& \cellcolor{green} & \cellcolor{blue} & \cellcolor{blue} &   &   & \cellcolor{blue} &   &   &   &   &   &   &   &   &   & \\\cline{2-18}
& \multicolumn{17}{|c}{}\\\cline{2-18}
&\cellcolor[gray]{0.7} $13$& \cellcolor{green} & \cellcolor{blue} & \cellcolor{blue} & \cellcolor{green} &  & \cellcolor{blue} & \cellcolor{blue} &   & \cellcolor{blue} &   &   & \cellcolor{blue} &   &  &   &   \\\cline{2-18}
&\cellcolor[gray]{0.7} $31$& \cellcolor{green} & \cellcolor{blue} & \cellcolor{blue} &   & \cellcolor{green}  & \cellcolor{blue} &   & \cellcolor{blue} &   & \cellcolor{blue} &   &   & \cellcolor{blue} &   &   &  \\\cline{2-18}
&\cellcolor[gray]{0.7} $33$& \cellcolor{green} & \cellcolor{blue} & \cellcolor{blue} &   &  & \cellcolor{blue} &   &   &   &   &   &   &   &   &   &  \\\cline{2-18}
&\cellcolor[gray]{0.7} $34$& \cellcolor{green} & \cellcolor{blue} & \cellcolor{blue} &   &  & \cellcolor{blue} &   &   &   &   &   &   &   &   &   &  \\\cline{2-18}
&\cellcolor[gray]{0.7} $32$& \cellcolor{green} & \cellcolor{blue} & \cellcolor{blue} &   & \cellcolor{green}  & \cellcolor{blue} &   & \cellcolor{blue} &   & \cellcolor{blue} &   &   & \cellcolor{blue} &   &   &  \\\cline{2-18}
&\cellcolor[gray]{0.7} $23$& \cellcolor{green} & \cellcolor{blue} & \cellcolor{blue} & \cellcolor{green} &  & \cellcolor{blue} & \cellcolor{blue} &   & \cellcolor{blue} &   &   & \cellcolor{blue} &   &   &   &  \\\hline
\multicolumn{18}{c}{}\\\hline
\multirow {13}*{$T_{(1,0)}$} & $ab=11$ & \cellcolor{green} & \cellcolor{green} & \cellcolor{green} & \cellcolor{green} & \cellcolor{blue} & \cellcolor{green} & \cellcolor{green} & \cellcolor{blue} & \cellcolor{green} & \cellcolor{blue} & \cellcolor{blue} & \cellcolor{green} & \cellcolor{blue} & \cellcolor{blue} & \cellcolor{blue} & \cellcolor{blue} \\\cline{2-18}
& $12$& \cellcolor{green} & \cellcolor{green} & \cellcolor{green} &   & \cellcolor{blue} & \cellcolor{green} &   & \cellcolor{blue} &   & \cellcolor{blue} &   &   & \cellcolor{blue} &   &   &   \\\cline{2-18}
& $21$& \cellcolor{green} & \cellcolor{green} &   & \cellcolor{green} & \cellcolor{blue} &   & \cellcolor{green} & \cellcolor{blue} &   &   & \cellcolor{blue} &   &   & \cellcolor{blue} &   &  \\\cline{2-18}
& $22$& \cellcolor{green} & \cellcolor{green} &   &   & \cellcolor{blue} &   &   & \cellcolor{blue} &   &   &   &   &   &   &   &  \\\cline{2-18}
& $41$& \cellcolor{green} &   & \cellcolor{green} & \cellcolor{green} & \cellcolor{blue} &   &   &   & \cellcolor{green} & \cellcolor{blue} & \cellcolor{blue} &   &   &   & \cellcolor{blue} &  \\\cline{2-18}
& $42$& \cellcolor{green} &   & \cellcolor{green} &   & \cellcolor{blue} &   &   &   &   & \cellcolor{blue} &   &   &   &   &   &  \\\cline{2-18}
& $31$& \cellcolor{green} &   &   & \cellcolor{green} & \cellcolor{blue} &   &   &   &   &   & \cellcolor{blue} &   &   &   &   &  \\\cline{2-18}
& $32$& \cellcolor{green} &   &   &   & \cellcolor{blue} &   &   &   &   &   &   &   &   &   &   &  \\\cline{2-18}
& \multicolumn{17}{|c}{}\\\cline{2-18}
&\cellcolor[gray]{0.7} $13$& \cellcolor{green} & \cellcolor{green} & \cellcolor{green} &   & \cellcolor{blue} & \cellcolor{green} &   & \cellcolor{blue} &   & \cellcolor{blue} &  &  & \cellcolor{blue} &  &  &  \\\cline{2-18}
&\cellcolor[gray]{0.7} $23$& \cellcolor{green} & \cellcolor{green} &   &   & \cellcolor{blue} &   &   & \cellcolor{blue} &   &   &  &   &   &  &   &   \\\cline{2-18}
&\cellcolor[gray]{0.7} $43$& \cellcolor{green} &   & \cellcolor{green} &   & \cellcolor{blue} &   &   &   &   & \cellcolor{blue} &  &   &   &   &  &  \\\cline{2-18}
&\cellcolor[gray]{0.7} $33$& \cellcolor{green} &   &   &   & \cellcolor{blue} &   &   &   &   &   &  &   &   &   &   &  \\\hline
\multicolumn{18}{c}{}\\\hline
\multirow {16}*{$T_{(1,1)}$, $T_{(1,3)}$} & $abc=111$& \cellcolor{green} & \cellcolor{green} & \cellcolor{green} & \cellcolor{green} & \cellcolor{green} & \cellcolor{green} & \cellcolor{green} & \cellcolor{green} & \cellcolor{green} & \cellcolor{green} & \cellcolor{green} & \cellcolor{green} & \cellcolor{green} & \cellcolor{green} & \cellcolor{green} & \cellcolor{green} \\\cline{2-18}
& $112$& \cellcolor{green} & \cellcolor{green} & \cellcolor{green} & \cellcolor{green} &  & \cellcolor{green} & \cellcolor{green} &  & \cellcolor{green} &  &  & \cellcolor{green} &  &  &  &  \\\cline{2-18}
& $121$& \cellcolor{green} & \cellcolor{green} & \cellcolor{green} &  & \cellcolor{green} & \cellcolor{green} &  & \cellcolor{green} &  & \cellcolor{green} &  &  & \cellcolor{green} &  &  &  \\\cline{2-18}
& $122$& \cellcolor{green} & \cellcolor{green} & \cellcolor{green} &  &  & \cellcolor{green} &  &  &  &  &  &  &  &  &  &  \\\cline{2-18}
& $211$& \cellcolor{green} & \cellcolor{green} &  & \cellcolor{green} & \cellcolor{green} &  & \cellcolor{green} & \cellcolor{green} &  &  & \cellcolor{green} &  &  & \cellcolor{green} &  &  \\\cline{2-18}
& $212$& \cellcolor{green} & \cellcolor{green} &  & \cellcolor{green} &  &  & \cellcolor{green} &  &  &  &  &  &  &  &  &  \\\cline{2-18}
& $221$& \cellcolor{green} & \cellcolor{green} &  &  & \cellcolor{green} &  &  & \cellcolor{green} &  &  &  &  &  &  &  &  \\\cline{2-18}
& $222$& \cellcolor{green} & \cellcolor{green} &  &  &  &  &  &  &  &  &  &  &  &  &  &  \\\cline{2-18}
& $411$& \cellcolor{green} &   & \cellcolor{green} & \cellcolor{green} & \cellcolor{green} &  &  &  & \cellcolor{green} & \cellcolor{green} & \cellcolor{green} &  &  &  & \cellcolor{green} &  \\\cline{2-18}
& $412$& \cellcolor{green} &   & \cellcolor{green} & \cellcolor{green} &  &  &  &  & \cellcolor{green} &  &  &  &  &  &  &  \\\cline{2-18}
& $421$& \cellcolor{green} &   & \cellcolor{green} &  & \cellcolor{green} &  &  &  &  & \cellcolor{green} &  &  &  &  &  &  \\\cline{2-18}
& $422$& \cellcolor{green} &   & \cellcolor{green} &  &  &  &  &  &  &  &  &  &  &  &  & \\\cline{2-18}
& $311$& \cellcolor{green} &   &  & \cellcolor{green} & \cellcolor{green} &  &  &  &  &  & \cellcolor{green} &  &  &  &  &  \\\cline{2-18}
& $312$& \cellcolor{green} &   &  & \cellcolor{green} &  &  &  &  &  &  &  &  &  &  &  &  \\\cline{2-18}
& $321$& \cellcolor{green} &   &  &  & \cellcolor{green} &  &  &  &  &  &  &  &  &  &  &  \\\cline{2-18}
& $322$& \cellcolor{green} &   &  &  &  &  &  &  &  &  &  &  &  &  &  &  \\\hline
\multicolumn{18}{c}{}\\\hline
\multirow {13}*{$T_{(1,2)}$} & $ac=11$ & \cellcolor{green} & \cellcolor{green} & \cellcolor{green} & \cellcolor{blue} & \cellcolor{green} & \cellcolor{green} & \cellcolor{blue} & \cellcolor{green} & \cellcolor{blue} & \cellcolor{green} & \cellcolor{blue} & \cellcolor{blue} & \cellcolor{green} & \cellcolor{blue} & \cellcolor{blue} & \cellcolor{blue} \\\cline{2-18}
& $12$& \cellcolor{green} & \cellcolor{green} & \cellcolor{green} & \cellcolor{blue} &   & \cellcolor{green} & \cellcolor{blue} &   & \cellcolor{blue} &   &   & \cellcolor{blue} &   &   &   &  \\\cline{2-18}
& $21$& \cellcolor{green} & \cellcolor{green} &   & \cellcolor{blue} & \cellcolor{green} &   & \cellcolor{blue} & \cellcolor{green} &   &   & \cellcolor{blue} &   &   & \cellcolor{blue} &   &  \\\cline{2-18}
& $22$& \cellcolor{green} & \cellcolor{green} &   & \cellcolor{blue} &   &   & \cellcolor{blue} &   &   &   &   &   &   &   &   &  \\\cline{2-18}
& $41$& \cellcolor{green} &   & \cellcolor{green} & \cellcolor{blue} & \cellcolor{green} &   &   &   & \cellcolor{blue} & \cellcolor{green} & \cellcolor{blue} &   &   &   & \cellcolor{blue} &  \\\cline{2-18}
& $42$& \cellcolor{green} &   & \cellcolor{green} & \cellcolor{blue} &   &   &   &   & \cellcolor{blue} &   &   &   &   &   &   &  \\\cline{2-18}
& $31$& \cellcolor{green} &   &   & \cellcolor{blue} & \cellcolor{green} &   &   &   &   &   & \cellcolor{blue} &   &   &   &   &  \\\cline{2-18}
& $32$& \cellcolor{green} &   &   & \cellcolor{blue} &   &   &   &   &   &   &   &   &   &   &   &  \\\cline{2-18}
& \multicolumn{17}{|c}{}\\\cline{2-18}
&\cellcolor[gray]{0.7} $13$& \cellcolor{green} & \cellcolor{green} & \cellcolor{green} & \cellcolor{blue} &  & \cellcolor{green} & \cellcolor{blue} &  & \cellcolor{blue} &  &  & \cellcolor{blue} &  &  &  &   \\\cline{2-18}
&\cellcolor[gray]{0.7} $23$& \cellcolor{green} & \cellcolor{green} &   & \cellcolor{blue} &  &   & \cellcolor{blue} &  &   &   &  &   &   &  &   &  \\\cline{2-18}
&\cellcolor[gray]{0.7} $43$& \cellcolor{green} &   & \cellcolor{green} & \cellcolor{blue} &  &   &   &   & \cellcolor{blue} &  &  &   &   &   &  &  \\\cline{2-18}
&\cellcolor[gray]{0.7} $33$& \cellcolor{green} &   &   & \cellcolor{blue} &  &   &   &   &   &   &  &   &   &   &   &  \\\hline
\end{tabular}}
\end{center}
\caption{Protected (green) and split (blue) fixed points under different Wilson line configurations. The matter representations we found in the absence of Wilson lines will completely survive when sitting at a protected fixed point/torus. At split singularities they will decompose according to the local gauge group, some of the pieces will be projected out by the Wilson lines.}
\label{protected}
\end{table}
The next simplest alternative is to have two complete families and the third one being a patchwork of states coming from different sectors. This is likely to occur under configurations 12-15 where one finds a sector containing two protected fixed points. The second brother in models $62$, $63$ and $96$, and the first of $84$ contain \st-plets ($\overline{\mathbf{16}}$) at the $T_{(1,1)}$ sector so that they can be compatible with this picture. Complete families transform as a doublet under the surviving $D_4$. This favors the heavy family to be the patchwork one. With regards to \es, model $29$ seems promising as well.

\subsection{Discrete Symmetries}
In addition to the family structure of the model one also needs to make sure that the possible interactions are realistic and for that one needs to review the selection rules for this particular orbifold. In section \ref{sec:zsixtwo} we pointed out that these rules could be treated as discrete symmetries of the field theory\footnote{It is worth to study the potential effects of rule 4 \cite{Font:1988tp} and the newly found rule 5 \cite{Kobayashi:2011cw}. In this work, however, we restrict ourselves to those selection rules which admit an interpretation as symmetries in the low energy effective field theory.} and now we attempt to discuss how does this work in the context of \ztf.\\

The relevant information on the boundary conditions leading to any string state $\Phi_i$ can be inferred from the conjugacy class $[g_i]$ associated to it. Couplings are allowed if the strings involved in it can be merged consistently, i.e. the coupling $\Phi_1 \Phi_2...\Phi_L\subset \mathcal{W}$ is non vanishing, provided the existence of a set of space group elements $g^\prime_i\in[g_i]$ such that
\begin{align}
\label{eq:selectionrule}
\prod_{i=1}^L g_i^\prime = \left(1,0 \right)\ .
\end{align}
This constraint is known as the \emph{space group selection rule} \cite{Dixon:1986qv}. For the case we are concerned, one can immediately see from the point group part that if each state $\Phi_i$ belongs to the sector $T_{(m_i,n_i)}$ one has the following restrictions
\begin{align}
\sum_{i=1}^L m_i=0\mod 2\,,\\
\sum_{i=1}^L n_i=0\mod 4\,,
\end{align}
implying that the superpotential possesses a $\mathbb{Z}_2$ and a $\mathbb{Z}_4$ discrete symmetry, where the charges of each state permit to determine which twisted sector it belongs to. One finds that the lattice part of equation \eqref{eq:selectionrule}, introduces additional discrete symmetries. More specifically, one can use factorizability to consider the problem planewise: The first plane contributes with two $\mathbb{Z}_2$ symmetries because the vectors $e_1$ and $e_2$ are not further identified under the point group. For the remaining planes one finds that each contributes with a $\mathbb{Z}_2$ discrete symmetry, where the charges of the fields depend on their corresponding location. (See table \ref{loc}.)\\
\begin{table}[H]
\centering
\renewcommand{\arraystretch}{1.4}
\begin{center}
\footnotesize{
\begin{tabular}{ccccccccc}
\cline{2-9}
\multicolumn{1}{c}{} & \multicolumn{1}{|c|}{$a=1$} & \multicolumn{1}{|c|}{$a=2$} & \multicolumn{1}{|c|}{$a=3$} & \multicolumn{1}{|c|}{$a=4$} & \multicolumn{1}{|c|}{$b=1,2$} & \multicolumn{1}{|c|}{$b=3$} & \multicolumn{1}{|c|}{$c=1,2$} & \multicolumn{1}{|c|}{$c=3,4$} \\
\hline
\multicolumn{1}{|c|}{$\mathbb{Z}_2^{1}$} & \multicolumn{1}{|c|}{$0$} & \multicolumn{1}{|c|}{$1$} & \multicolumn{1}{|c|}{$1$} & \multicolumn{1}{|c|}{$0$} &  &  & & \\
\cline{1-5}
\multicolumn{1}{|c|}{$\mathbb{Z}_2^{1^\prime}$} & \multicolumn{1}{|c|}{$0$} & \multicolumn{1}{|c|}{$0$} & \multicolumn{1}{|c|}{$1$} & \multicolumn{1}{|c|}{$1$} &  &  & & \\
\cline{1-7}
\multicolumn{1}{|c|}{$\mathbb{Z}_2^{2}$} &  &  &  &  & \multicolumn{1}{|c|}{$0$} & \multicolumn{1}{|c|}{$1$} &  &  \\
\cline{1-1}\cline{6-9}
\multicolumn{1}{|c|}{$\mathbb{Z}_2^{3}$} &  &  &  &  &  &  & \multicolumn{1}{|c|}{$0$} & \multicolumn{1}{|c|}{$1$}  \\
\cline{1-1}\cline{8-9}
\end{tabular}}
\end{center}
\caption{Symmetries resulting from the lattice parts of the space group selection rule. $\mathbb{Z}_2^{1}$ and $\mathbb{Z}_2^{1^\prime}$ result from the first complex plane, $\mathbb{Z}_2^{2}$ and $\mathbb{Z}_2^{3}$ from the second and third, respectively. The indices $a,b,c$ label the fixed points in each complex plane according to the notation adopted in figures \ref{T01} to \ref{T12}.}
\label{loc}
\end{table}

The discrete symmetries we have just presented describe a manner to track the location of a certain field on the orbifold. There are, in addition, extra symmetries related to the presence of singularities which are indistinguishable from each other. In contrast to those descending from the space group selection rule, the symmetries arising from the degeneracy of the singularities are very sensitive to the Wilson lines. For the case of interest we start from the building blocks proposed in \cite{Kobayashi:2006wq} to find the complete location symmetry in the absence of WL: Let us start by considering the first plane: Since only a $\mathbb{Z}_2$ identification acts there, it can be regarded as the direct product of two orbicircles $S^1/\mathbb{Z}_2$, as long as one keeps in mind that the $\mathbb{Z}_2$ identification should act simultaneously on both of them. On each orbicircle, one finds that there is a freedom to exchange between the fixed points i.e. a $S_2$ symmetry, whose multiplicative closure with the $\mathbb{Z}_2\times \mathbb{Z}_2$ space group selection rule leads to a $D_4$ symmetry. The resulting symmetry on the complex plane, ends up being the direct product of two $D_4$ factors, divided by the simultaneous $\mathbbm{Z}_2$ orbifold action we have mentioned previously. Under this flavor symmetry, untwisted states transform in the trivial representation $(A_1,A_1)$ where $A_1$ denotes the invariant singlet of $D_4$, while twisted states sitting at the four fixed points furnish a four dimensional representation $(D,D)$, with $D$ being the $D_4$ doublet\footnote{The notation we use to denote the $D_4$ representation is the same as in ref. \cite{Kobayashi:2006wq}.}. \\

For the remaining planes consider a $\mathbb{T}^2/\mathbb{Z}_4$, in the first twisted sector one finds two fixed points along the diagonal in the fundamental domain (see e.g. the second plane in figure \ref{T01}), for which one finds again an $S_2$, which combined to the $\mathbb{Z}_4\times\mathbb{Z}_2$ space group symmetries leads to $(D_4\times \mathbb{Z}_4)/\mathbb{Z}_2$. When it comes to the representations one can easily see that states from the first twisted sector come in doublet representations and the fixed points of the second twisted sector allow us to realize all one dimensional representations of the $D_4$. Consider for instance the fixed points in the third plane of figure \ref{T02}, and take $|1\rangle$ and $|2\rangle$ as identical states sitting at $c=1$ and 2 (Note that these states are invariant under the $\mathbb{Z}_4$ lattice automorphism). The symmetric and antisymmetric combinations transform as $A_1$ and $A_2$ under $D_4$. The remaining fixed points $c=3,4$ get identified under the $\mathbb{Z}_4$ rotation. Even and odd combinations under the rotation gives rise to the representations $B_1$ and $B_2$ of $D_4$. The third twisted sector is the inverse of the first one, so that states there transform also as doublets. The charge of a given state under the $\mathbb{Z}_4$ is just the order of the twisted sector it belongs to.\\

In order to obtain the complete symmetry group for the six dimensional orbifold we have to take the direct product of the previous factors and mod out the point group identifications. Finally one obtains
\begin{equation}
G_{\text{Flavor}}=\frac{\left(\frac{D_4\times D_4}{\mathbb{Z}_2}\right)\times\left(\frac{D_4\times \mathbb{Z}_4}{\mathbb{Z}_2}\right)\times\left(\frac{D_4\times \mathbb{Z}_4}{\mathbb{Z}_2}\right)}{\mathbb{Z}_2\times\mathbb{Z}_4}=\frac{D_4^4\times \mathbb{Z}_4}{\mathbb{Z}_2^4}.
\label{flavor}
\end{equation}
This can be used to see how different twisted sectors transform under this flavor symmetry:
\begin{itemize}
 \item The bulk states are all flavor singlets.
\item For $T_{(0,1)}$ and $T_{(0,3)}$ the four fixed points form states transforming as $(A_1,A_1,D,D)_{1}$ and $(A_1,A_1,D,D)_{3}$, respectively. This notation indicates just that the states transform as doublets under the latter two $D_4$ factors in eq. \eqref{flavor}, while the subindices are the charges under the $\mathbb{Z}_4$.
\item For $T_{(0,2)}$ the states are of the form $(A_1,A_1,A_i,A_j)_{2}$ with $i,j=1,2,3,4$. The states with $i,j=1,2$ correspond to the ordinary fixed tori. For the special ones one has six alternatives. Depending on how the other pieces in the physical state transform under the $\mathbbm{Z}_4$ generator of the point group, one has to choose between the invariant (even) combinations
\begin{align*}
 &(A_1,A_1,A_1,A_3)_{2}, \quad  (A_1,A_1,A_2,A_3)_{2}, \quad  (A_1,A_1,A_3,A_3)_{2}, \\
 &(A_1,A_1,A_3,A_1)_{2}, \quad  (A_1,A_1,A_3,A_2)_{2}, \quad  (A_1,A_1,A_4,A_4)_{2},
\end{align*}
or the odd ones
\begin{align*}
 &(A_1,A_1,A_1,A_4)_{2}, \quad  (A_1,A_1,A_2,A_4)_{2}, \quad  (A_1,A_1,A_3,A_4)_{2}, \\
 &(A_1,A_1,A_4,A_1)_{2}, \quad  (A_1,A_1,A_4,A_2)_{2}, \quad  (A_1,A_1,A_4,A_3)_{2},
\end{align*}
in order to build a state which is space group invariant.
\item For $T_{(1,0)}$, the states sitting at ordinary tori are of the form $(D,D,A_i,A_1)_{0}$ $i=1,2$. At the special ones we have either $(D,D,A_3,A_1)_{0}$ for even or $(D,D,A_4,A_1)_{0}$ for odd states.
\item A similar situation occurs in the $T_{(1,2)}$, where ordinary tori transform according to $(D,D,A_1,A_i)_{0}$ $i=1,2$. For the special singularities one has either $(D,D,A_1,A_3)_{0}$ for even or $(D,D,A_1,A_4)_{0}$ for odd states.
\item States in $T_{(1,1)}$ or $T_{(1,3)}$ transform as $(D,D,D,D)_{1}$ or $(D,D,D,D)_{3}$, respectively.
\end{itemize}
 The breakdown of the flavor group induced by the Wilson lines occurs blockwise: note that each of the Wilson lines $W_1$ or $W_2$ is associated to each of the orbicircles in the first plane. The effect of such Wilson lines is to break the permutation symmetry so that the $D_4$ factor gets broken to the space group selection rule on the orbicircle.\\

For the case of the Wilson lines $W_3$ and $W_4$, the flavor group on the $\mathbbm{T}^2 /\mathbbm{Z}_4 $ block gets broken according to

\begin{equation*}
(D_4 \times \mathbbm{Z}_4)/\mathbbm{Z}_2 = S_2 \ltimes (\mathbbm{Z}^p_4 \times \mathbbm{Z}^l_2) \rightarrow (\mathbbm{Z}_4 \times \mathbbm{Z}_2) \, .
\end{equation*}

\noindent The $R$-symmetries for this model arise from discrete remnants of the $10D$ Lorentz group, in particular a rotation of $\pi$ in the first complex plane is a symmetry of the orbifold space and the same holds for independent rotations of $\pi/2$ on the second and third complex planes. If $\Phi_1\Phi_2...\Phi_L$ is a coupling invariant under such rotations, it should meet the following requirements
\begin{equation}
\sum_{\alpha=1}^{L}R_\alpha^1=-1\hspace{1mm}\mathrm{mod}\hspace{1mm}2, \quad\quad \sum_{\alpha=1}^{L}R_\alpha^2=-1\hspace{1mm}\mathrm{mod}\hspace{1mm}4,\quad\quad\sum_{\alpha=1}^{L}R_\alpha^3=-1\hspace{1mm}\mathrm{mod}\hspace{1mm}4.
\end{equation}
Be reminded that the $R$-charge is the picture invariant combination of the right moving weights and the number operators \cite{Kobayashi:2004ya}, for the generic state defined in eq. \ref{eq:phys}, the $R$-charges on each plane are given by
\begin{equation}
 R^i=q_{\rm sh}^i-N^i+N^{i*}
\end{equation}
\subsection{Top-Yukawa Coupling at the GUT Level}
We have gained knowledge about the matter representations at protected and split singularities as well as relevant constraints on the allowed couplings. The remaining question is how can one use the Mini--Landscape criteria to determine which shift embeddings can be of relevance for model building. Since we do not have any control on the unshielded singularities, we can not check whether the hypercharge is anomalous or not. A $\mu$ term of the order of the gravitino mass can only be explicitly determined at the level of the VEV configuration. All we can do is to partially check for the presence of a renormalizable, trilinear top-Yukawa. Since we have the spectrum at the level of \SO{10} or \es\ we can only check for situations in which the coupling of interest descends from allowed interactions at the GUT level.\\

\begin{table}[H]
\begin{center}
\begin{tabular}{|l||l|}\hline
Couplings involving untwisted Fields & Couplings involving twisted fields only \\ \hline
$1. \quad T_{(0,2)} T_{(0,2)} U_{1}    $ & $6. \quad T_{(0,2)} T_{(1,2)} T_{(1,0)} $  \\
$2. \quad T_{(1,0)} T_{(1,0)} U_3 $ &$7. \quad  T_{(1,1)} T_{(1,1)} T_{(0,2)}$ \\
$3. \quad T_{(1,2)} T_{(1,2)} U_2 $ &$8. \quad T_{(1,1)} T_{(0,1)} T_{(1,2)} $\\
$4. \quad T_{(0,1)} T_{(0,3)} U_1  $ &$9. \quad T_{(1,1)}T_{(0,3)}T_{(1,0)}$ \\
$5. \quad U_1 U_2 U_3 $& \\ \hline
\end{tabular}
\end{center}
\caption{Table of all trilinear couplings in the $\mathbb{Z}_2 \times \mathbb{Z}_4$ orbifold.}
\label{tab:Yukawas}
\end{table}

In table \ref{tab:Yukawas} we present all allowed trilinear couplings for \ztf. Assuming that the ingredients for such a coupling do not appear at unshielded singularities, we will be able to draw partial statements on the fertility of the shifts. If that is the case, we have to demand that the GUT state, giving rise to the Higgs lies at a singularity where the Wilson lines could serve to project out undesired color triplets out of the multiplet (doublet-triplet splitting).\\

In our search for appealing models we focus on shifts and Wilson line configurations where one has two complete and one patchwork family. The requirement of two complete families implies that not all Wilson lines are switched on (see table \ref{protected}). We assume that the coupling $Q\overline{U}H_u$ is unique in the sense that the fields giving rise to SM matter are non degenerate. This is justified because having more than one up-type Yukawa of order one might spoil the mass hierarchy of the families. This criterion rules out couplings involving $T_{(1,1)}$, since the states there are always degenerate. We can study the possibility for the remaining alternatives to be realized by our shifts. For each of them one can look at table \ref{protected}, to find configurations with non degenerate split multiplets in the relevant sectors. Then one can see, out of these configurations, where the two complete families are located, and finally one should search for the existence of a shift embedding with the desired content. The results of our analysis are summarized in table \ref{tab:IncompleteSO10}.
\begin{table}[H]
\begin{center}
\renewcommand{\arraystretch}{1.2}
\begin{tabular}{|c |c| c | c |}\hline
Coupling & WL config. & Sectors for Families & Shift Embedding \\ \hline
$T_{(0,1)}T_{(0,3)}U_1 $ & 14,15 & $T_{(1,3)}$ & No Model available \\ \hline
$T_{(0,2)}T_{(0,2)}U_1 $ & 14,15 & $T_{(1,3)}$ & No Model available \\ \hline
$T_{(1,0)}T_{(1,0)}U_3 $ & None & $ - $ & - \\ \hline
$T_{(1,2)}T_{(1,2)}U_2 $ & None & - & - \\ \hline
$T_{(1,0)}T_{(0,2)}T_{(1,2)} $ & None & - & - \\ \hline
\multirow{2}*{$U_1 U_2 U_3 $} & \multirow{2}*{6-15} & Depends on  & At least one \\
& & WL config. & per WL config.\\
\hline
\end{tabular}
\end{center}
\caption{WL configurations and shift embeddings consistent with a unique renormalizable coupling. All criteria can only be fulfilled in the case of three untwisted fields.}
\label{tab:IncompleteSO10}
\end{table}
A purely untwisted coupling is supported by many Wilson line configurations and shifts, while other cases are ruled out. We went one step further and explored the existence of untwisted trilinear couplings at the GUT level for all \SO{10} and $E_6$ shifts (see appendix \ref{app:mattercontent}).
 Interestingly enough, roughly $~75\%$ of all \SO{10} and $~50 \%$ of all $E_6$ embeddings permit gauge-top unification scenarios for an appropriate choice of the WLs.\\\\
To conclude let us briefly comment on the presence of Higgs bilinears. Note that the $R$-charge of each of the fields involved is related to each of the complex planes. If one takes the up-type Higgs to arise from the state with $R$-charge $(0,-1,0,0)$ there is always a multiplet transforming in the conjugate representation with the same R-charge, this would-be down-type Higgs helps to build a neutral bilinear. The neutrality of this bilinear is due to the presence of the $\mathbbm{Z}_2$ $R$-Symmetry\footnote{In the notation of ref. \cite{Lee:2011dya} all charges are integer, the mentioned $\mathbbm{Z}_2$ has to be regarded as a $\mathbbm{Z}_4$ $R$-Symmetry.} supported by the $\mathbbm{Z}_2$ orbifold plane \cite{Lebedev:2007hv}.
\section{Comparison to other models}
\label{sec:ComparisonModels}
Finally we want to compare the rules for viable model building we have identified in \ztf\ with those available from other orbifolds, where specific models have been constructed. In particular, we discuss the possibilities to obtain a large top-Yukawa coupling, in order to find the favored locations for some of the SM fields. \\
\subsection{Comparison to \zsix\ }
The \zsix\ picture has been already introduced in section \ref{sec:zsixtwo} and \ref{sec:LessonsfromZ6}, where it was shown that the appearance of the three families as complete GUT multiplets is not possible. An alternative is to have one of the families as a patchwork of twisted and untwisted states whereas the other two are complete. Similarly as in the Mini--Landscape, the \ztf\ favors an identical scenario. In both cases we see that most of the shifts permit an untwisted top-Yukawa at the GUT level. One can think that the presence of this coupling is the most probable reason for the hierarchical structure of the SM, given the statistical preference for the up-type Higgs as well as the other relevant pieces for the coupling to live in the bulk.\\

An additional feature of these models is the presence of a $\mathbb{Z}_2$ orbifold plane that supports an vector-like up- and down-Higgs pair that can form a neutral bilinear under all selection rules. This provides a solution to the $\mu$-problem as discussed in sec. \ref{sec:higgs}.
\subsection{Comparison to \ztt\ }
Another well studied orbifold is \ztt\ \cite{Forste:2004ie}. It results from taking the torus spanned by the root lattice of $SU(2)^6$, and dividing it by the $\mathbb{Z}_2$ generators associated to the shifts $v_1 = (0,0,\frac12,-\frac12)$ and $v_2 = (0,\frac12,0,-\frac12)$. This leads to three twisted sectors with 16 fixed tori each. The root lattice admits six Wilson lines of order two (one per lattice vector). In \ztt\ only the following trilinear couplings are allowed
\begin{equation*}
U_1U_2U_3 \quad T_{(1,0)}T_{(1,0)}U_1 \quad T_{(0,1)}T_{(0,1)}U_3 \quad T_{(1,1)}T_{(1,1)}U_2 \quad T_{(1,0)}T_{(0,1)}T_{(1,1)} \, .
\end{equation*}
A substantial difference between this and the orbifolds we have previously considered is that in this model the untwisted spectrum is completely vector-like. It then follows that the left- and right-chiral quarks should be twisted fields, ruling out the possibility for an entirely untwisted coupling. \\
On the other hand, the coupling $T_{(1,0)}T_{(0,1)}T_{(1,1)}$ is a valid possibility but it involves all available twisted sectors. This makes it impossible to retain a unique trilinear coupling while having two light families as complete GUT multiplets.\\

The remaining couplings are then $T_{(1,0)}T_{(1,0)}U$,\, $T_{(0,1)}T_{(0,1)}U$ and $T_{(1,1)}T_{(1,1)}U$. By the reasoning above, top-quark and all other chiral matter must be located at the twisted sectors, thus the up-Higgs must be in the untwisted sector. In order to attain the right degeneracies one needs at least five Wilson lines. After turning on these Wilson lines there are no protected tori. This makes it unlikely to get two complete families at the level of \SO{10}. The only feature shared by the other two orbifolds and \ztt\ is the possibility for untwisted Higgs fields.\\
 
The classification of the gauge embeddings has been performed in ref. \cite{Forste:2004ie}. It is found, that there are only five inequivalent embeddings out of which three lead to an \es\ factor. Those models suffer from the complications we have previously mentioned. The remaining ones contain an \SU{8} factor. Such models have shown to be a fertile patch when one mods out an additional freedom of this orbifold. In fact, the \ztt\ geometry permits a freely acting involution $\mathbb{Z}_{2-{\rm free}}$ \cite{Blaszczyk:2009in} whose action can be embedded into the gauge coordinates. This procedure results in a pairwise identification of the fixed points which halves the multiplicities and hence, eases the search for three family models. The projection induced by the $\mathbb{Z}_{2-{\rm free}}$ acts on bulk fields only, so that the local gauge enhancements remain unaffected.\\

The models were constructed in the following way: Starting from a pair of shifts leading to an \SU{8} factor, the Wilson lines\footnote{The identification $W_2 = W_4 = W_6$ is required in order to further mod out the involution.} $W_2 = W_4 = W_6$, $W_3$ and $W_5$ were turned on in such a way, that the mentioned factor is broken to \SU{5}. At the sector $T_{(1,0)}$ one finds a single protected torus where the third family is located and the remaining two come from $T_{(0,1)}$. The pattern of flavor symmetries is very similar to that of \ztf\ and it turns out that the mentioned families transform under a $D_4$ symmetry which survives as a consequence of $W_1=0$. The $\mathbb{Z}_{2-{\rm free}}$ action breaks the \SU{5} down to the SM gauge group. This has the advantage that all twisted fields remain as complete \SU{5} representations so that the standard hypercharge is unbroken.\\

There has been successful model building within the free fermionic formulation of the heterotic string as well. It has been argued that this corresponds to \ztt\ orbifold compactifications in the bosonic formulation \cite{Faraggi:1995yd} at the self-dual point. In these models all families come from twisted sectors and only the top-quark couples trilinearly to the up-type Higgs, which in turn resides in the bulk \cite{Faraggi:1992fa}.

\section{Conclusions and Outlook}
\label{sec:Conclusion}
In this work we wanted to explore the possibilities for localization of the MSSM fields in the extra dimensions in order to arrive at models of physical relevance within heterotic orbifold compactifications. The major ingredients for this exploration were inspired by the \zsix\ Mini--Landscape, where the field location plays a crucial role in solving field theory issues such as the hierarchy, flavor and $\mu$- problems. \\

Motivated by the lessons from the Mini--Landscape, our attempt was to study the alternatives provided by the \ztf\ orbifold model, which has not been used for model building so far. We studied the geometry and classified the fixed points/tori in terms of the enhancements of the gauge symmetry that occur at some of them. The inequivalent shift embeddings were computed and out of them 35 were found to contain an \SO{10} and 26 an \es\ factor. For these models, the matter spectrum was computed and the interplay between the Wilson lines and the orbifold topography was analyzed. With all those ingredients we discussed possible strategies to obtain three families with a hierarchy for their masses.\\

The picture appears very similar to that in \zsix\:
\begin{enumerate}
\item A scenario in which the three families arise from complete \SO{10} multiplets is not consistent with a hierarchy for the mass of the SM fields.
\item A completely untwisted top-Yukawa coupling seems to be the most favored situation, leading to the familiar gauge-top unification scheme.
\item The presence of a $\mathbbm{Z}_2$ torus guarantees a down-Higgs in the bulk as well, if the Higgs pair remains massless in the low energy, the model will enjoy of gauge-Higgs unification.
\item The two light families usually arise from the twisted sectors. They can appear as complete multiplets of the underlying local GUT, if so, they will transform as a doublet under a $D_4$ flavor symmetry.
\end{enumerate}
The \ztt\ orbifold was considered as well, where it is necessary for all chiral SM fields to arise from twisted sectors. One has to point out that the construction of the class of \ztt\ models which are of physical relevance, differs structurally from those based on \zsix\ and \ztf\ due to the free involution one has to mod out. Whereas the gauge group in the bulk emerges as the intersection of all local gauge symmetries, in the Blasczcyk \emph{et. al.} model \cite{Blaszczyk:2009in} the \SU{5} enhancement is common to all fixed tori. The breakdown to the Standard Model is induced in the bulk by virtue of the $\mathbbm{Z}_{2,\text{free}}$. It is still surprising that most of the locations are similar, except for the left- and right-chiral top-quark.\\

These common features can serve as guidelines for future model building. For the concrete case of \ztf\ we did not construct explicit models yet. We thus do not know the potential influence of fields at the unshielded singularities. The  construction of specific models could bring some insights on the pattern of discrete symmetries, specially now that codes have been developed to compute spectra and vacuum configurations \cite{Nilles:2011aj}. In particular, a crucial ingredient for the solution of the $\mu$-problem to work is the survival of a meaningful $R$-symmetry. The \ztf\ orbifold seems to be a promising ground to find realistic MSSM candidates, even compared to the \zsix\ case.\\

\acknowledgments

We would like to thank Michael Blaszczyk, Sven Krippendorf, Christoph L\"udeling, Matthias Schmitz and Fabian R\"uhle for helpful discussions and suggestions. This work was partially supported by the SFB-Transregio TR33 The
Dark Universe (Deutsche Forschungsgemeinschaft) and the European Union 7th network
program Unification in the LHC era (PITN-GA-2009-237920). We thank the Simons Center for Geometry and Physics in Stony Brook (NY) and the Corfu Summer Institute for Particle Physics and Gravity for their hospitality during the completion of this work.
\clearpage
\appendix
\section{Shift Embeddings}
\label{app:embeddings}
In this appendix we want to explain the method we employed to construct all inequivalent embeddings for $\mathbbm{Z}_2\times\mathbbm{Z}_4$ in more detail.
Let us denote a certain embedding as $\{V_N,V_M\}$, meaning that $V_N$ and $V_M$ reproduce the effects of the point group generators $\theta$ and $\omega$ ($\theta^N = \omega^M = \mathbbm{1}$) as translations in the gauge coordinates.\\
Assume that for the $T_{(k_1,k_2)}$ twisted sector one can find a weight $p\in\Gamma_{16}$ (with $\Gamma_{16}$ being the sixteen dimensional gauge lattice), such that up to a certain oscillator combination, the shifted weight $p_{\rm sh}=p+k_1 V_N+k_2V_M$ permits constructing a state with mass eigenvalue
\begin{align}
\label{eq:masslessL}
\frac{M^2_L}{8}=\frac{\left(p_{\rm sh}\right)^2}{2} + 1 + N + \delta_c \, .
\end{align}
For the embedding $\{V_N+\lambda,V_M+\lambda^\prime\}$ with $\lambda,\lambda^\prime\in\Gamma_{16}$, one can show that $p_{\rm sh}$ is also a shifted weight of this new model. This means that with both of these embeddings one finds the same left-moving states. Similarly one can use the isometries of $\Gamma_{16}$ to map any $p_{\rm sh}$ from $\{V_N,V_M\}$ to $\sigma p_{\rm sh}$ of $\{\sigma V_N,\sigma V_M\}$, provided $\sigma\in\text{Aut}(\Gamma_{16})$. From these simple arguments it follows that the embeddings which differ by lattice vectors, as well as those which are related via automorphisms acting simultaneously on both $V_N$ and $V_M$, contain the same massless left-moving states\footnote{When comparing the weights $p_{\rm sh}$ and $\sigma p_{\rm sh}$ which lead to massless left movers under the embeddings $\{V_N,V_M\}$ and $\{\sigma V_N,\sigma V_M\}$ respectively, one can observe that $\sigma$ also maps between the simple roots of the gauge groups induced by these embeddings, meaning that $p_{\rm sh}$ and $\sigma p_{\rm sh}$ have the same Dynkin labels, i.e. the representations these weights lead to are identical for  both models.}. This is, however, not sufficient to guarantee that these models will have the same physical spectrum. For that we also have to ensure that the orbifold projectors have the same effects in both models. Assume now that the space group element $h=(\theta^{n_1}\omega^{n_2},\tilde{\lambda} )$ commutes with the generating element of a given fixed point $z_f$ of $T_{(k_1,k_2)}$. The only part of the projection induced by $h$ which is embedding dependent, is given by
\begin{equation}
\delta^{(k_1,k_2)}(n_1,n_2)=\exp\left\{2\pi i\left[p_{\rm sh}-{\textstyle \frac{ 1}{ 2}}(k_1V_N+k_2V_M)\right]\cdot(n_1V_N+n_2V_M)\right\}\,,
\label{phase}
\end{equation}
for a given string state with gauge momentum $p_{\rm sh}$. When choosing the embedding to be   
$\{\sigma (V_N+\lambda),\sigma (V_M+\lambda^\prime)\}$, the weight $\sigma p_{\rm sh}$ leads to the same left-moving state as before. It can be shown that in this new model the transformation phase for  $\sigma p_{\rm sh}$ under $h$ is given by
\begin{equation}
\delta^{\prime(k_1,k_2)}(n_1,n_2)=\exp\{2\pi {\rm i}(k_1n_2-k_2n_1)\Phi\}\delta^{(k_1,k_2)}(n_1,n_2),
\label{difference}
\end{equation}
with the \emph{brother's phase} $\Phi$ being \cite{Ploger:2007iq}
\begin{equation}
\Phi=\frac{1}{2}\left(V_M\cdot\lambda_2-V_2\cdot\lambda_M+\lambda_N\cdot\lambda_M\right)\,.
\label{bro}
\end{equation}
Note that the above equation is independent of the automorphism $\sigma$. This is a very powerful result since we can now mod out the automorphism group from the set of all possible embeddings. The further addition of lattice vectors to the embedding leads to an equivalent model if $\Phi$ is an integer number. Note also that from eqs. \eqref{eq:Phys} and \eqref{proy} and from the fact that $\Gamma_{16}$ is integral, gauge embeddings which differ by lattice vectors have the same untwisted matter content. However, the twisted matter can change depending on the \itshape brothers \normalfont phase, so that each inequivalent embedding has a fixed amount of independent \itshape brother \normalfont models. \\\\
Now we specify to the particular case of $\mathbbm{Z}_2\times\mathbbm{Z}_4$. We have chosen the gauge lattice $\Gamma_{16}$ as that of $E_8\times E_8$. This enables us to decompose the embedding vectors as
\begin{equation}
V_4=A_4\oplus B_4,\quad V_2=A_2\oplus B_2 \, ,
\label{decomp}
\end{equation}
where $4A_4$, $4B_4$, $2A_2$ and $2B_2$ are vectors from the lattice of one single $E_8$ factor. The decomposition takes advantage of the fact that the isometries of $\Gamma_{16}$ can all be written as the direct product of inner automorphisms of each $E_8$ root lattice. This implies the equivalence of the embeddings $\{A_2\oplus B_2,A_4\oplus B_4\}$ and $\{\sigma_1A_2\oplus \sigma_2 B_2,\sigma_1A_4\oplus\sigma_2 B_4\}$, with $\sigma_1,\sigma_2\in\text{Aut}(\mathfrak{e}_8)$.
The above arguments permit us to consider just one $E_8$ factor and restrict the elements in $\text{Aut}(\mathfrak{e}_8)$ to act simultaneously on the combination $(A_2,A_4)$. Since we have neglected inequivalences which can be cured upon addition of lattice vectors, we take the basis $\{\alpha_k\}_{k=1,...,8}$ and focus on vectors from the following minimal sets
\begin{align}
A_4=\frac{1}{4}\sum_{k=1}^{8}a_k\alpha_k&,\quad a_k=0,1,2,3 \, , \\
A_2=\frac{1}{2}\sum_{k=1}^{8}b_k\alpha_k&,\quad b_k=0,1 \label{a2} \, .
\end{align}
Then we can construct all combinations of the form $(A_2,A_4)$ and act with automorphisms to find all those which are inequivalent. All possible embeddings will result from pairings of such combinations if the modular invariance conditions (\ref{1mod}), (\ref{2mod}) and (\ref{3mod}) are satisfied. Though this procedure seems relatively straightforward, we will avoid constructing $\text{Aut}(\mathfrak{e}_8)$ explicitly and follow an approach which is more efficient from the computational point of view. We make use of the inequivalent vectors found for the $\mathbbm{Z}_4$ orbifold \cite{Katsuki1989}. These correspond to all inequivalent vectors $A_4$ after modding out the whole $\text{Aut}(\mathfrak{e}_8)$. Following this approach, our task simplifies to only look for equivalence relations of the $\mathbbm{Z}_2$ vectors $A_2$.\\\\
By completely modding out all isometries and lattice shifts from the $\mathbbm{Z}_4$ vectors, we have to mo- dify the equivalence relations for their $\mathbbm{Z}_2$ companions. For a given $A_4$, two combinations $(A_2,A_4)$ and $(A^{\prime}_2,A_4)$ are equivalent, only if one can find an element $\sigma$, from the subgroup
\begin{equation}
\text{Aut}(\mathfrak{e}_8|A_4)=\left\{\sigma\in\text{Aut}(\mathfrak{e}_8)~|~(\sigma A_4-A_4)\in\Gamma(\mathfrak{e}_8)\right\}\,,
\end{equation}
such that $(\sigma A_2-A^\prime_2)\in\Gamma(\mathfrak{e}_8)$. Generating the subgroups $\text{Aut}(\mathfrak{e}_8|A_4)$ is still far from feasible. To avoid such exhaustive construction, we can use the fact that any element in the Weyl-Coxeter group can be written as a product of Weyl-reflections. For each inequivalent $A_4$ vector we construct the set of Weyl-reflections which leave it invariant (up to lattice vectors)
\begin{equation}
\mathcal{W}(A_4)=\left\{\sigma_{\alpha}=\mathbbm{1}-\frac{\alpha\alpha^T}{\left<\alpha,\alpha\right>}~|~\alpha^2=2;\,\, \alpha,(\sigma_{\alpha}A_4-A_4)\in\Gamma(\mathfrak{e}_8)\right\}\,.
\end{equation}
The equivalence is checked by taking the elements from \eqref{a2} and merging them to equivalence classes under the action of  $\mathcal{W}(A_4)$.  The algorithm implemented in this way accounts for the equivalence relations induced by all elements of the automorphism group that can be written as products of the Weyl reflections in $\mathcal{W}(A_4)$. It can happen that there are some isometries which belong to $\text{Aut}(\mathfrak{e}_8)$ but can not be written as a product of elements in $\mathcal{W}(A_4)$. The equivalences associated to such isometries are not testable by our means. Nevertheless, the output is reduced enough so that one just needs to take further care of those combinations $(A_2,A_4)$ and $(A^\prime_2,A_4)$ which lead to the same gauge structure. In that case the equivalence is proven by the explicit construction of the automorphism that relates them.
Since the mass equation for both $E_8$ factors decouples in the untwisted sector, one can calculate the group decomposition and untwisted matter contribution for a combination $(A_2,A_4)$. The results realize all possible breaking structures expected for point groups of order eight \cite{Carter70}.\\\\
The combinations we previously found need to be paired to form sixteen dimensional vectors consistent with modular invariance.
Consider two vectors $\lambda_2,\lambda_4\in\Gamma_{16}$ and the embedding $\{V_2+\lambda_2,V_4+\lambda_4\}$, induced by pairing $(A_2,A_4)$ and $(A_2^\prime,A_4^\prime)$, such that $V_2=A_2\otimes A_2^\prime$ and $V_4=A_4\otimes A_4^\prime$. This embedding is only valid if the conditions $(\ref{1mod})-(\ref{3mod})$ hold.
Note that the first two equations, in contrast to the third, do not receive any contribution from the lattice shifts. These conditions are called strong, since they can be used to decide which combinations $(A_4,A_2)$ and $(A_4^\prime,A_2^\prime)$ can be paired together in a consistent way. In many cases one can satisfy the third modularity condition by introducing some lattice vectors. Nevertheless, the properties of $\Gamma_{16}$ only make this possible if
\begin{equation}
4\left(V_2\cdot V_4+\frac{1}{8}\right)=0\mod 1 \, .
\label{weak}
\end{equation}
This weaker condition permits us to avoid the explicit search for lattice vectors, especially at this stage in which we are only interested in the gauge group decomposition introduced by each embedding. Clearly there is not only one single choice of lattice vectors for which the consistency conditions for modular invariance are satisfied. One can deduce from eqs.\ (\ref{bro}) and (\ref{weak}), that for each valid embedding, adding lattice vectors leads to two inequivalent models. \\\\
We found 144 valid embeddings which satisfy eqs. (\ref{1mod}), (\ref{2mod}) and (\ref{weak}). Those embeddings and their corresponding brother phase can be found in table \ref{tab:ModInvarPair} of appendix \ref{app:Shiftembbedings}. Even though the embeddings are not in agreement with the strong modular invariance conditions, they can be used to compute the spectrum by taking the brother phases into account. Projections of the from 
\begin{equation}
\left(p_{sh}-\frac{1}{2}V_g\right)\cdot V_h-\left(R-\frac{1}{2}v_g\right)\cdot v_h+(k_1n_2-k_2n_1)\Phi=0\mod 1\,.
\end{equation}
lead to the same spectrum as the one we would find for an entirely modular invariant model. 
Note that the presented algorithm realizes inequivalent embeddings for the whole massive tower of string states and not only the massless ones. \\
The classification algorithm we have developed can be straightforwardly extended to other $\mathbbm{Z}_N \times \mathbbm{Z}_M$ orbifolds.
The simplifications we have made (by weakening the modular invariance conditions) can be extended to the Wilson lines as well. This might simplify the search for physical interesting backgrounds.

\section{Table of Shift Embeddings}
\label{app:Shiftembbedings}

\begin{table}[H]
\caption{The 144 inequivalent models that have been constructed. The \emph{brother} phases $\Phi$ (ses eq. \eqref{bro}) that cure the mismatch from \ref{weak} are shown in the third column. The first column gives the model number while the second gives the compatible $V_2$ shift vector to $V_{4}$ which is always given in the first line of each subtable.}
\vspace{3mm}
\label{tab:ModInvarPair}
  \centering
{\scriptsize
  \begin{tabular}{|cr@{}l|r|c|}
\hline
 &\multicolumn{2}{c|}{$2\cdot V_2$}&\multicolumn{1}{c|}{~$\Phi$}& Group Decomposition \\
\hline
\hline
\multicolumn{5}{|c|}{$4\cdot V_4=(0,0,0,0,0,0,0,0)\oplus(1,1,0,0,0,0,0,0)$} \\
\hline
1&$(0,0,0,0,0,0,0,0)\oplus$&$(-1,0,0,0,0,0,1,0)$&$0,\frac{1}{2}$&E$_8|$E$_6\times$U(1)$^2$\\
2&$(-1,0,0,0,0,0,1,0)\oplus$&$(-1,0,-1,-1,0,0,0,1)$&$0,\frac{1}{2}$&E$_7\times$SU(2)$|$SU(8)$\times$U(1)\\
3&$(-1,-1,-1,0,0,0,0,1)\oplus$&$(-1,0,0,0,0,0,1,0)$&$0,\frac{1}{2}$&SO(16)$|$E$_6\times$U(1)$^2$\\
\hline\hline
\multicolumn{5}{|c|}{$4\cdot V_4=(0,0,0,0,0,0,0,0)\oplus(3,1,0,0,0,0,0,0)$} \\
\hline
4&$(0,0,0,0,0,0,0,0)\oplus$&$(-1,0,0,0,0,0,1,0)$&$-\frac{1}{8},\frac{3}{8}$&E$_8|$SO(10)$\times$U(1)$^3$\\
5&$(0,0,0,0,0,0,0,0)\oplus$&$-\frac{1}{2}(1,-1,1,1,1,1,1,-1)$&$0,\frac{1}{2}$&E$_8|$SU(6)$\times$SU(2)$\times$U(1)$^2$\\
6&$(-1,0,0,0,0,0,1,0)\oplus$&$\frac{1}{2}(-3,-1,-1,1,1,1,1,1)$&$-\frac{1}{4},\frac{1}{4}$&E$_7\times$SU(2)$|$SU(6)$\times$SU(2)$\times$U(1)$^2$\\
7&$(-1,0,0,0,0,0,1,0)\oplus$&$(-1,0,-1,-1,0,0,0,1)$&$-\frac{1}{8},\frac{3}{8}$&E$_7\times$SU(2)$|$SU(4)$^2\times$U(1)$^2$\\
8&$(-1,-1,-1,0,0,0,0,1)\oplus$&$(-1,0,0,0,0,0,1,0)$&$-\frac{1}{8},\frac{3}{8}$&SO(16)$|$SO(10)$\times$U(1)$^3$\\
9&$(-1,-1,-1,0,0,0,0,1)\oplus$&$-\frac{1}{2}(1,-1,1,1,1,1,1,-1)$&$0,\frac{1}{2}$&SO(16)$|$SU(6)$\times$SU(2)$\times$U(1)$^2$\\
\hline\hline
\multicolumn{5}{|c|}{$4\cdot V_4=(2,2,0,0,0,0,0,0)\oplus(1,1,0,0,0,0,0,0)$} \\
\hline
10&$(0,0,0,0,0,0,0,0)\oplus$&$(-1,0,0,0,0,0,1,0)$&$0,\frac{1}{2}$&E$_7\times$SU(2)$|$E$_6\times$U(1)$^2$\\
11&$(-1,-1,0,0,0,0,0,0)\oplus$&$(-1,0,-1,-1,0,0,0,1)$&$-\frac{1}{4},\frac{1}{4}$&E$_7\times$SU(2)$|$SU(8)$\times$U(1)\\
12&$(-1,0,0,0,0,0,1,0)\oplus$&$(-1,0,-1,-1,0,0,0,1)$&$-\frac{1}{8},\frac{3}{8}$&E$_6\times$U(1)$^2|$SU(8)$\times$U(1)\\
13&$(-1,1,0,0,0,0,0,0)\oplus$&$(-1,0,-1,-1,0,0,0,1)$&$0,\frac{1}{2}$&SO(12)$\times$SU(2)$^2|$SU(8)$\times$U(1)\\
14&$(-1,-1,-1,0,0,0,0,1)\oplus$&$(-1,0,0,0,0,0,1,0)$&$-\frac{1}{4},\frac{1}{4}$&SO(12)$\times$SU(2)$^2|$E$_6\times$U(1)$^2$\\
15&$(-1,0,-1,-1,0,0,0,1)\oplus$&$(-1,0,0,0,0,0,1,0)$&$-\frac{1}{8},\frac{3}{8}$&SU(8)$\times$U(1)$|$E$_6\times$U(1)$^2$\\
\hline\hline
\multicolumn{5}{|c|}{$4\cdot V_4=(2,2,0,0,0,0,0,0)\oplus(3,1,0,0,0,0,0,0)$}\\
\hline
16&$(0,0,0,0,0,0,0,0)\oplus$&$(-1,0,0,0,0,0,1,0)$&$-\frac{1}{8},\frac{3}{8}$&E$_7\times$SU(2)$|$SO(10)$\times$U(1)$^3$\\
17&$(0,0,0,0,0,0,0,0)\oplus$&$-\frac{1}{2}(1,-1,1,1,1,1,1,-1)$&$0,\frac{1}{2}$&E$_7\times$SU(2)$|$SU(6)$\times$SU(2)$\times$U(1)$^2$\\
18&$(-1,-1,0,0,0,0,0,0)\oplus$&$\frac{1}{2}(-3,-1,-1,1,1,1,1,1)$&$0,\frac{1}{2}$&E$_7\times$SU(2)$|$SU(6)$\times$SU(2)$\times$U(1)$^2$\\
19&$(-1,-1,0,0,0,0,0,0)\oplus$&$(-1,0,-1,-1,0,0,0,1)$&$-\frac{3}{8},\frac{1}{8}$&E$_7\times$SU(2)$|$SU(4)$^2\times$U(1)$^2$\\
 20&$(-1,0,0,0,0,0,1,0)\oplus$&$\frac{1}{2}(-3,-1,-1,1,1,1,1,1)$&$-\frac{3}{8},\frac{1}{8}$&E$_6\times$U(1)$^2|$SU(6)$\times$SU(2)$\times$U(1)$^2$\\
21&$(-1,0,0,0,0,0,1,0)\oplus$&$(-1,0,-1,-1,0,0,0,1)$&$-\frac{1}{4},\frac{1}{4}$&E$_6\times$U(1)$^2|$SU(4)$^2\times$U(1)$^2$\\
22&$(-1,1,0,0,0,0,0,0)\oplus$&$\frac{1}{2}(-3,-1,-1,1,1,1,1,1)$&$-\frac{1}{4},\frac{1}{4}$&SO(12)$\times$SU(2)$^2|$SU(6)$\times$SU(2)$\times$U(1)$^2$\\
23&$(-1,1,0,0,0,0,0,0)\oplus$&$(-1,0,-1,-1,0,0,0,1)$&$-\frac{1}{8},\frac{3}{8}$&SO(12)$\times$SU(2)$^2|$SU(4)$^2\times$U(1)$^2$\\
24&$(-1,-1,-1,0,0,0,0,1)\oplus$&$(-1,0,0,0,0,0,1,0)$&$-\frac{3}{8},\frac{1}{8}$&SO(12)$\times$SU(2)$^2|$SO(10)$\times$U(1)$^3$\\
25&$(-1,-1,-1,0,0,0,0,1)\oplus$&$-\frac{1}{2}(1,-1,1,1,1,1,1,-1)$&$-\frac{1}{4},\frac{1}{4}$&SO(12)$\times$SU(2)$^2|$SU(6)$\times$SU(2)$\times$U(1)$^2$\\
26&$(-1,0,-1,-1,0,0,0,1)\oplus$&$(-1,0,0,0,0,0,1,0)$&$-\frac{1}{4},\frac{1}{4}$&SU(8)$\times$U(1)$|$SO(10)$\times$U(1)$^3$\\
27&$(-1,0,-1,-1,0,0,0,1)\oplus$&$-\frac{1}{2}(1,-1,1,1,1,1,1,-1)$&$-\frac{1}{8},\frac{3}{8}$&SU(8)$\times$U(1)$|$SU(6)$\times$SU(2)$\times$U(1)$^2$\\
\hline
\hline
\multicolumn{5}{|c|}{$4\cdot V_4=(1,1,0,0,0,0,0,0)\oplus(4,0,0,0,0,0,0,0)$}\\
\hline
28&$(-1,0,0,0,0,0,1,0)\oplus$&$(0,0,0,0,0,0,0,0)$&$0,\frac{1}{2}$&E$_6\times$U(1)$^2|$SO(16)\\
29&$(-1,0,0,0,0,0,1,0)\oplus$&$(0,0,0,0,0,-2,0,0)$&$0,\frac{1}{2}$&E$_6\times$U(1)$^2|$SO(16)\\
30&$(-1,0,0,0,0,0,1,0)\oplus$&$\frac{1}{2}(-3,-1,-1,1,1,1,1,1)$&$-\frac{3}{8},\frac{1}{8}$&E$_6\times$U(1)$^2|$SU(8)$\times$U(1)\\
31&$(-1,0,0,0,0,0,1,0)\oplus$&$(-1,-1,-1,0,0,0,0,1)$&$-\frac{1}{4},\frac{1}{4}$&E$_6\times$U(1)$^2|$SO(8)$\times$SO(8)\\
32&$(-1,0,-1,-1,0,0,0,1)\oplus$&$(-1,0,0,0,0,0,1,0)$&$-\frac{1}{4},\frac{1}{4}$&SU(8)$\times$U(1)$|$SO(12)$\times$SU(2)$2$\\
33&$(-1,0,-1,-1,0,0,0,1)\oplus$&$-\frac{1}{2}(1,1,1,1,1,1,-1,-1)$&$-\frac{1}{8},\frac{3}{8}$&SU(8)$\times$U(1)$|$SU(8)$\times$U(1)\\
\hline
\hline
\multicolumn{5}{|c|}{$4\cdot V_4=(1,1,0,0,0,0,0,0)\oplus(1,1,1,1,1,1,1,-1)$} \\
\hline
34&$(0,0,0,0,0,0,0,0)\oplus$&$-\frac{1}{2}(1,1,1,1,1,1,-1,-1)$&$-\frac{1}{8},\frac{3}{8}$&E$_7\times$U(1)$|$SU(7)$\times$U(1)$^2$ \\
35&$(0,0,0,0,0,0,0,0)\oplus$&$\frac{1}{2}(-1,-1,-1,-1,1,1,1,1)$&$0,\frac{1}{2}$&E$_7\times$U(1)$|$SU(5)$\times$SU(3)$\times$U(1)$^2$\\
36&$(-1,-1,0,0,0,0,0,0)\oplus$&$\frac{1}{2}(-1,1,1,1,1,-3,1,-1)$&$0,\frac{1}{2}$&E$_7\times$U(1)$|$SU(7)$\times$U(1)$^2$\\
37&$(-1,-1,0,0,0,0,0,0)\oplus$&$-\frac{1}{2}(1,1,1,1,-1,3,-1,-1)$&$-\frac{1}{4},\frac{1}{4}$&E$_7\times$U(1)$|$SU(5)$\times$SU(3)$\times$U(1)$^2$\\
38&$(-1,0,0,0,0,0,1,0)\oplus$&$(0,0,0,0,0,0,0,0)$&$0,\frac{1}{2}$&E$_6\times$U(1)$^2|$SU(8)$\times$U(1)\\
39&$(-1,0,0,0,0,0,1,0)\oplus$&$(0,0,0,0,0,-2,0,0)$&$-\frac{1}{8},\frac{3}{8}$&E$_6\times$U(1)$^2|$SU(8)$\times$U(1)\\
40&$(-1,0,0,0,0,0,1,0)\oplus$&$(-1,-1,-1,0,0,0,0,1)$&$-\frac{1}{4},\frac{1}{4}$&E$_6\times$U(1)$|$SU(4)$^2\times$U(1)$^2$\\
41&$(-1,1,0,0,0,0,0,0)\oplus$&$\frac{1}{2}(-1,1,1,1,1,-3,1,-1)$&$-\frac{3}{8},\frac{1}{8}$&SO(12)$\times$SU(2)$\times$U(1)$|$SU(7)$\times$U(1)$^2$\\

\hline
 \end{tabular}
}
\end{table}
\begin{table}[H]
  \centering
{\scriptsize
  \begin{tabular}{|cr@{}l|r|c|}
\hline
 &\multicolumn{2}{c|}{$2\cdot V_2$}&\multicolumn{1}{c|}{~$\Phi$}& Group Decomposition \\
\hline
\hline
\multicolumn{5}{|c|}{$4\cdot V_4=(1,1,0,0,0,0,0,0)\oplus(1,1,1,1,1,1,1,-1)$} \\
\hline
42&$(-1,1,0,0,0,0,0,0)\oplus$&$-\frac{1}{2}(1,1,1,1,-1,3,-1,-1)$&$-\frac{1}{8},\frac{3}{8}$&SO(12)$\times$SU(2)$\times$U(1)$|$SU(5)$\times$SU(3)$\times$U(1)$^2$\\
43&$(-1,-1,-1,0,0,0,0,1)\oplus$&$-\frac{1}{2}(1,1,1,1,1,1,-1,-1)$&$-\frac{1}{4},\frac{1}{4}$&SO(12)$\times$SU(2)$\times$U(1)$|$SU(7)$\times$U(1)$^2$\\
44&$(-1,-1,-1,0,0,0,0,1)\oplus$&$\frac{1}{2}(-1,-1,-1,-1,1,1,1,1)$&$-\frac{1}{8},\frac{3}{8}$&SO(12)$\times$SU(2)$\times$U(1)$|$SU(5)$\times$SU(3)$\times$U(1)$^2$\\
45&$(-1,0,-1,-1,0,0,0,1)\oplus$&$(0,0,0,0,-1,-1,0,0)$&$-\frac{1}{8},\frac{3}{8}$&SU(8)$\times$U(1)$|$SU(6)$\times$SU(2)$\times$U(1)$^2$\\
46&$(-1,0,-1,-1,0,0,0,1)\oplus$&$(-1,0,0,0,0,0,1,0)$&$0,\frac{1}{2}$&SU(8)$\times$U(1)$|$SU(6)$\times$SU(2)$\times$U(1)$^2$\\
\hline\hline
\multicolumn{5}{|c|}{$4\cdot V_4=(2,1,1,0,0,0,0,0)\oplus(2,0,0,0,0,0,0,0)$}\\
\hline
47&$(0,0,0,0,0,0,0,0)\oplus$&$-\frac{1}{2}(1,1,1,1,1,1,-1,-1)$&$0,\frac{1}{2}$&E$_6\times$SU(2)$\times$U(1)$|$SU(7)$\times$U(1)$^2$\\
48&$(-1,-1,0,0,0,0,0,0)\oplus$&$(0,0,0,0,0,0,0,0)$&$-\frac{1}{8},\frac{3}{8}$&E$_6\times$U(1)$^2|$SO(14)$\times$U(1)\\
49&$(-1,-1,0,0,0,0,0,0)\oplus$&$(0,0,0,0,0,-2,0,0)$&$-\frac{1}{8},\frac{3}{8}$&E$_6\times$U(1)$^2|$SO(14)$\times$U(1)\\
50&$(-1,-1,0,0,0,0,0,0)\oplus$&$(-1,-1,-1,0,0,0,0,1)$&$-\frac{1}{4},\frac{1}{4}$&E$_6\times$U(1)$^2|$SO(8)$\times$SU(4)$\times$U(1)\\
51&$(-1,0,0,0,0,0,1,0)\oplus$&$\frac{1}{2}(-3,-1,-1,1,1,1,1,1)$&$-\frac{1}{4},\frac{1}{4}$&SO(10)$\times$SU(2)$\times$U(1)$^2|$SU(7)$\times$U(1)$^2$\\
52&$(-1,0,1,0,0,0,0,0)\oplus$&$(0,0,0,0,0,0,0,0)$&$0,\frac{1}{2}$&SO(10)$\times$U(1)$^3|$SO(14)$\times$U(1)\\
53&$(-1,0,1,0,0,0,0,0)\oplus$&$(0,0,0,0,0,-2,0,0)$&$0,\frac{1}{2}$&SO(10)$\times$U(1)$^3|$SO(14)$\times$U(1)\\
54&$(-1,0,1,0,0,0,0,0)\oplus$&$(-1,-1,-1,0,0,0,0,1)$&$-\frac{1}{8},\frac{3}{8}$&SO(10)$\times$U(1)$^3|$SO(8)$\times$SU(4)$\times$U(1)\\
55&$(0,0,0,-1,0,0,1,0)\oplus$&$\frac{1}{2}(-3,-1,-1,1,1,1,1,1)$&$-\frac{1}{8},\frac{3}{8}$&SU(6)$\times$SU(2)$^2\times$U(1)$|$SU(7)$\times$U(1)$^2$\\
56&$(0,-1,1,0,0,0,0,0)\oplus$&$\frac{1}{2}(-3,-1,-1,1,1,1,1,1)$&$-\frac{1}{8},\frac{3}{8}$&E$_6\times$SU(2)$\times$U(1)$|$SU(7)$\times$U(1)$^2$\\
57&$(-1,-1,-1,0,0,0,0,1)\oplus$&$-\frac{1}{2}(1,1,1,1,1,1,-1,-1)$&$-\frac{1}{4},\frac{1}{4}$&SO(10)$\times$SU(2)$\times$U(1)$^2|$SU(7)$\times$U(1)$^2$\\
58&$(-1,-1,0,-1,0,0,0,1)\oplus$&$(-1,0,0,0,0,0,1,0)$&$-\frac{1}{4},\frac{1}{4}$&SU(6)$\times$SU(2)$\times$U(1)$^2|$SO(12)$\times$U(1)$^2$\\
59&$(-1,-1,0,-1,0,0,0,1)\oplus$&$(0,-1,0,0,0,0,1,0)$&$-\frac{1}{8},\frac{3}{8}$&SU(6)$\times$SU(2)$^2\times$U(1)$|$SO(10)$\times$SU(2)$^2\times$U(1)\\
60&$(-1,0,0,-1,-1,0,0,1)\oplus$&$-\frac{1}{2}(1,1,1,1,1,1,-1,-1)$&$-\frac{1}{8},\frac{3}{8}$&SU(6)$\times$SU(2)$^2\times$U(1)$|$SU(7)$\times$U(1)$^2$\\
\hline\hline
\multicolumn{5}{|c|}{$4\cdot V_4=(2,1,1,0,0,0,0,0)\oplus(2,2,2,0,0,0,0,0)$}\\
\hline
61&$(0,0,0,0,0,0,0,0)\oplus$&$-\frac{1}{2}(1,1,1,1,1,1,-1,-1)$&$-\frac{1}{8},\frac{3}{8}$&E$_6\times$SU(2)$\times$U(1)$|$SU(5)$\times$SU(3)$\times$U(1)$^2$\\
62&$(-1,-1,0,0,0,0,0,0)\oplus$&$(0,0,0,0,0,0,0,0)$&$-\frac{1}{8},\frac{3}{8}$&E$_6\times$U(1)$^2|$SO(10)$\times$SU(4)\\
63&$(-1,-1,0,0,0,0,0,0)\oplus$&$(0,0,0,0,0,-2,0,0)$&$-\frac{1}{8},\frac{3}{8}$& E$_6\times$U(1)$^2|$SO(10)$\times$SU(4)\\
64&$(-1,-1,0,0,0,0,0,0)\oplus$&$(-1,-1,-1,0,0,0,0,1)$&$0,\frac{1}{2}$&E$_6\times$U(1)$^2|$SO(8)$\times$SU(4)$\times$U(1)\\
65&$(-1,-1,0,0,0,0,0,0)\oplus$&$(-1,-1,0,-1,0,0,0,1)$&$-\frac{3}{8},\frac{1}{8}$&E$_6\times$U(1)$^2|$SU(4)$\times$SU(2)$^4\times$U(1)\\
66&$(-1,0,0,0,0,0,1,0)\oplus$&$\frac{1}{2}(-3,-1,-1,1,1,1,1,1)$&$-\frac{3}{8},\frac{1}{8}$&SO(10)$\times$SU(2)$\times$U(1)$^2|$SU(5)$\times$SU(3)$\times$U(1)$^2$\\
67&$(-1,0,1,0,0,0,0,0)\oplus$&$(0,0,0,0,0,0,0,0)$&$0,\frac{1}{2}$&SO(10)$\times$U(1)$^3|$SO(10)$\times$SU(4)\\
68&$(-1,0,1,0,0,0,0,0)\oplus$&$(0,0,0,0,0,-2,0,0)$&$0,\frac{1}{2}$&SO(10)$\times$U(1)$^3|$SO(10)$\times$SU(4)\\
69&$(-1,0,1,0,0,0,0,0)\oplus$&$(-1,-1,-1,0,0,0,0,1)$&$-\frac{1}{8},\frac{3}{8}$&SO(10)$\times$U(1)$^3|$SO(8)$\times$SU(4)$\times$U(1)\\
70&$(-1,0,1,0,0,0,0,0)\oplus$&$(-1,-1,0,-1,0,0,0,1)$&$-\frac{3}{8},\frac{1}{8}$&SO(10)$\times$U(1)$^3|$SU(4)$\times$SU(2)$^4\times$U(1)\\
71&$(0,0,0,-1,0,0,1,0)\oplus$&$\frac{1}{2}(-3,-1,-1,1,1,1,1,1)$&$-\frac{1}{4},\frac{1}{4}$&SU(6)$\times$SU(2)$^2\times$U(1)$|$SU(5)$\times$SU(3)$\times$U(1)$^2$\\
72&$(0,-1,1,0,0,0,0,0)\oplus$&$\frac{1}{2}(-3,-1,-1,1,1,1,1,1)$&$-\frac{1}{4},\frac{1}{4}$&E$_6\times$SU(2)$\times$U(1)$|$SU(5)$\times$SU(3)$\times$U(1)$^2$\\
73&$(-1,-1,-1,0,0,0,0,1)\oplus$&$-\frac{1}{2}(1,1,1,1,1,1,-1,-1)$&$-\frac{3}{8},\frac{1}{8}$&SO(10)$\times$SU(2)$\times$U(1)$^2|$SU(5)$\times$SU(3)$\times$U(1)$^2$\\
74&$(-1,-1,0,-1,0,0,0,1)\oplus$&$(-1,-1,0,0,0,0,0,0)$&$-\frac{3}{8},\frac{1}{8}$&SU(6)$\times$SU(2)$\times$U(1)$^2|$SO(10)$\times$SU(2)$^2\times$U(1)\\
75&$(-1,-1,0,-1,0,0,0,1)\oplus$&$(-1,0,0,0,0,0,1,0)$&$-\frac{1}{4},\frac{1}{4}$&SU(6)$\times$SU(2)$\times$U(1)$^2|$SO(8)$\times$SU(2)$^2\times$U(1)$^2$\\
76&$(-1,-1,0,-1,0,0,0,1)\oplus$&$(0,0,0,-1,0,0,1,0)$&$-\frac{1}{8},\frac{3}{8}$&SU(6)$\times$SU(2)$\times$U(1)$^2|$SU(4)$^2\times$SU(2)$^2$\\
77&$(-1,0,0,-1,-1,0,0,1)\oplus$&$-\frac{1}{2}(1,1,1,1,1,1,-1,-1)$&$-\frac{1}{4},\frac{1}{4}$&SU(6)$\times$SU(2)$^2\times$U(1)$|$SU(5)$\times$SU(3)$\times$U(1)$^2$\\
\hline
\hline
\multicolumn{5}{|c|}{$4\cdot V_4=(4,0,0,0,0,0,0,0)\oplus(3,1,0,0,0,0,0,0)$}\\
\hline
78&$(0,0,0,0,0,0,0,0)\oplus$&$(-1,0,0,0,0,0,1,0)$&$-\frac{1}{8},\frac{3}{8}$&SO(16)$|$SO(10)$\times$U(1)$^3$\\
79&$(0,0,0,0,0,0,0,0)\oplus$&$-\frac{1}{2}(1,-1,1,1,1,1,1,-1)$&$0,\frac{1}{2}$&SO(16)$|$SU(6)$\times$SU(2)$\times$U(1)$^2$\\
80&$(-1,0,0,0,0,0,1,0)\oplus$&$\frac{1}{2}(-3,-1,-1,1,1,1,1,1)$&$0,\frac{1}{2}$&SO(12)$\times$SU(2)$^2|$SU(6)$\times$SU(2)$\times$U(1)$^2$\\
81&$(-1,0,0,0,0,0,1,0)\oplus$&$(-1,0,-1,-1,0,0,0,1)$&$-\frac{3}{8},\frac{1}{8}$&SO(12)$\times$SU(2)$^2|$SU(4)$^2\times$U(1)$^2$\\
82&$-\frac{1}{2}(1,1,1,1,1,1,-1,-1)\oplus$&$\frac{1}{2}(-3,-1,-1,1,1,1,1,1)$&$-\frac{3}{8},\frac{1}{8}$&SU(8)$\times$U(1)$|$SU(6)$\times$SU(2)$\times$U(1)$^2$\\
83&$-\frac{1}{2}(1,1,1,1,1,1,-1,-1)\oplus$&$(-1,0,-1,-1,0,0,0,1)$&$-\frac{1}{4},\frac{1}{4}$&SU(8)$\times$U(1)$|$SU(4)$^2\times$U(1)$^2$\\
84&$(0,0,0,0,0,-2,0,0)\oplus$&$(-1,0,0,0,0,0,1,0)$&$-\frac{1}{8},\frac{3}{8}$&SO(14)$\times$U(1)$|$SO(10)$\times$U(1)$^3$\\
85&$(0,0,0,0,0,-2,0,0)\oplus$&$-\frac{1}{2}(1,-1,1,1,1,1,1,-1)$&$0,\frac{1}{2}$&SO(14)$\times$U(1)$|$SU(6)$\times$SU(2)$\times$U(1)$^2$\\
86&$\frac{1}{2}(-3,-1,-1,1,1,1,1,1)\oplus$&$(-1,0,0,0,0,0,1,0)$&$0,\frac{1}{2}$&SU(7)$\times$U(1)$^2|$SO(10)$\times$U(1)$^3$\\
87&$\frac{1}{2}(-3,-1,-1,1,1,1,1,1)\oplus$&$-\frac{1}{2}(1,-1,1,1,1,1,1,-1)$&$-\frac{3}{8},\frac{1}{8}$&SU(7)$\times$U(1)$^2|$SU(6)$\times$SU(2)$\times$U(1)$^2$\\
88&$(-1,-1,-1,0,0,0,0,1)\oplus$&$(-1,0,0,0,0,0,1,0)$&$-\frac{3}{8},\frac{1}{8}$&SO(8)$\times$SU(4)$\times$U(1)$|$SO(10)$\times$U(1)$^3$\\
89&$(-1,-1,-1,0,0,0,0,1)\oplus$&$-\frac{1}{2}(1,-1,1,1,1,1,1,-1)$&$-\frac{1}{4},\frac{1}{4}$&SO(8)$\times$SU(4)$\times$U(1)$|$SU(6)$\times$SU(2)$\times$U(1)$^2$\\
\hline\hline
\multicolumn{5}{|c|}{$4\cdot V_4=(2,0,0,0,0,0,0,0)\oplus(3,1,1,1,1,1,0,0)$}\\
\hline
90&$(0,0,0,0,0,0,0,0)\oplus$&$(-1,0,0,0,0,0,1,0)$&$-\frac{1}{8},\frac{3}{8}$&SO(14)$\times$U(1)$|$SU(6)$\times$SU(2)$\times$U(1)$^2$\\
91&$(-1,0,0,0,0,0,1,0)\oplus$&$\frac{1}{2}(-1,-3,1,1,1,1,-1,1)$&$-\frac{1}{8},\frac{3}{8}$&SO(12)$\times$U(1)$^2|$SU(8)$\times$U(1)\\
92&$(-1,0,0,0,0,0,1,0)\oplus$&$(-1,-1,-1,0,0,0,0,1)$&$-\frac{3}{8},\frac{1}{8}$&SO(12)$\times$U(1)$^2|$SU(4)$^2\times$U(1)$^2$\\
93&$-\frac{1}{2}(1,1,1,1,1,1,-1,-1)\oplus$&$(0,0,0,0,0,0,0,0)$&$0,\frac{1}{2}$&SU(7)$\times$U(1)$^2|$SU(8)$\times$SU(2)\\
94&$-\frac{1}{2}(1,1,1,1,1,1,-1,-1)\oplus$&$(-1,0,0,1,-1,1,0,0)$&$-\frac{1}{8},\frac{3}{8}$&SU(7)$\times$U(1)$^2|$SU(6)$\times$SU(2)$^2\times$U(1)\\
95&$-\frac{1}{2}(1,1,1,1,1,1,-1,-1)\oplus$&$(0,-1,-1,0,0,0,-1,1)$&$-\frac{1}{8},\frac{3}{8}$&SU(7)$\times$U(1)$^2|$SU(4)$^2\times$SU(2)$\times$U(1)\\
96&$(0,-1,0,0,0,0,1,0)\oplus$&$\frac{1}{2}(-1,-3,1,1,1,1,-1,1)$&$0,\frac{1}{2}$&SO(10)$\times$SU(2)$^2\times$U(1)$|$SU(8)$\times$U(1)\\
97&$(0,-1,0,0,0,0,1,0)\oplus$&$(-1,-1,-1,0,0,0,0,1)$&$-\frac{1}{4},\frac{1}{4}$&SO(10)$\times$SU(2)$^2\times$U(1)$|$SU(4)$^2\times$U(1)$^2$\\
98&$(0,0,0,0,0,-2,0,0)\oplus$&$(-1,0,0,0,0,0,1,0)$&$-\frac{1}{8},\frac{3}{8}$&SO(14)$\times$U(1)$|$SU(6)$\times$SU(2)$\times$U(1)$^2$\\
99&$\frac{1}{2}(-3,-1,-1,1,1,1,1,1)\oplus$&$(-1,-1,0,0,0,0,0,0)$&$-\frac{3}{8},\frac{1}{8}$&SU(7)$\times$U(1)$^2|$SU(6)$\times$SU(2)$^2\times$U(1)\\
100&$\frac{1}{2}(-3,-1,-1,1,1,1,1,1)\oplus$&$(-1,0,0,0,0,1,0,0)$&$-\frac{1}{4},\frac{1}{4}$&SU(7)$\times$U(1)$^2|$SU(4)$^2\times$SU(2)$\times$U(1)\\
101&$\frac{1}{2}(-3,-1,-1,1,1,1,1,1)\oplus$&$(0,0,0,0,0,0,-1,1)$&$-\frac{1}{8},\frac{3}{8}$&SU(7)$\times$U(1)$^2|$SU(8)$\times$SU(2)\\
102&$(-1,-1,-1,0,0,0,0,1)\oplus$&$(-1,0,0,0,0,0,1,0)$&$-\frac{1}{4},\frac{1}{4}$&SO(8)$\times$SU(4)$\times$U(1)$|$SU(6)$\times$SU(2)$\times$U(1)$^2$\\
\hline
\end{tabular}
}
\end{table}
\begin{table}[H]
  \centering
{\scriptsize
  \begin{tabular}{|cr@{}l|r|c|}
\hline
 &\multicolumn{2}{c|}{$2\cdot V_2$}&\multicolumn{1}{c|}{~$\Phi$}& Group Decomposition \\

\hline
\hline
\multicolumn{5}{|c|}{$4\cdot V_4=(3,1,0,0,0,0,0,0)\oplus(1,1,1,1,1,1,1,-1)$}\\
\hline
103&$(0,0,0,0,0,0,0,0)\oplus$&$-\frac{1}{2}(1,1,1,1,1,1,-1,-1)$&$-\frac{1}{8},\frac{3}{8}$&SO(12)$\times$SU(2)$\times$U(1)$|$SU(7)$\times$U(1)$^2$\\
104&$(0,0,0,0,0,0,0,0)\oplus$&$\frac{1}{2}(-1,-1,-1,-1,1,1,1,1)$&$0,\frac{1}{2}$&SO(12)$\times$SU(2)$\times$U(1)$|$SU(5)$\times$SU(3)$\times$U(1)$^2$\\
105&$(-1,-1,0,0,0,0,0,0)\oplus$&$\frac{1}{2}(-1,1,1,1,1,-3,1,-1)$&$-\frac{1}{8},\frac{3}{8}$&SO(12)$\times$SU(2)$\times$U(1)$|$SU(7)$\times$U(1)$^2$\\
106&$(-1,-1,0,0,0,0,0,0)\oplus$&$-\frac{1}{2}(1,1,1,1,-1,3,-1,-1)$&$-\frac{3}{8},\frac{1}{8}$&SO(12)$\times$SU(2)$\times$U(1)$|$SU(5)$\times$SU(3)$\times$U(1)$^2$\\
107&$(-1,0,0,0,0,0,1,0)\oplus$&$(0,0,0,0,0,0,0,0)$&$-\frac{1}{8},\frac{3}{8}$&SO(10)$\times$U(1)$^3|$SU(8)$\times$U(1)\\
108&$(-1,0,0,0,0,0,1,0)\oplus$&$(0,0,0,0,0,-2,0,0)$&$-\frac{1}{4},\frac{1}{4}$&SO(10)$\times$U(1)$^3|$SU(8)$\times$U(1)\\
109&$(-1,0,0,0,0,0,1,0)\oplus$&$(-1,-1,-1,0,0,0,0,1)$&$-\frac{3}{8},\frac{1}{8}$&SO(10)$\times$U(1)$^3|$SU(4)$^2\times$U(1)$^2$\\
110&$(-1,1,0,0,0,0,0,0)\oplus$&$\frac{1}{2}(-1,1,1,1,1,-3,1,-1)$&$0,\frac{1}{2}$&SO(12)$\times$SU(2)$\times$U(1)$|$SU(7)$\times$U(1)$^2$\\
111&$(-1,1,0,0,0,0,0,0)\oplus$&$-\frac{1}{2}(1,1,1,1,-1,3,-1,-1)$&$-\frac{1}{4},\frac{1}{4}$&SO(12)$\times$SU(2)$\times$U(1)$|$SU(5)$\times$SU(3)$\times$U(1)$^2$\\
112&$-\frac{1}{2}(1,1,1,1,1,1,-1,-1)\oplus$&$\frac{1}{2}(-1,1,1,1,1,-3,1,-1)$&$0,\frac{1}{2}$&SU(6)$\times$U(1)$^3|$SU(7)$\times$U(1)$^2$\\
113&$-\frac{1}{2}(1,1,1,1,1,1,-1,-1)\oplus$&$-\frac{1}{2}(1,1,1,1,-1,3,-1,-1)$&$-\frac{1}{4},\frac{1}{4}$&SU(6)$\times$U(1)$^3|$SU(5)$\times$SU(3)$\times$U(1)$^2$\\
114&$-\frac{1}{2}(1,-1,1,1,1,1,1,-1)\oplus$&$(0,0,0,0,0,0,0,0)$&$0,\frac{1}{2}$&SU(6)$\times$SU(2)$\times$U(1)$^2|$SU(8)$\times$U(1)\\
115&$-\frac{1}{2}(1,-1,1,1,1,1,1,-1)\oplus$&$(0,0,0,0,0,-2,0,0)$&$-\frac{1}{8},\frac{3}{8}$&SU(6)$\times$SU(2)$\times$U(1)$^2|$SU(8)$\times$U(1)\\
116&$-\frac{1}{2}(1,-1,1,1,1,1,1,-1)\oplus$&$(-1,-1,-1,0,0,0,0,1)$&$-\frac{1}{4},\frac{1}{4}$&SU(6)$\times$SU(2)$\times$U(1)$^2|$SU(4)$^2\times$U(1)$^2$\\
117&$(0,0,-1,0,0,0,1,0)\oplus$&$\frac{1}{2}(-1,1,1,1,1,-3,1,-1)$&$-\frac{3}{8},\frac{1}{8}$&SO(8)$\times$SU(2)$^3\times$U(1)$|$SU(7)$\times$U(1)$^2$\\
118&$(0,0,-1,0,0,0,1,0)\oplus$&$-\frac{1}{2}(1,1,1,1,-1,3,-1,-1)$&$-\frac{1}{8},\frac{3}{8}$&SO(8)$\times$SU(2)$^3\times$U(1)$|$SU(5)$\times$SU(3)$\times$U(1)$^2$\\
119&$(0,0,0,0,0,-2,0,0)\oplus$&$-\frac{1}{2}(1,1,1,1,1,1,-1,-1)$&$-\frac{1}{8},\frac{3}{8}$&SO(12)$\times$SU(2)$\times$U(1)$|$SU(7)$\times$U(1)$^2$\\
120&$(0,0,0,0,0,-2,0,0)\oplus$&$\frac{1}{2}(-1,-1,-1,-1,1,1,1,1)$&$0,\frac{1}{2}$&SO(12)$\times$SU(2)$\times$U(1)$|$SU(5)$\times$SU(3)$\times$U(1)$^2$\\
121&$\frac{1}{2}(-3,-1,-1,1,1,1,1,1)\oplus$&$(0,0,0,0,-1,-1,0,0)$&$-\frac{3}{8},\frac{1}{8}$&SU(6)$\times$SU(2)$\times$U(1)$^2|$SU(6)$\times$SU(2)$\times$U(1)$^2$\\
122&$\frac{1}{2}(-3,-1,-1,1,1,1,1,1)\oplus$&$(-1,0,0,0,0,0,1,0)$&$-\frac{1}{4},\frac{1}{4}$&SU(6)$\times$SU(2)$\times$U(1)$^2|$SU(6)$\times$SU(2)$\times$U(1)$^2$\\
123&$\frac{1}{2}(-3,1,-1,-1,1,1,1,1)\oplus$&$-\frac{1}{2}(1,1,1,1,1,1,-1,-1)$&$-\frac{3}{8},\frac{1}{8}$&SU(6)$\times$U(1)$^3|$SU(7)$\times$U(1)$^2$\\
124&$\frac{1}{2}(-3,1,-1,-1,1,1,1,1)\oplus$&$\frac{1}{2}(-1,-1,-1,-1,1,1,1,1)$&$-\frac{1}{4},\frac{1}{4}$&SU(6)$\times$U(1)$^3|$SU(5)$\times$SU(3)$\times$U(1)$^2$\\
125&$(-1,-1,-1,0,0,0,0,1)\oplus$&$-\frac{1}{2}(1,1,1,1,1,1,-1,-1)$&$-\frac{3}{8},\frac{1}{8}$&SO(8)$\times$SU(2)$^3\times$U(1)$|$SU(7)$\times$U(1)$^2$\\
126&$(-1,-1,-1,0,0,0,0,1)\oplus$&$\frac{1}{2}(-1,-1,-1,-1,1,1,1,1)$&$-\frac{1}{4},\frac{1}{4}$&SO(8)$\times$SU(2)$^3\times$U(1)$|$SU(5)$\times$SU(3)$\times$U(1)$^2$\\
127&$(-1,0,-1,-1,0,0,0,1)\oplus$&$(0,0,0,0,-1,-1,0,0)$&$-\frac{1}{4},\frac{1}{4}$&SU(4)$^2\times$U(1)$^2|$SU(6)$\times$SU(2)$\times$U(1)$^2$\\
128&$(-1,0,-1,-1,0,0,0,1)\oplus$&$(-1,0,0,0,0,0,1,0)$&$-\frac{1}{8},\frac{3}{8}$&SU(4)$^2\times$U(1)$^2|$SU(6)$\times$SU(2)$\times$U(1)$^2$\\
\hline
\hline
\multicolumn{5}{|c|}{$4\cdot V_4=(2,2,2,0,0,0,0,0)\oplus(3,1,1,1,1,1,0,0)$} \\
\hline
129&$(0,0,0,0,0,0,0,0)\oplus$&$(-1,0,0,0,0,0,1,0)$&$-\frac{1}{8},\frac{3}{8}$&SO(10)$\times$SU(4)$|$SU(6)$\times$SU(2)$\times$U(1)$^2$\\
130&$(-1,-1,0,0,0,0,0,0)\oplus$&$\frac{1}{2}(-1,-3,1,1,1,1,-1,1)$&$-\frac{1}{4},\frac{1}{4}$&SO(10)$\times$SU(2)$^2\times$U(1)$|$SU(8)$\times$U(1)\\
131&$(-1,-1,0,0,0,0,0,0)\oplus$&$(-1,-1,-1,0,0,0,0,1)$&$0,\frac{1}{2}$&SO(10)$\times$SU(2)$^2\times$U(1)$|$SU(4)$^2\times$U(1)$^2$\\
132&$(-1,0,0,0,0,0,1,0)\oplus$&$\frac{1}{2}(-1,-3,1,1,1,1,-1,1)$&$-\frac{1}{8},\frac{3}{8}$&SO(8)$\times$SU(2)$^2\times$U(1)$^2|$SU(8)$\times$U(1)\\
133&$(-1,0,0,0,0,0,1,0)\oplus$&$(-1,-1,-1,0,0,0,0,1)$&$-\frac{3}{8},\frac{1}{8}$&SO(8)$\times$SU(2)$^2\times$U(1)$^2|$SU(4)$^2\times$U(1)$^2$\\
134&$-\frac{1}{2}(1,1,1,1,1,1,-1,-1)\oplus$&$(0,0,0,0,0,0,0,0)$&$-\frac{1}{8},\frac{3}{8}$&SU(5)$\times$SU(3)$\times$U(1)$^2|$SU(8)$\times$SU(2)\\
135&$-\frac{1}{2}(1,1,1,1,1,1,-1,-1)\oplus$&$(-1,0,0,1,-1,1,0,0)$&$-\frac{1}{4},\frac{1}{4}$&SU(5)$\times$SU(3)$\times$U(1)$^2|$SU(8)$\times$SU(2)\\
136&$-\frac{1}{2}(1,1,1,1,1,1,-1,-1)\oplus$&$(0,-1,-1,0,0,0,-1,1)$&$-\frac{1}{4},\frac{1}{4}$&SU(5)$\times$SU(3)$\times$U(1)$^2|$SU(4)$^2\times$SU(2)$\times$U(1)\\
137&$(0,0,0,-1,0,0,1,0)\oplus$&$\frac{1}{2}(-1,-3,1,1,1,1,-1,1)$&$0,\frac{1}{2}$&SU(4)$^2\times$SU(2)$^2|$SU(8)$\times$U(1)\\
138&$(0,0,0,-1,0,0,1,0)\oplus$&$(-1,-1,-1,0,0,0,0,1)$&$-\frac{1}{4},\frac{1}{4}$&SU(4)$^2\times$SU(2)$^2|$SU(4)$^2\times$U(1)$^2$\\
139&$(0,0,0,0,0,-2,0,0)\oplus$&$(-1,0,0,0,0,0,1,0)$&$-\frac{1}{8},\frac{3}{8}$&SO(10)$\times$SU(4)$|$SU(6)$\times$SU(2)$\times$U(1)$^2$\\
140&$\frac{1}{2}(-3,-1,-1,1,1,1,1,1)\oplus$&$(-1,-1,0,0,0,0,0,0)$&$0,\frac{1}{2}$&SU(5)$\times$SU(3)$\times$U(1)$^2|$SU(6)$\times$SU(2)$^2\times$U(1)\\
141&$\frac{1}{2}(-3,-1,-1,1,1,1,1,1)\oplus$&$(-1,0,0,0,0,1,0,0)$&$-\frac{3}{8},\frac{1}{8}$&SU(5)$\times$SU(3)$\times$U(1)$^2|$SU(4)$^2\times$SU(2)$\times$U(1)\\
142&$\frac{1}{2}(-3,-1,-1,1,1,1,1,1)\oplus$&$(0,0,0,0,0,0,-1,1)$&$-\frac{1}{4},\frac{1}{4}$&SU(5)$\times$SU(3)$\times$U(1)$^2|$SU(8)$\times$SU(2)\\
143&$(-1,-1,-1,0,0,0,0,1)\oplus$&$(-1,0,0,0,0,0,1,0)$&$0,\frac{1}{2}$&SO(8)$\times$SU(4)$\times$U(1)$|$SU(6)$\times$SU(2)$\times$U(1)$^2$\\
144&$(-1,-1,0,-1,0,0,0,1)\oplus$&$(-1,0,0,0,0,0,1,0)$&$-\frac{3}{8},\frac{1}{8}$&SU(4)$\times$SU(2)$^4\times$U(1)$|$SU(6)$\times$SU(2)$\times$U(1)$^2$\\
\hline
 \end{tabular}
}
\end{table}

\subsection{Complete Modular Invariant Embeddings}
The modular invariant embeddings presented before need to fulfill the strong modular invariant conditions. Here we present some fully modular versions of the embeddings which we found promising for phenomenology. The embeddings are given according to the following choice for the twists
\begin{equation*}
  v_2= (0,\frac12 , 0 , -\frac12)\, \quad v_4 =(0,0,\frac14 , -\frac14)
\end{equation*}
which is the convention used by the C++ orbifolder.
\begin{align*}   \text{Model 21, brother 1: }  &V_2= \left(0,\frac12,0,0,0,0,\frac12,0\right)\left(2,0,0,0,0,-1,0,0\right)\\
 &V_4=\left(\frac14,\frac14,0,0,0,0,0,0\right)\left(1,0,0,0,0,1,1,0\right)
\end{align*}
\begin{align*}
 \text{Model 62, brother 2: } &V_2 = \left(\frac12,0,\frac12,0,0,0,0,0\right)\left(\frac32, \frac12, \frac12, -\frac12,-\frac12,-\frac12,-\frac12,\frac12\right) \\
   &V_4 = \left(\frac12,\frac14,\frac14,0,0,0,0,0\right)\left(1,0,0,-\frac12,-\frac12,-\frac12,\frac12,-\frac12\right) \\\\
\text{Model 63, brother 2: } &V_2=\left(\frac12,0,\frac12,0,0,0,0,0\right)\left(\frac32,\frac12,\frac12,-\frac12,-\frac12,-\frac32,-\frac12,\frac12\right)  \\
  &V_4 = \left(\frac12,\frac14,\frac54,-1,0,0,0,0\right)\left(\frac12,\frac12,\frac12,0,0,0,0,0\right) \\\\
 \text{Model  67, brother 1: }  &V_2 = \left(\frac12, \frac12,1,0,0,0,0,0\right)\left(1,1,1,0,0,0,0,0\right) \\
   &V_4=\left(\frac12, \frac14, \frac14, 0,0,0,0,0\right)\left(\frac12,\frac12,\frac12,0,0,0,0,0\right) \\\\
 \text{Model 84, brother 1: }  &V_2 = \left(2,0,0,0,0,-1,0,0\right)\left(\frac32,0,-\frac12,-\frac12,-\frac12,-\frac12,0,\frac12\right) \\
  &V_4= \left(1,0,0,0,0,0,0,0\right)\left(\frac34,\frac14,0,0,0,0,0,0\right)\\\\
\text{Model 96, brother 2: }  & V_2=\left(1,-\frac12,0,0,0,-\frac12,0,0\right)\left(\frac54,-\frac14,\frac34,\frac34,\frac34,\frac34,-\frac14,\frac14\right) \\
  &V_4 = \left(\frac12,0,0,0,0,0,0,0\right)\left(\frac54,-\frac14,-\frac14,-\frac14,-\frac14,-\frac14,\frac12,-\frac12\right)
\end{align*}
\section{Matter Content of \SO{10} and \es\ models}
\label{app:mattercontent}
\begin{table}[H]
\centering 
\caption{Left chiral spectrum for shifts containing an $SO(10)$ factor. Only the fields charged under \SO{10} are shown, for each model the splitting in two rows is made to distinguish between the brother models. When one refers to model $67_1$, for example, the subindex is used to distinguish between the two \SO{10} factors contained in the gauge group of model 67. The check mark indicates that Gauge Top Unification is possible. At the sectors $T_{(0,2)}$, $T_{(1,0)}$ and $T_{(1,2)}$ we put two columns, to display the content of ordinary and special tori separately. The number ``2'' after the checkmark in some of the models indicates that two trilinear untwisted couplings are possible.}\label{matterso10} \vspace{0.2cm}
\tiny{
\renewcommand{\arraystretch}{1.4}
\begin{tabular}{|c|c|c|cccccccccc|} 
\hline 
Model & Untwisted &GTU &(0,3)  & \multicolumn{2}{c}{(0,2)} & (0,1) & \multicolumn{2}{c}{(1,0)} & (1,3) & \multicolumn{2}{c}{(1,2)} & (1,1) \\
\hline\hline
    \multirow{2}*{4}    &    \multirow {2}*{4($\mathbf{10}$), 2($\overline{\mathbf{16}}$)}     & \multirow{2}*{$\checkmark$} &  $\mathbf{10}$,$\mathbf{16}$    & $\mathbf{10}$ & $\mathbf{10}$,$\mathbf{16}$ & $\mathbf{10}$,$\mathbf{16}$& $\mathbf{16}$ & $\mathbf{10}$,$\mathbf{16}$ & & $\mathbf{16}$ & $\mathbf{10}$,$\mathbf{16}$ & $\mathbf{10}$\\
\cline{4-13}
& & & $\mathbf{10}$,$\overline{\mathbf{16}}$ & $\mathbf{10}$ & $\mathbf{10}$,$\mathbf{16}$ & $\mathbf{10}$,$\overline{\mathbf{16}}$ & $\mathbf{10}$ & $\mathbf{10}$,$\mathbf{16}$ & & $\mathbf{10}$ & $\mathbf{10}$,$\mathbf{16}$ & $\overline{\mathbf{16}}$\\
\hline\hline
    \multirow {2}*{8}    &    \multirow {2}*{4($\mathbf{10}$), 2($\overline{\mathbf{16}}$)}&\multirow{2}*{$\checkmark$}     &$\mathbf{10}$,$\mathbf{16}$ & $\mathbf{10}$ & $\mathbf{10}$,$\mathbf{16}$ & $\mathbf{10}$,$\mathbf{16}$&  &  & &  &  & \\
\cline{4-13}
&     & &$\mathbf{10}$,$\overline{\mathbf{16}}$ & $\mathbf{10}$ & $\mathbf{10}$,$\mathbf{16}$ & $\mathbf{10}$,$\overline{\mathbf{16}}$&  & & &  &  & \\
\hline\hline
    \multirow {2}*{16}    &    \multirow{2}*{4($\mathbf{10}$), 2($\overline{\mathbf{16}}$)}     & \multirow{2}*{$\checkmark$} &     & $\mathbf{16}$ & $\mathbf{10}$,$\mathbf{16}$ & $\mathbf{10}$& $\mathbf{16}$ & $\mathbf{10}$,$\mathbf{16}$ &  & $\mathbf{10}$ & $\mathbf{10}$,$\mathbf{16}$ & \\
\cline{4-13}
& & & $\mathbf{10}$ & $\mathbf{16}$ & $\mathbf{10}$,$\mathbf{16}$ & & $\mathbf{10}$ & $\mathbf{10}$,$\mathbf{16}$ & & $\mathbf{16}$ & $\mathbf{10}$,$\mathbf{16}$ &  \\
\hline\hline
    \multirow {2}*{24}    &    \multirow{2}*{4($\mathbf{10}$), 2($\overline{\mathbf{16}}$)} & \multirow{2}*{\checkmark} &    & $\mathbf{16}$ & $\mathbf{10}$,$\mathbf{16}$ & $\mathbf{10}$&  & & &  &  & \\
\cline{4-13}
&    & & $\mathbf{10}$ & $\mathbf{16}$ & $\mathbf{10}$,$\mathbf{16}$ & & &  & &  & & \\
\hline\hline
    \multirow {1}*{26}    &4($\mathbf{10}$), 2($\overline{\mathbf{16}}$) & \multirow{1}*{$\checkmark$} & $\mathbf{10}$ & $\overline{\mathbf{16}}$ & $\mathbf{10}$,$\overline{\mathbf{16}}$ & $\mathbf{10}$&  &  & &  &  & \\
\hline\hline
    \multirow {2}*{51}    &($\mathbf{1},\mathbf{16}$), ($\mathbf{1},\overline{\mathbf{16}}$),     & \multirow{2}*{$\checkmark$} &    ($\mathbf{1},\mathbf{10}$)&  &  & ($\mathbf{1},\overline{\mathbf{16}}$) &  &  & &  & ($\mathbf{1},\mathbf{10}$) & \\
\cline{4-13}
& ($\mathbf{2},\mathbf{10}$), ($\mathbf{2},\overline{\mathbf{16}}$)    & & ($\mathbf{1},\mathbf{16}$) &  &  & ($\mathbf{1},\mathbf{10}$) &  &  & & ($\mathbf{1},\mathbf{10}$) & ($\mathbf{1},\mathbf{10}$) & \\
\hline\hline
    \multirow{2}*{52}    &    2($\mathbf{10}$), 2($\mathbf{16}$),     &\multirow{2}*{$\checkmark2$}& $\mathbf{16}$    &  &  &  $\mathbf{10}$ & $\mathbf{16}$ & $\mathbf{10}$,$\mathbf{16}$ & &  &  & $\mathbf{10}$\\
\cline{4-13}
&    2($\overline{\mathbf{16}}$) &  &  $\mathbf{10}$  &  &  & $\overline{\mathbf{16}}$& $\mathbf{10}$ & $\mathbf{10}$,$\mathbf{16}$ & & &  & \\
\hline\hline
    \multirow{2}*{53}    &    2($\mathbf{10}$), 2($\mathbf{16}$),     & \multirow{2}*{$\checkmark2$} &$\mathbf{16}$ &  &  & $\mathbf{10}$&  &  & & $\mathbf{10}$ & $\mathbf{10}$,$\overline{\mathbf{16}}$ & \\
\cline{4-13}
& 2($\overline{\mathbf{16}}$) & & $\mathbf{10}$ &  &  & $\overline{\mathbf{16}}$ &  &  & & $\overline{\mathbf{16}}$ & $\mathbf{10}$,$\overline{\mathbf{16}}$ & $\mathbf{10}$\\
\hline\hline
    \multirow {2}*{54}    &    2($\mathbf{10}$), 2($\mathbf{16}$),     & \multirow {2}*{$\checkmark$} &  $\mathbf{16}$    &  & & $\mathbf{10}$ & & & & &  & \\
\cline{4-13}
& 2($\overline{\mathbf{16}}$)  && $\mathbf{10}$ &  &  & $\overline{\mathbf{16}}$ &  &  & &  &  & \\
\hline\hline
    \multirow {2}*{57}    & ($\mathbf{1},\mathbf{16}$), ($\mathbf{1},\overline{\mathbf{16}}$),      & \multirow {2}*{$\checkmark$} & ($\mathbf{1},\mathbf{10}$)    & &  & ($\mathbf{1},\mathbf{16}$)& ($\mathbf{1},\mathbf{10}$) & ($\mathbf{1},\mathbf{10}$) & &  &  & \\
\cline{4-13}
& ($\mathbf{2},\mathbf{10}$), ($\mathbf{2},\overline{\mathbf{16}}$)  && ($\mathbf{1},\overline{\mathbf{16}}$) &  &  & ($\mathbf{1},\mathbf{10}$)& & ($\mathbf{1},\mathbf{10}$) & &  &  & \\
\hline\hline
    \multirow {2}*{59}    &    ($\mathbf{2},\mathbf{2},\mathbf{10}$),($\mathbf{1},\mathbf{2},\mathbf{16}$),  & \multirow {2}*{$\checkmark$}& ($\mathbf{1},\mathbf{1},\mathbf{10}$) &  & ($\mathbf{1},\mathbf{1},\mathbf{10}$) & &  &  & &  &  & \\
\cline{4-13}
& ($\mathbf{2},\mathbf{1},\mathbf{16}$)  &&  &  & ($\mathbf{1},\mathbf{1},\mathbf{10}$) & ($\mathbf{1},\mathbf{1},\mathbf{10}$) &  &  & & &  & \\
\hline\hline
    \multirow {2}*{62}    &    \multirow {2}*{($\mathbf{4},\overline{\mathbf{16}}$)}      &\multirow {2}*{$X$} & ($\mathbf{1},\mathbf{16}$)    & ($\mathbf{1},\mathbf{10}$) & ($\mathbf{1},\mathbf{10}$) & &  &  & & & ($\mathbf{1},\mathbf{10}$) & \\
\cline{4-13}
&  & &  & ($\mathbf{1},\mathbf{10}$) & ($\mathbf{1},\mathbf{10}$) & ($\mathbf{1},\overline{\mathbf{16}}$)&  &  & & ($\mathbf{1},\mathbf{10}$) & ($\mathbf{1},\mathbf{10}$) & ($\mathbf{1},\overline{\mathbf{16}}$)\\
\hline\hline
    \multirow {2}*{63}    &    \multirow {2}*{($\mathbf{4},\mathbf{16}$)}      &\multirow {2}*{$X$}&     & ($\mathbf{1},\mathbf{10}$) & ($\mathbf{1},\mathbf{10}$) & ($\mathbf{1},\overline{\mathbf{16}}$) &  & ($\mathbf{1},\mathbf{10}$) & & &  & \\
\cline{4-13}
&  & & ($\mathbf{1},\mathbf{16}$)  & ($\mathbf{1},\mathbf{10}$) & ($\mathbf{1},\mathbf{10}$) &  & ($\mathbf{1},\mathbf{10}$)  & ($\mathbf{1},\mathbf{10}$)  &  &  &  & ($\mathbf{1},\mathbf{16}$) \\
\hline\hline
    \multirow {2}*{66}    & ($\mathbf{1},\mathbf{16}$), ($\mathbf{1},\overline{\mathbf{16}}$),&\multirow {2}*{$\checkmark$} &    & &  & &  & & & ($\mathbf{1},\mathbf{10}$) & ($\mathbf{1},\mathbf{10}$) & \\
\cline{4-13}
& ($\mathbf{2},\mathbf{10}$), ($\mathbf{2},\overline{\mathbf{16}}$) & &  &  &  & & &  & &  & ($\mathbf{1},\mathbf{10}$)  & \\
\hline\hline
    \multirow {2}*{67$_1$}    &    2($\mathbf{10}$), 2($\mathbf{16}$),&\multirow {2}*{$\checkmark2$}     &  &  &  &  & $\mathbf{16}$ & $\mathbf{10}$,$\mathbf{16}$ & &  &  & \\
\cline{4-13}
&    2($\overline{\mathbf{16}}$)   & &  &  &  &  & $\mathbf{10}$ & $\mathbf{10}$,$\mathbf{16}$ & &  &  & \\
\hline\hline
    \multirow {2}*{67$_2$}    &    \multirow {2}*{($\mathbf{4},\overline{\mathbf{16}}$)}  &\multirow {2}*{$X$} & ($\mathbf{1},\mathbf{16}$)    & ($\mathbf{1},\mathbf{10}$) & ($\mathbf{1},\mathbf{10}$) & &  &  & & ($\mathbf{1},\mathbf{10}$) & ($\mathbf{1},\mathbf{10}$) & \\
\cline{4-13}
& &  &  & ($\mathbf{1},\mathbf{10}$) & ($\mathbf{1},\mathbf{10}$) & ($\mathbf{1},\overline{\mathbf{16}}$)&  &  & &  & ($\mathbf{1},\mathbf{10}$) & \\
\hline\hline
    \multirow {2}*{68$_1$}    &    2($\mathbf{10}$), 2($\mathbf{16}$), &\multirow {2}*{$\checkmark2$} & &  & & &  &  & & $\overline{\mathbf{16}}$ & $\mathbf{10}$,$\overline{\mathbf{16}}$ & \\
\cline{4-13}
& 2($\overline{\mathbf{16}}$) & & &  &  & &  &  & & $\mathbf{10}$ & $\mathbf{10}$,$\overline{\mathbf{16}}$ & \\
\hline\hline
    \multirow {2}*{68$_2$}    &    \multirow {2}*{($\mathbf{4},\mathbf{16}$)} &\multirow {2}*{$X$} &    & ($\mathbf{1},\mathbf{10}$) & ($\mathbf{1},\mathbf{10}$) & ($\mathbf{1},\overline{\mathbf{16}}$)& ($\mathbf{1},\mathbf{10}$) & ($\mathbf{1},\mathbf{10}$) & &  &  & \\
\cline{4-13}
& & & ($\mathbf{1},\mathbf{16}$) & ($\mathbf{1},\mathbf{10}$) & ($\mathbf{1},\mathbf{10}$) & & & ($\mathbf{1},\mathbf{10}$) & &  & & \\
\hline\hline
    \multirow {2}*{69}    &    2($\mathbf{10}$), 2($\mathbf{16}$),&\multirow {2}*{$\checkmark2$}     &  &  &  &  &  &  & &  &  & \\
\cline{4-13}
&    2($\overline{\mathbf{16}}$)   & & &  &  &  &  &  & &  &  & $\mathbf{10}$ \\
\hline \hline
    \multirow {2}*{70}    &    2($\mathbf{10}$), 2($\mathbf{16}$),& \multirow {2}*{$\checkmark2$} &  &  &  &  &  &  & &  &  & \\
    \cline{4-13}
&    2($\overline{\mathbf{16}}$) &  &  &  &  &  &  &  & &  &  & \\
\hline\hline
    \multirow {2}*{73}    &    ($\mathbf{1},\mathbf{16}$), ($\mathbf{1},\overline{\mathbf{16}}$),&\multirow {2}*{$\checkmark$}  & &  &  & &  & ($\mathbf{1},\mathbf{10}$) &  &  &  & \\
\cline{4-13}
& ($\mathbf{2},\mathbf{10}$), ($\mathbf{2},\overline{\mathbf{16}}$)& &  &  & & & ($\mathbf{1},\mathbf{10}$) & ($\mathbf{1},\mathbf{10}$) & &  &  & \\
\hline
\end{tabular}   }
\end{table}
\begin{table}[H]
\centering 
\tiny{
\renewcommand{\arraystretch}{1.4}
\begin{tabular}{|c|c|c|cccccccccc|} 
\hline 
Model & Untwisted & GTU &(0,3)  & \multicolumn{2}{c}{(0,2)} & (0,1) & \multicolumn{2}{c}{(1,0)} & (1,3) & \multicolumn{2}{c}{(1,2)} & (1,1) \\
\hline\hline
    \multirow {2}*{74}    &    ($\mathbf{2},\mathbf{1},\mathbf{16}$)&\multirow {2}*{$X$}  &  & ($\mathbf{1},\mathbf{1},\mathbf{10}$) & ($\mathbf{1},\mathbf{1},\mathbf{10}$) & ($\mathbf{1},\mathbf{1},\overline{\mathbf{16}}$)&  &  & &  &  & \\
\cline{4-13}
& ($\mathbf{1},\mathbf{2},\overline{\mathbf{16}}$) & & ($\mathbf{1},\mathbf{1},\mathbf{16}$) & ($\mathbf{1},\mathbf{1},\mathbf{10}$) & ($\mathbf{1},\mathbf{1},\mathbf{10}$) &  &  &  &  &  &  & \\
\hline\hline
    \multirow {2}*{78}    &    \multirow {2}*{4($\mathbf{10}$), 2($\overline{\mathbf{16}}$)} &\multirow {2}*{$\checkmark$} & & $\mathbf{10}$ & $\mathbf{10}$,$\mathbf{16}$ & & $\mathbf{16}$ & $\mathbf{10}$,$\mathbf{16}$ & & $\mathbf{16}$ & $\mathbf{10}$,$\mathbf{16}$ & \\
\cline{4-13}
& &  &  & $\mathbf{10}$ & $\mathbf{10}$,$\mathbf{16}$ & & $\mathbf{10}$ & $\mathbf{10}$,$\mathbf{16}$ & & $\mathbf{10}$ & $\mathbf{10}$,$\mathbf{16}$ & \\
\hline\hline
    \multirow {2}*{84}    &    \multirow {2}*{4($\mathbf{10}$), 2($\overline{\mathbf{16}}$)} &\multirow {2}*{$\checkmark$} & & $\mathbf{10}$ & $\mathbf{10}$,$\mathbf{16}$ & &  & & & & & $\overline{\mathbf{16}}$\\
\cline{4-13}
& &  &  & $\mathbf{10}$ & $\mathbf{10}$,$\mathbf{16}$   & &  &  & &  &  & $\mathbf{10}$\\
\hline\hline
    \multirow {2}*{86}    &    \multirow {2}*{4($\mathbf{10}$), 2($\overline{\mathbf{16}}$)} &\multirow {2}*{$\checkmark$} & & $\mathbf{10}$ & $\mathbf{10}$,$\overline{\mathbf{16}}$ & &  & & & & & \\
\cline{4-13}
& &  &  & $\mathbf{10}$ & $\mathbf{10}$,$\overline{\mathbf{16}}$   & &  &  & &  &  & \\
\hline\hline
    \multirow {2}*{88}    &    \multirow {2}*{4($\mathbf{10}$), 2($\overline{\mathbf{16}}$)} &\multirow {2}*{$\checkmark$} & & $\mathbf{10}$ & $\mathbf{10}$,$\mathbf{16}$ & &  & & & & & \\
\cline{4-13}
&  & &  & $\mathbf{10}$ & $\mathbf{10}$,$\mathbf{16}$   & &  &  & &  &  & \\
\hline\hline
    \multirow {2}*{96}    &    ($\mathbf{2},\mathbf{2},\mathbf{10}$),($\mathbf{1},\mathbf{2},\mathbf{16}$), &\multirow {2}*{$\checkmark$} &   & ($\mathbf{1},\mathbf{1},\mathbf{10}$) & ($\mathbf{1},\mathbf{1},\mathbf{10}$) & &  &  & &  &  & \\
\cline{4-13}
& ($\mathbf{2},\mathbf{1},\mathbf{16}$)  &&  & ($\mathbf{1},\mathbf{1},\mathbf{10}$) & ($\mathbf{1},\mathbf{1},\mathbf{10}$) & &  &  & & &  &  ($\mathbf{1},\mathbf{1},\mathbf{16}$) \\
\hline\hline
    \multirow {2}*{97}    &    ($\mathbf{2},\mathbf{2},\mathbf{10}$),($\mathbf{1},\mathbf{2},\mathbf{16}$), &\multirow {2}*{$\checkmark$} &   & ($\mathbf{1},\mathbf{1},\mathbf{10}$) & ($\mathbf{1},\mathbf{1},\mathbf{10}$) & &  &  & &  &  & \\
\cline{4-13}
& ($\mathbf{2},\mathbf{1},\mathbf{16}$)  & &  & ($\mathbf{1},\mathbf{1},\mathbf{10}$) & ($\mathbf{1},\mathbf{1},\mathbf{10}$) & &  &  & & &  &  \\
\hline\hline
    \multirow {2}*{107}    &    \multirow {2}*{4($\mathbf{10}$), 2($\overline{\mathbf{16}}$)} &\multirow {2}*{$\checkmark$}  & &  &  & $\mathbf{10}$ & $\mathbf{16}$ & $\mathbf{10}$,$\mathbf{16}$ & & & & \\
\cline{4-13}
& &  & $\mathbf{10}$ &  &    & & $\mathbf{10}$ & $\mathbf{10}$,$\mathbf{16}$  & &  &  & \\
\hline\hline
    \multirow {2}*{108}    &    \multirow {2}*{4($\mathbf{10}$), 2($\overline{\mathbf{16}}$)} &\multirow {2}*{$\checkmark$} & &  &  & $\mathbf{10}$ &  &  & & $\mathbf{10}$ & $\mathbf{10}$,$\mathbf{16}$ & \\
\cline{4-13}
&  &  & $\mathbf{10}$ &  &    & &  &   & &  $\mathbf{16}$ & $\mathbf{10}$,$\mathbf{16}$  & \\
\hline\hline
    \multirow {2}*{109}    &    \multirow {2}*{4($\mathbf{10}$), 2($\overline{\mathbf{16}}$)} &\multirow {2}*{$\checkmark$}  & &  &  & $\mathbf{10}$ &  &  & &  &  & \\
\cline{4-13}
&   & & $\mathbf{10}$ &  &    & &  &   & &   &   & \\
\hline\hline
    \multirow {2}*{129}    &    \multirow {2}*{($\mathbf{4},\overline{\mathbf{16}}$)} & \multirow {2}*{$X$} &     &  & ($\mathbf{1},\mathbf{10}$) & &  &  & & ($\mathbf{1},\mathbf{10}$) & ($\mathbf{1},\mathbf{10}$) & \\
\cline{4-13}
& &  &  &  & ($\mathbf{1},\mathbf{10}$) & &  &  & &  & ($\mathbf{1},\mathbf{10}$) & \\
\hline\hline
    \multirow {2}*{130}    &    ($\mathbf{2},\mathbf{1},\mathbf{16}$) &\multirow {2}*{$X$} &  &  & ($\mathbf{1},\mathbf{1},\mathbf{10}$) & &  &  & &  &  & ($\mathbf{1},\mathbf{1},\mathbf{10}$)\\
\cline{4-13}
& ($\mathbf{1},\mathbf{2},\overline{\mathbf{16}}$)&  &  &  & ($\mathbf{1},\mathbf{1},\mathbf{10}$) &  &  &  &  &  &  & \\
\hline\hline
    \multirow {2}*{131}    &    ($\mathbf{2},\mathbf{1},\mathbf{16}$)&\multirow {2}*{$X$}  &  &  & ($\mathbf{1},\mathbf{1},\mathbf{10}$) & &  &  & &  &  & \\
\cline{4-13}
& ($\mathbf{1},\mathbf{2},\overline{\mathbf{16}}$)&  &  &  & ($\mathbf{1},\mathbf{1},\mathbf{10}$) &  &  &  &  &  &  & \\
\hline\hline
    \multirow {2}*{139}    &    \multirow {2}*{($\mathbf{4},\mathbf{16}$)} &\multirow {2}*{$X$} &     &  & ($\mathbf{1},\mathbf{10}$) & &  & ($\mathbf{1},\mathbf{10}$) & &  &  & \\
\cline{4-13}
& &  &  &  & ($\mathbf{1},\mathbf{10}$) & & ($\mathbf{1},\mathbf{10}$) & ($\mathbf{1},\mathbf{10}$) & &  &  & \\
\hline
\end{tabular}   }
\end{table}

\begin{table}[H]
\centering 
\caption{Left chiral spectrum for shifts containing an $E_6$ factor. Only the fields charged under \es\ are shown, for each model the splitting in two rows is made to distinguish between the brother models. The check mark indicates that Gauge Top Unification is possible. At the sectors $T_{(0,2)}$, $T_{(1,0)}$ and $T_{(1,2)}$ we put two columns, to display the content of ordinary and special tori separately.}\label{mattere6} \vspace{0.2cm} \vspace{0.2cm}
\tiny{
\renewcommand{\arraystretch}{1.4}
\begin{tabular}{|c|c|c|cccccccccc|} 
\hline  
Model & Untwisted & GTU &(0,3) & \multicolumn{2}{c}{(0,2)} & (0,1) & \multicolumn{2}{c}{(1,0)} & (1,3) & \multicolumn{2}{c}{(1,2)} & (1,1) \\
\hline\hline
\multirow {2}*{1} & \multirow {2}*{$3(\mathbf{27})$, $\overline{\mathbf{27}}$} & \multirow{2}*{$\checkmark$} & $\mathbf{27}$ & $\mathbf{27}$ & $\mathbf{27}$ & $\mathbf{27}$ & $\mathbf{27}$ & $\mathbf{27}$ &  & $\mathbf{27}$ & $\mathbf{27}$ & $\mathbf{27}$ \\
\cline{4-13}
&  & & $\overline{\mathbf{27}}$ & $\mathbf{27}$ & $\mathbf{27}$ & $\overline{\mathbf{27}}$ &  & $\mathbf{27}$ &  &  & $\mathbf{27}$ &  \\
\hline\hline
\multirow {2}*{3} & \multirow {2}*{$3(\mathbf{27})$, $\overline{\mathbf{27}}$} & \multirow{2}*{$\checkmark$} & $\mathbf{27}$ & $\mathbf{27}$ & $\mathbf{27}$ & $\mathbf{27}$ & & &  & &  & \\
\cline{4-13}
&  & & $\overline{\mathbf{27}}$ & $\mathbf{27}$ & $\mathbf{27}$ & $\overline{\mathbf{27}}$ &  & &  &  & &  \\
\hline\hline
\multirow{2}*{10} & \multirow {2}*{$3(\mathbf{27})$, $\overline{\mathbf{27}}$} & \multirow{2}*{$\checkmark$}  &  &  & $\mathbf{27}$ &  & $\mathbf{27}$ & $\mathbf{27}$ &  &  & $\mathbf{27}$ & \\
\cline{4-13}
& & &  &  & $\mathbf{27}$ &  &  & $\mathbf{27}$ &  & $\mathbf{27}$ & $\mathbf{27}$ &  \\
\hline\hline
\multirow {2}*{12} & \multirow {2}*{$\mathbf{27}$, $\overline{\mathbf{27}}$}& \multirow{2}*{$X$} & $\mathbf{27}$ & & & $\mathbf{27}$ & & &  & &  & \\
\cline{4-13}
& & & $\overline{\mathbf{27}}$ & & & $\overline{\mathbf{27}}$ &  & &  &  & &  \\
\hline\hline
\multirow {2}*{14} & \multirow {2}*{$3(\mathbf{27})$, $\overline{\mathbf{27}}$} & \multirow{2}*{$\checkmark$} &  & & $\mathbf{27}$ &  & & &  & &  & \\
\cline{4-13}
& & & & & $\mathbf{27}$ &  &  & &  &  & &  \\
\hline\hline
\multirow {2}*{15} & \multirow {2}*{$3(\mathbf{27})$, $\overline{\mathbf{27}}$}& \multirow{2}*{$\checkmark$} &  & & $\overline{\mathbf{27}}$ &  & & &  & &  & \\
\cline{4-13}
& & & & & $\overline{\mathbf{27}}$ &  &  & &  &  & &  \\
\hline\hline
20 & $\mathbf{27}$, $\overline{\mathbf{27}}$ & $X$ & & & &  & & &  & &  & \\
\hline\hline
21 & $\mathbf{27}$, $\overline{\mathbf{27}}$& $X$ & & & &  & & &  & &  & \\
\hline\hline
\multirow {2}*{28} & \multirow {2}*{$3(\mathbf{27})$, $\overline{\mathbf{27}}$}& \multirow{2}*{$\checkmark$} &  & $\mathbf{27}$ & $\mathbf{27}$ &  & $\mathbf{27}$ & $\mathbf{27}$ &  & $\mathbf{27}$ & $\mathbf{27}$ & \\
\cline{4-13}
& & &  & $\mathbf{27}$ & $\mathbf{27}$ &  &  & $\mathbf{27}$ &  &  & $\mathbf{27}$ &  \\
\hline\hline
\multirow {2}*{29} & \multirow {2}*{$3(\mathbf{27})$, $\overline{\mathbf{27}}$} & \multirow{2}*{$\checkmark$} &  & $\mathbf{27}$ & $\mathbf{27}$ &  &  &  &  &  &  & \\
\cline{4-13}
& & &  & $\mathbf{27}$ & $\mathbf{27}$ &  &  &  &  &  &  & $\mathbf{27}$ \\
\hline\hline
\multirow {2}*{30} & \multirow {2}*{$3(\mathbf{27})$, $\overline{\mathbf{27}}$} & \multirow{2}*{$\checkmark$} &  & $\overline{\mathbf{27}}$ & $\overline{\mathbf{27}}$ &  &  &  &  &  &  & \\
\cline{4-13}
& & &  & $\overline{\mathbf{27}}$ & $\overline{\mathbf{27}}$ &  &  &  &  &  &  &  \\
\hline\hline
\multirow {2}*{31} & \multirow {2}*{$3(\mathbf{27})$, $\overline{\mathbf{27}}$}&\multirow{2}*{$\checkmark$} &  & $\mathbf{27}$ & $\mathbf{27}$ &  &  &  &  &  &  & \\
\cline{4-13}
& & &  & $\mathbf{27}$ & $\mathbf{27}$ &  &  &  &  &  &  &  \\
\hline\hline
\multirow {2}*{38} & \multirow {2}*{$3(\mathbf{27})$, $\overline{\mathbf{27}}$}&\multirow{2}*{$\checkmark$} &  &  &  &  & $\mathbf{27}$ & $\mathbf{27}$ &  &  &  & \\
\cline{4-13}
& & &  &  &  &  &  & $\mathbf{27}$ &  &  &  &  \\
\hline\hline
\multirow {2}*{39} & \multirow {2}*{$3(\mathbf{27})$, $\overline{\mathbf{27}}$}&\multirow{2}*{$\checkmark$} &  &  &  &  &  &  &  & $\mathbf{27}$ & $\mathbf{27}$ & \\
\cline{4-13}
& & &  &  &  &  &  &  &  &  & $\mathbf{27}$ &  \\
\hline\hline
40 & $3(\mathbf{27})$, $\overline{\mathbf{27}}$ & $\checkmark $ &  &  &  &  &  &  &  &  &  & \\
\hline\hline
\multirow {2}*{47} & \multirow {2}*{$(\mathbf{2},\mathbf{27})$}&\multirow{2}*{$X$} & $(\mathbf{1},\overline{\mathbf{27}})$ &  &  &  &  &  &  &  &  & \\
\cline{4-13}
& & &  &  &  &  $(\mathbf{1},\mathbf{27})$ &  &  &  &  &  &  \\
\hline\hline
\multirow {2}*{48} & \multirow {2}*{$\mathbf{27}$, $\overline{\mathbf{27}}$} &\multirow{2}*{$X$} &  &  &  & $\mathbf{27}$ & $\mathbf{27}$ & $\mathbf{27}$ &  &  &  &  \\
\cline{4-13}
& & & $\overline{\mathbf{27}}$ &  &  &  &  & $\mathbf{27}$ &  &  &  &  \\
\hline\hline
\multirow {2}*{49} & \multirow {2}*{$\mathbf{27}$, $\overline{\mathbf{27}}$}&\multirow{2}*{$X$} & & & & $\mathbf{27}$ & & & & & $\overline{\mathbf{27}}$ & \\
\cline{4-13}
& & & $\overline{\mathbf{27}}$ & & & & & & & $\overline{\mathbf{27}}$  & $\overline{\mathbf{27}}$ & \\
\hline\hline
\multirow {2}*{50} & \multirow {2}*{$\mathbf{27}$, $\overline{\mathbf{27}}$} &\multirow{2}*{$X$} & & & & $\mathbf{27}$ & & & & &  & \\
\cline{4-13}
& & & $\overline{\mathbf{27}}$ & & & & & & & & & \\
\hline\hline
56 & \multirow {2}*{$(\mathbf{2},\overline{\mathbf{27}})$}&\multirow{2}*{$X$} & $(\mathbf{1},\overline{\mathbf{27}})$ & & & & & & & &  & \\
\cline{4-13}
& & & & & & $(\mathbf{1},\mathbf{27})$ & & & & & & \\
\hline\hline
61 & $(\mathbf{2},\mathbf{27})$ & $X$  & & & & & & & & &  & \\
\hline\hline
\multirow {2}*{62} & \multirow {2}*{$\mathbf{27}$, $\overline{\mathbf{27}}$} &\multirow{2}*{$X$} & & & & & $\mathbf{27}$ & $\mathbf{27}$ & & &  & \\
\cline{4-13}
&  & & & & & & & $\mathbf{27}$ & & & & \\
\hline\hline
\multirow {2}*{63} & \multirow {2}*{$\mathbf{27}$, $\overline{\mathbf{27}}$}& \multirow{2}*{$X$} & & & & & & & & $\overline{\mathbf{27}}$ & $\overline{\mathbf{27}}$ & \\
\cline{4-13}
& & & & & & & & & & & $\overline{\mathbf{27}}$ & \\
\hline\hline
64 & $\mathbf{27}$, $\overline{\mathbf{27}}$& $X$ & & & & & & & & & & \\
\hline\hline
65 & $\mathbf{27}$, $\overline{\mathbf{27}}$ & $X$  & & & & & & & & & & \\
\hline\hline
72 & $(\mathbf{2},\overline{\mathbf{27}})$ & $X$ & & & & & & & & & & \\
\hline\hline
\end{tabular}   }
\end{table}

\vskip 1cm
\bibliographystyle{paper}
\small{
\providecommand{\href}[2]{#2}\begingroup\raggedright

\endgroup}

\end{document}